\documentclass[12pt]{article}

\usepackage{color}
\usepackage{amsmath}
\usepackage{amsfonts}
\usepackage{amssymb}
\usepackage{caption}
\usepackage{graphicx}
\usepackage{slashed}            % for slashed characters in math mode
\usepackage{subfig}             % for sub figures
\usepackage{xspace}				% For spacing after commands
\usepackage{cite}

\usepackage[english]{babel}
\usepackage{fancyhdr}
\usepackage{amsmath}
\usepackage{amssymb}
\usepackage{amsfonts}
\usepackage{psfrag}
\usepackage[applemac]{inputenc}
\usepackage{graphicx}
\newcommand{\arXivold}[1]{\href{http://arxiv.org/pdf/#1}{{\tt #1}}}

\renewcommand{\tilde}{\widetilde} % dinky tildes look silly

\newcommand{\beq}{\begin{eqnarray}}
\newcommand{\eeq}{\end{eqnarray}}

\newcommand{\bag}{\begin{align}}
\newcommand{\eag}{\end{align}}

\usepackage[hypertexnames=false]{hyperref}		% kills links in index (if exists)

\begin{document}

\begin{titlepage}
\begin{flushright}
\bf{KIAS-P17028}
\end{flushright}
\vskip.5cm

\begin{center} %TITLE HERE
{\huge \bf{A Perturbative RS I Cosmological Phase Transition}}
\vspace*{0.3cm}

\end{center}

\begin{center} % AUTHORS HERE
{\bf \ Don Bunk$^a$, Jay Hubisz$^b$,  Bithika Jain$^c$} 
\end{center}
%\vskip 4pt

\begin{center} 
% PLACES HERE

$^{a}$ {\it  Department of Physics, Skidmore College, Saratoga Springs, NY  12866} \\
$^{b}$ {\it  Department of Physics, Syracuse University, Syracuse, NY  13244} \\
$^{c}$ {\it School of Physics, Korea Institute for Advanced Study, Seoul 130-722, Korea} \\

\vspace*{0.1cm}

{\tt  

\href{mailto:dbunk@skidmore.edu}{dbunk@skidmore.edu} \href{mailto:jhubisz@syr.edu}{jhubisz@syr.edu},   \href{mailto:bjain@kias.re.kr}{bjain@kias.re.kr}
}

\end{center}

\vglue 0.3truecm

\centerline{\large\bf Abstract}
\begin{quote}
We identify a class of Randall-Sundrum type models with a successful first order cosmological phase transition during which a 5D dual of approximate conformal symmetry is spontaneously broken.  Our focus is on soft-wall models that naturally realize a light radion/dilaton and suppressed dynamical contribution to the cosmological constant.  We discuss phenomenology of the phase transition after developing a theoretical and numerical analysis of these models both at zero and finite temperature.  We demonstrate a model with a TeV-Planck hierarchy and with a successful cosmological phase transition where the UV value of the curvature corresponds, via AdS/CFT, to an $N$ of $20$, where 5D gravity is expected to be firmly in the perturbative regime. 

\end{quote}

\end{titlepage}

%\tableofcontents
%\newpage

\setcounter{equation}{0}
\setcounter{footnote}{0}
%%%%%%%%%%%%%%%%%%%%%%%%%%%%%%%%%%%%%%%%%%%%%%%%%%
%%%%%%%%%%%%%%%%%%%%%%%%%%%%%%%%%%%%%%%%%%%%%%%%%%
%%%%%%%%%%%%%%%%%%%%%%%%%%%%%%%%%%%%%%%%%%%%%%%%%%
%%%%%%%%%%%%%%%%%%%%%%%%%%%%%%%%%%%%%%%%%%%%%%%%%%
%%%%%%%%%%%%%%%%%%%%%%%%%%%%%%%%%%%%%%%%%%%%%%%%%%
%%%%%%%%%%%%%%%%%%%%%%%%%%%%%%%%%%%%%%%%%%%%%%%%%%
%%%%%%%%%%%%%%%%%%%%%%%%%%%%%%%%%%%%%%%%%%%%%%%%%%
%%%%%%%%%%%%%%%%%%%%%%%%%%%%%%%%%%%%%%%%%%%%%%%%%%

\section{Introduction}
%%%%%%%%%%%%%%%%%%%%%%%%%%%%%%%%%%%%%%%%%%%%%%%%%%%%%
%%%%%%%%%%%%%%%%%%%%%%%%%%%%%%%%%%%%%%%%%%%%%%%%%%%%%
Randall-Sundrum (RS) models~\cite{Randall:1999ee} offer an attractive solution to the hierarchy problem, and put the cosmological constant problem~\cite{Weinberg:1988cp} into a new perspective~\cite{Rubakov:1983bz,Sundrum:2003yt}.  In RS models, the warping of AdS space geometrically generates large hierarchies.  Interestingly, the effective 4D value of the cosmological constant is a \emph{sum} of terms involving the bulk 5D cosmological constant, and two brane tensions associated with the UV and IR branes.  The tiny observed value of the cosmological constant is obtained by separately tuning the UV brane tension against the bulk cosmological constant, and the IR brane tension against the same bulk cosmological constant.

This ``double" fine-tuning in the 5D theory is necessary to force a flat direction for the location of the branes, for which the potential would otherwise cause either collapse of the geometry or a run-away.  The Goldberger-Wise stabilization mechanism offers a solution to this tuning problem, with a 5D scalar field developing a spatially varying vacuum expectation value in the bulk of the extra dimension, and leading to a non-trivial potential for the location of the IR brane~\cite{Goldberger:1999uk,Goldberger:1999un}, stabilizing the ``radion."  However, this solution relied upon the mistune in the brane tensions being small to begin with, so that the bulk scalar field vev did not deform the geometry significantly from AdS and that the backreaction of the scalar field on the geometry remained small.  Thus a degree of tuning remained, as naive dimensional analysis (NDA) from consideration of graviton loops suggests that the mistune be parametrically larger, with natural values for a  quartic coupling for the radion being $\lambda \sim {\mathcal O}\left[ (4 \pi)^2 \right]$.

In addition to this naturalness issue, a more phenomenological and pressing problem plagues these models:  The phase transition during which the vacuum expectation value for the radion develops is first order, and estimates of bubble nucleation rates in early universe cosmology strongly suggested that a RS phase transition would not proceed to completion due to Hubble expansion out-pacing true-vacuum bubble creation.  In the region of parameter space where nucleation is fast enough, perturbativity of the 5D gravity theory is right on or past the threshold of being lost~\cite{Creminelli:2001th,Kaplan:2006yi,Randall:2006py}.  In this work, we  address whether a recently studied class of geometries which contain a light radion mode but deviate far from AdS achieve a better transition rate while remaining perturbative.

There is strong motivation for considering such models. In terms of the AdS/CFT correspondence, the double tuning of RS in the absence of a stabilization mechanism has a natural interpretation~\cite{Gubser:1999vj,Verlinde:1999fy,Rattazzi:2000hs,ArkaniHamed:2000ds,Csaki:1999mp}.  The tuning of the UV brane tension against the bulk cosmological constant is viewed as a tuning of the bare cosmological constant in a non-supersymmetric CFT very close to zero.  This is required as a cosmological constant term would explicitly break conformal invariance, yet there is no supersymmetry to enforce this cancellation.  The second tuning of the IR brane tension is interpreted as a tuning associated with the scale invariant quartic associated with an order parameter associated with spontaneous breaking of the CFT.  The flat direction for the ``radion" degree of freedom in RS appears as a tuning of this allowed parameter in the CFT to zero.  If non-zero, such a quartic coupling would forbid the generation of a condensate that spontaneously breaks the CFT~\cite{Fubini:1976jm}.  A solution to this problem appears if one allows a deformation of the CFT, i.e. by the introduction of a near-marginal operator.   This small scale dependence effectively deforms the scale invariant quartic into a more generic potential that may have non-trivial minima away from the origin.   The Goldberger-Wise stabilization mechanism is a dual to this prescription, but, as noted above, has tuning issues as well as cosmology problems.

It had long been thought that this fine tuning is unavoidable, as it reflects a coincidence problem in the 4D CFT dual - a flat direction in the theory that happens to coincide with a near-zero in the $\beta$ functions for the theory~\cite{Holdom,Kutasov:2011fr,Bellazzini:2012vz}.  However, it has been shown that if the scalar potential has only a soft dependence on $\phi$, with the coefficients of the higher order interaction terms in the GW bulk potential being small, then the scalar field enters a significant back-reaction regime before the higher curvature terms come to dominate and perturbative control is lost.  It has been shown that despite this large back-reaction, the dual theory is still conformal, and there is still a light dilaton that realizes scale invariance non-linearly.  These ``soft-wall" scenarios are the models that are of interest in this work.  This ansatz for this type of bulk scalar potential is equivalent, via the AdS/CFT dictionary, to having a beta function in the CFT that remains small for a large range of the coupling.  With this type of presumed dynamics, the coupling explores a large range of values during the running, and the scale invariant quartic could potentially find a zero, essentially finding a flat direction dynamically, and permitting a condensate that spontaneously breaks the approximate conformal invariance without fine tuning~\cite{CPR,Bellazzini:2013fga,Coradeschi:2013gda}.  Other holographic studies of this scenario show that the dilaton mass in such models is suppressed relative to the breaking scale, and the cosmological constant is also parametrically suppressed~\cite{Chacko:2012sy,Chacko:2013dra,Megias:2014iwa,Cox:2014zea}. 

In this work, we explore aspects of the cosmological phase transition in these soft-wall models.  Efforts first are focused on a clear exposition of the theory of the dilaton effective potential at vanishing temperature.  We then perform numerical calculations of the zero temperature potential for various ranges of the free parameters.  Next, we study the theory of the model at finite temperature, and again perform detailed numerics of the finite temperature potential.  Finally, we put the above results to work on the problem of the early universe conformal phase transition, finding an enhanced nucleation rate in soft-wall dilaton scenarios, and a phase transition that completes for much smaller (and thus perturbative) values of the curvature corresponding to a larger $N$ dual CFT.  Finally, we comment on the potential for the early universe conformal phase transition to be observed as a stochastic gravitational wave background signal due to the dynamics of bubble collisions.

%%%%%%%%%%%%%%%%%%%%%%%%%%
%%%%%%%%%%%%%%%%%%%%%%%%%%
%%%%%%%%%%%%%%%%%%%%%%%%%%
%%%%%%%%%%%%%%%%%%%%%%%%%%
%%%%%%%%%%%%%%%%%%%%%%%%%%
\section{Zero-Temperature Dilaton Effective Theory}
\label{sec:zeroT}
We consider classical solutions to theories with a real $5D$ scalar field minimally coupled to gravity.  The action we consider has a bulk contribution given by:
\begin{equation}
S_\text{bulk} = \int d^5 x \sqrt{g} \left[ \frac{1}{2}\left(\partial_M \phi \right)^2 -V(\phi) - \frac{1}{2 \kappa^2} R \right]
\end{equation}
where $\kappa^{-2} \equiv 2 M_*^3$, with $M_*$ being the 5D planck scale.~\footnote{The 5D theory is taken to be compactified on an $S_1/Z_2$ orbifold, with branes at the fixed points, and the integral in the action is taken to be over the full circle, including a double copy of the action.}  We consider metric solutions with flat $4D$ slices:
\begin{equation}
ds^2 = e^{-2 A(\tilde{y})} \eta_{\mu\nu} dx^\mu dx^\nu -d\tilde{y}^2
\end{equation}
which can equivalently be expressed in coordinates $y = A (\tilde{y})$ that we find particularly convenient for this work:
\begin{equation}
ds^2 = e^{-2 y}  \eta_{\mu\nu} dx^\mu dx^\nu - \frac{dy^2}{G(y)}
\end{equation}
where $G(y)= \left[ A'(\tilde{y}(y)) \right]^2$.

There are branes at orbifold fixed points taken to reside at $y = y_0$, and $y = y_1$.  The scalar field has brane localized potentials at these points:
\begin{equation}
S_\text{brane} = -\int d^4x \left[ \sqrt{g_\text{ind} (y_0)} V_0 (\phi(y_0)) + \sqrt{g_\text{ind} (y_1)} V_1 (\phi(y_1)) \right].
\end{equation}

Utilizing $\dot{\ }$ to represent derivatives with respect to $y$, the Einstein and scalar field equations can be written as:
\begin{align}
G &= \frac{\frac{-\kappa^2}{6} V(\phi)}{1- \frac{\kappa^2}{12} \dot{\phi}^2} \label{eq:GEOM} \\
\frac{\dot{G}}{G} &= \frac{2 \kappa^2}{3} \dot{\phi}^2 \label{eq:dotGEOM} \\
\ddot{\phi} &= \left( 4 - \frac{1}{2} \frac{\dot{G}}{G} \right) \dot{\phi} + \frac{1}{G} \frac{\partial V}{\partial \phi}.
\end{align}
The Einstein equations can be used to eliminate $G$ in the scalar field equation of motion:
\begin{equation}
\ddot{\phi} = 4 \left( \dot{\phi} - \frac{3}{2\kappa^2} \frac{\partial \log V(\phi)}{\partial \phi} \right) \left( 1 - \frac{\kappa^2}{12} \dot{\phi}^2 \right).
\label{eq:PHIEOM}
\end{equation}
 The total value of the classical action can be expressed as a pure boundary term.   In particular, after substituting for the kinetic and potential terms for $\phi$ using the Einstein field equations, and taking into account contributions from singular terms in the scalar curvature at the orbifold fixed points,  the resulting 4D effective potential is given by~\cite{Bellazzini:2013fga}
 \begin{equation}
 V_\text{eff} = e^{-4 y_0} \left[ V_0 (\phi (y_0)) - \frac{6}{\kappa^2} \sqrt{G(y_0)} \right] + e^{-4 y_1} \left[ V_1 (\phi (y_1)) + \frac{6}{\kappa^2} \sqrt{G(y_1)} \right].
 \end{equation}
Since the effective action is a pure boundary term, the 4D potential depends only on the asymptotic behavior of the geometry and the scalar field.
 
In the next two subsections we discuss the application of these equations first to the case of constant bulk potential $V(\phi) = -\frac{6k^2}{\kappa^2}$, which is a review of previous results in the literature placed in the context of the motivation for this work.  This is then extended to more general potentials that correspond to our weak $\phi$ dependence ansatz where we perform a numerical analysis over a broad range of parameter space.  The constant potential case corresponds via AdS/CFT to an undeformed CFT.  In the section on general potentials, we add a term to the 5D action that corresponds to sourcing a marginally relevant operator that stabilizes the pure scale-invariant dilaton quartic coupling typical for conformal field theories.

\subsection{Example:  Constant Bulk Potential}
The case of constant potential can be solved analytically~\cite{Csaki:2000wz}, and the result for $\phi$ is given by
\begin{equation}
\phi = \phi_0 \pm \frac{1}{4}\sqrt{\frac{12}{\kappa^2}} \log \left[ e^{4 (y - y_c)} \left( 1+ \sqrt{ 1 + e^{8 (y_c - y)} } \right) \right].
\end{equation}
The integration constant $y_c$ is chosen so as to correspond to the value of $y$ for which the behavior of $\phi$ changes qualitatively from $\phi \approx $ constant $= \phi_0$ to a behavior that is linear in $y$:  $\phi \approx \phi_0 \pm \left( \log 2+ \sqrt{\frac{12}{\kappa^2}} (y-y_c) \right)$.  

We can also evaluate the expression for $G(y)$ exactly.  Taking $V = -\frac{6 k^2}{\kappa^2}$, and defining $f \equiv k e^{- y_c}$, and $\mu \equiv k e^{-y}$ we have
\begin{equation}
G = k^2 \left[ 1 + \left( \frac{f}{\mu} \right)^8  \right].
\end{equation}

With the above information we can extract the dilaton potential.  We take the ``stiff wall" limit where we presume that the boundary potential fixes $\phi$ at particular values on the branes:  $\phi(y_0) = \phi_\text{UV}$ and $\phi(y_1) = \phi_\text{IR}$.  We also take the potentials in this limit to be pure tensions:  $V_0 (\phi_\text{UV}) = \Lambda_0$ and $V_1 (\phi_\text{IR}) = \Lambda_1$.  Defining $\mu_0 = k e^{-y_0}$, $\mu_1 = k e^{-y_1}$ we have:
\begin{align}
 V_\text{eff} &= \left(\frac{\mu_0}{k}\right)^4 \left[ \Lambda_0  - \frac{6k}{\kappa^2} \sqrt{1+ \left( \frac{f}{\mu_0}\right)^8} \right] \nonumber \\
 &+ \left(\frac{\mu_1}{k}\right)^4 \left[ \Lambda_1+ \frac{6k}{\kappa^2} \sqrt{1+ \left( \frac{f}{\mu_1}\right)^8} \right].
\end{align}

The scalar boundary conditions determine a combination of the free parameters $\mu_0$, $\mu_1$, $f$, and $\phi_0$.  We fix $\mu_0$ by matching the 4D observed Planck scale, and for the purposes of this discussion, we will hold $f$ fixed.  The UV boundary condition essentially sets $\phi_0 \approx \phi_\text{UV}$, up to terms of order $(f/\mu_0)^4$.  The IR boundary condition fixes the ratio 
\begin{equation}
\left( \frac{f}{\mu_1} \right)^4 = \frac{1}{2} \exp \left[ \sqrt{\frac{4 \kappa^2}{3}} \left( \phi_\text{IR} - \phi_\text{UV} \right) \right],
\end{equation}
so that $\mu_1$ might be replaced by a function of $f$ in the expression for the effective potential.

Neglecting terms of order $\left( f/\mu_0 \right)^8$ induced by the explicit breaking of conformal invariance associated with sourcing 4D gravity at the scale $\mu_0$, the effective potential as a function of $f$ can be written as
\begin{equation}
 V_\text{eff} \approx \mu_0^4 \left( \frac{\Lambda_0}{k^4}  - \frac{6}{\kappa^2k^3} \right)  +f^4 \left( \frac{2\Lambda_1}{k^4} \exp \left[ -\sqrt{\frac{4 \kappa^2}{3}} \left( \phi_\text{IR} - \phi_\text{UV} \right) \right] + \frac{6}{\kappa^2 k^3}   \right).
\end{equation}
The first term in this expression is the contribution to the bare cosmological constant.  This is expected to be either tuned to zero by choosing $\Lambda_0 = \frac{6k}{\kappa^2}$, or made vanishing by the introduction of additional UV symmetries such as supersymmetry.  The second is the contribution to the cosmological constant via the spontaneous breaking of conformal symmetry, or in other words, the dilaton quartic.   As before, the quartic is a sum of two terms, one from the IR brane tension and the other from the bulk cosmological constant.  The former is suppressed by the hierarchy in $\phi$, such that for very large separation between $\phi_\text{UV}$ and $\phi_\text{IR}$, a very large negative value of $\Lambda_1$ is required to cancel the positive contribution from the bulk geometry.~\footnote{As the final value of the total vacuum energy is non-zero in this procedure, our assumption of a static solution is only an approximation, and a non-trivial cosmology would typically be produced~\cite{Nihei:1999mt,Kaloper:1999sm}.  We neglect this effect here, assuming that the bare term is adjusted so that the final vacuum energy is vanishing.}

Note that we expect higher curvature operators that are expected to be induced by quantum corrections to give contributions to the dilaton potential, but the form is still be that of a scale invariant quartic plus derivative terms unless a non-trivial scalar potential is included.

\begin{figure}[!htbp]
	\center
\includegraphics[width=0.6\textwidth]{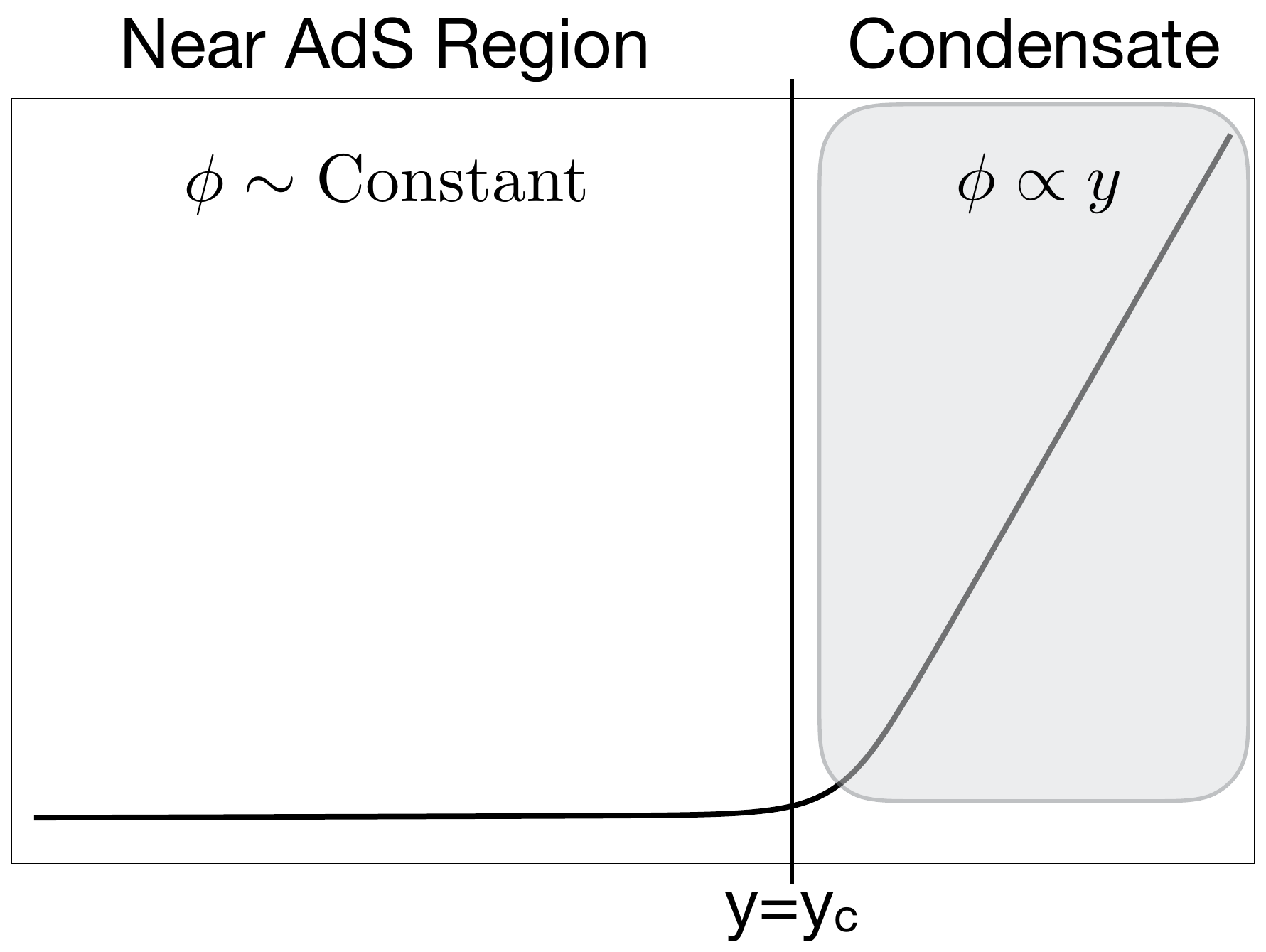}
\label{fig:phiplt}
\caption{This cartoon shows the evolution of the field $\phi$.  It begins in a region where $\phi$ is nearly constant, and the geometry is nearly AdS.  At a critical value of the coordinate, $y = y_c$ which is close in proper distance to a curvature singularity, $\phi$ then begins linear evolution, as shown and the curvature quickly grows large.  The gap of the theory is set by the position of this ``soft-wall."}
\label{fig:phiplot}
\end{figure}

The interpretation of this result is that even when the IR brane does not play a major role, there is the notion of a breaking scale of conformal symmetry given by $f \equiv k e^{-y_c}$.  This scale corresponds in 5D to a position in the extra dimension at which the leading behavior of the curvature, or equivalently the scalar field evolution, makes a transition from one type of behavior to another.  In Figure~\ref{fig:phiplot}, 
we show schematically the behavior of the scalar field evolution along the $y$ coordinate.  The behavior begins with slow evolution where the geometry is nearly AdS and then transitions to linear behavior where the curvature becomes large.   Without an IR brane, a singularity at finite proper distance from the UV brane terminates the geometry.  We term this region in $y$ where $\phi$ is linear the ``condensate" region.  The effective potential for $f$ is precisely what is expected for an approximately conformal theory with explicit breaking manifest in the form of a bare CC, and from the introduction of the Planck brane itself, making the position of this turnover of the 5D behavior of the scalar-gravity background a candidate for the dilaton.  

Note that this ``soft-wall dilaton" is unstable, just as in the original RS I model.  Unless $\Lambda_1$ is large and negative quartic coupling is large and positive, driving $f$ to zero in the absence of a stabilization mechanism.   This means that the effective potential is minimized when the conformal symmetry is unbroken.  Alternatively $\Lambda_1$ can be tuned to make the quartic vanish, and give the dilaton a flat direction.   Further, the ansatz of flat 4D metric slices is only valid in the case that the total cosmological constant vanishes, or when all terms in the effective potential are arranged so as to exactly cancel each other.  The tuning of the bare CC, and the tuning of the dilaton quartic are precisely the two tunings that are required in two-brane RS models. 

That there is a lack of stability of the constant potential case with soft wall breaking of conformal symmetry comes as little surprise.  Typical conformal theories without supersymmetry do not support spontaneous conformal breaking due to the presence of the scale invariant quartic (in other words, the lack of scalar flat directions).~\footnote{In superconformal theories, while there are potentially vacua that spontaneously break conformality, they are degenerate along supersymmetric flat directions, and there is no unique vacuum for the theory.  This is also the case of the original un-stabilized RS model, where tuning of the brane tensions is necessary to create a static geometry, but then there is no dynamical selection of the inter-brane separation.} A deformation of the CFT, or in other words a departure from conformality, is required to stabilize a VEV against the scale invariant quartic.  In the next two sub-sections, we demonstrate how deformations of the CFT (introduced in the AdS dual by considering a nontrivial bulk scalar potential) can stabilize the soft-wall dilaton.

%%%%%%%%%%%%%%%%%%%%%%%%%%%%%%%%%%%%
%%%%%%%%%%%%%%%%%%%%%%%%%%%%%%%%%%%%
%%%%%%%%%%%%%%%%%%%%%%%%%%%%%%%%%%%%
%%%%%%%%%%%%%%%%%%%%%%%%%%%%%%%%%%%%
%%%%%%%%%%%%%%%%%%%%%%%%%%%%%%%%%%%%
\subsection{Non-Constant Bulk Potentials}
To stabilize the dilaton, we consider adding a deformation to the bulk scalar potential.  For example, a mass term for $\phi$ could be added:
\begin{equation}
V(\phi) = - \frac{6 k^2}{\kappa^2} \left( 1 - \frac{\kappa^2}{3} \epsilon \phi^2 \right)
\label{eq:bulkpot}
\end{equation}
Note that $\epsilon$ is defined to be dimensionless.  The non-zero mass term for $\phi$ corresponds via AdS/CFT to a non-trivial quantum scaling dimension for the CFT operator that maps to $\phi$ via the duality~\cite{Witten:1998qj}.  If $\phi$ takes a non-trivial value on the boundary of AdS, then this operator is sourced in the dual (approximate) CFT, contributing as a small explicit violation of conformal invariance.  In the constant potential limit, this operator is precisely marginal, and does not deform the CFT.  When $\epsilon$ is negative, which is the ansatz we will take in our work, the scalar field is tachyonic, and tends to grow with increasing $y$.  This is dual to sourcing a near-marginal relevant deformation of the CFT.  This is somewhat similar to what occurs in QCD or technicolor-like theories, although in that case conformal invariance is badly broken in the infrared when the coupling becomes strong.  We make the assumption that the $\phi$-dependent terms in the potential remain small in the region of large $\phi$, or in the approximate CFT dual, that the $\beta$-functions remain small even when the coupling becomes large.  Condensation is triggered not by strong coupling, as happens in QCD, but rather by a coincidence of the coupling constant having a value associated with a near zero in the effective dilaton quartic.  The slow running of the coupling over a large range of coupling values allows the theory to explore the landscape of quartics until a near zero is found and the theory condenses.

In the region of large back-reaction, the equations of motion do not admit analytic solutions to the equations of motion when $\epsilon$ is non-zero.  The equations can be solved approximately using the method of boundary layer matching, however this approximation begins to break down when $\phi$ is large, but its behavior is not yet governed by the IR condensate asymptotics.  Due to these difficulties, we resort to numerical solutions to the equations of motion to study the behavior of this system.

We presume throughout the rest of this work that the brane localized potentials enforce the stiff wall limit, and the branes are thus localized at positions where the scalar field $\phi$ is equal to $\phi_0$ and $\phi_1$ in the UV and IR, respectively.  We do not expect any aspects of the analysis to change much if this condition is relaxed.  

The UV brane tension is fine tuned to enforce the condition that the cosmological constant vanish in the limit when the conformal breaking scale goes to zero.  This is a tuning of the bare cosmological constant - the remaining cosmological constant at non-zero values of $f$ is due purely to the conformal symmetry breaking condensate.  It is this remaining dynamical contribution which is suppressed by the small value of $\epsilon$.  

The value of $f$ depends then depends on initial conditions for the derivative of $\phi$, with the behavior of $\phi$ in the UV encoding the information regarding the scale of symmetry breaking.  The brane potentials select the value of $\phi$ at the position of the branes, but this derivative of $\phi$ is not fixed.  Varying this derivative is equivalent to varying over the value of the breaking scale, $f$.  

For a given IR brane tension, there is a given value for $f$ at the minimum of the effective potential that we derived above - equivalently, minimization of the action fixes the derivative of $\phi$ at the position of the UV brane. The value for $f$ in terms of the geometry is
\begin{equation}
f^{-1} = \int_{y_0}^{y_1} \frac{e^{y}}{\sqrt{G}} dy
\label{eq:feq}
\end{equation}
We have verified numerically that the masses of resonances obtaining their mass from conformal symmetry breaking (e.g. masses of gauge boson KK modes) track almost exactly linearly with the above definition for $f$, even in cases where the backreaction is very large.  This definition agrees in the limit of vanishing backreaction with the usual definition, $f \approx k e^{-y_1}$, but differs substantially from it in the regions of interest in this study.  

We discuss results in terms of the dimensionless quantity $N$, where $N$ is expressed in terms of the 5D curvature and Planck scale:
\begin{equation}
N^2 = \frac{8 \pi^2}{\kappa^2 k^3}
\end{equation}
For perturbativity of the 5D gravity model, $N$ must be taken to be somewhat large.  Note that the effective 4D Planck scale, $M_\text{Pl}^2 = \frac{N^2}{16 \pi^2} k^2$, should be held fixed, so that a particular value of $N$ corresponds to a given value of $k$.  It is also convenient to work in terms of a rescaled $G$:  $G = k^2 \tilde{G}$.  Finally, we also rescale both of the brane tensions: $\Lambda_{0,1} = 6 \frac{k}{\kappa^2} \tilde{\Lambda}_{0,1}$.  

With these rescaled parameters and functions, and with the expression for $f$ in Eq.~(\ref{eq:feq}), we can express the effective dilaton potential in terms of dimensionless quantities as
\begin{equation}
V_\text{dilaton} = \frac{192 \pi^2}{N^2} M_\text{Pl}^4  \left\{ \left[ \tilde{\Lambda}_0 -\sqrt{\tilde{G}_0} \right] + e^{-4 y_1} \left[ \tilde{\Lambda}_1 + \sqrt{\tilde{G}_1} \right] \right\},
\end{equation}
and we can write
\begin{equation}
f = \frac{4 \pi M_\text{Pl}}{N} \left[ \int_{y_0}^{y_1} \frac{e^{y}}{\sqrt{\tilde{G}}} dy \right]^{-1}.
\end{equation}
For numerical evolution of the scalar equation of motion in a manner that is independent of $N$, we define a dimensionless scalar field $\tilde{\phi} = \kappa \phi$, in which case the scalar equation of motion is
\begin{equation}
\ddot{\tilde{\phi}} = 4 \left( \dot{\tilde{\phi}} - \frac{3}{2} \frac{ \partial \log V(\tilde{\phi}) }{\partial \tilde{\phi}} \right) \left( 1 - \frac{1}{12} \dot{\tilde{\phi}}^2 \right)
\end{equation}
For initial and final values of the scalar field $\phi_0$ and $\phi_1$, we employ a rescaling that is common in the literature: $\phi_{0,1}^2 = k^3 v_{0,1}^2 = \left( \frac{4 \pi M_\text{Pl}}{N} \right)^3 v_{0,1}^2$, where $v_{0,1}$ are the dimensionless values of the field on the branes, set in the stiff wall limit.  We report results as a function of $v_{0,1}$, $N$, $\epsilon$, and $\tilde{\Lambda}_1$.

\begin{figure}[!htbp]
	\center
\includegraphics[width=0.49\textwidth]{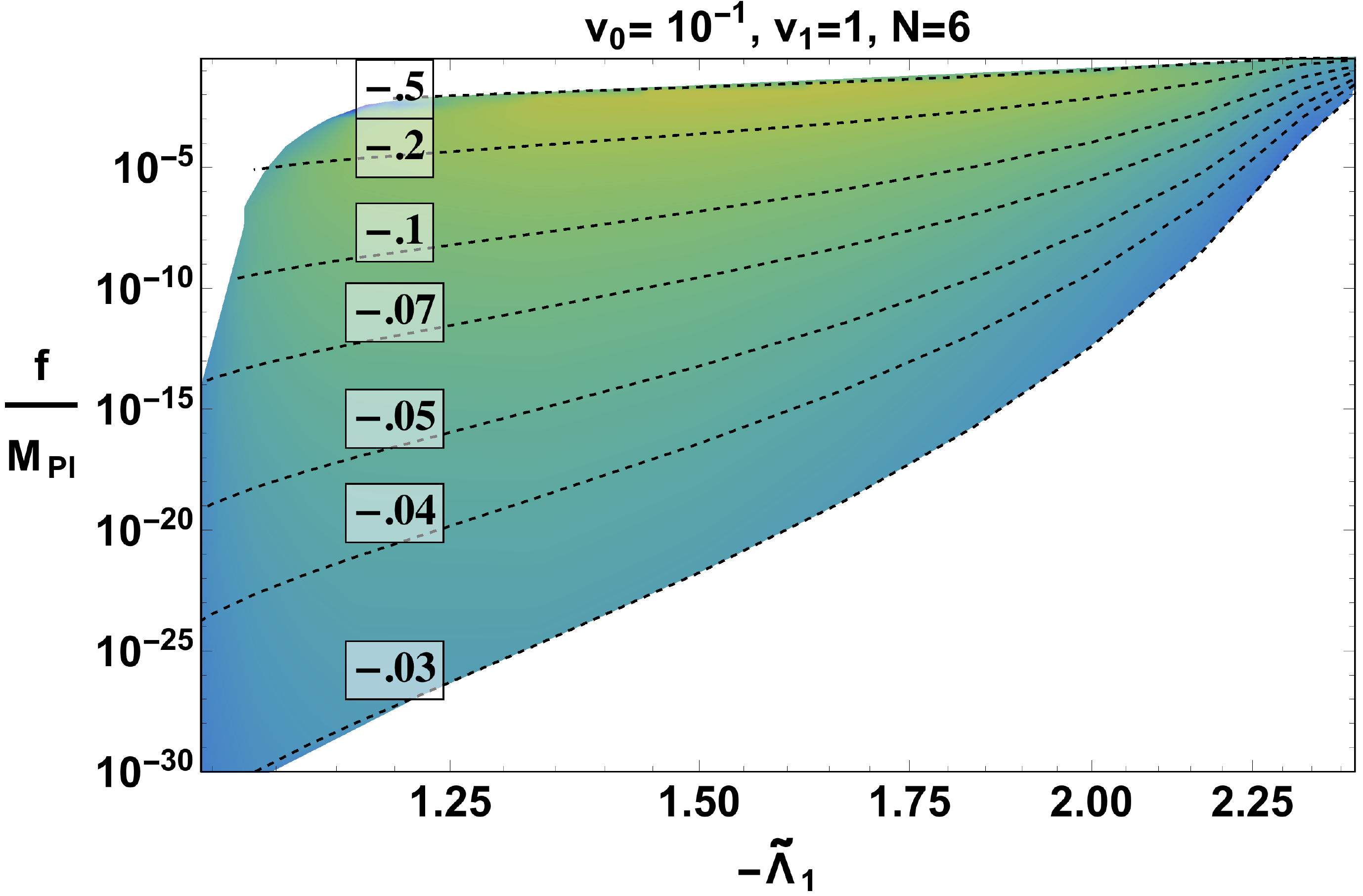}
\includegraphics[width=0.49\textwidth]{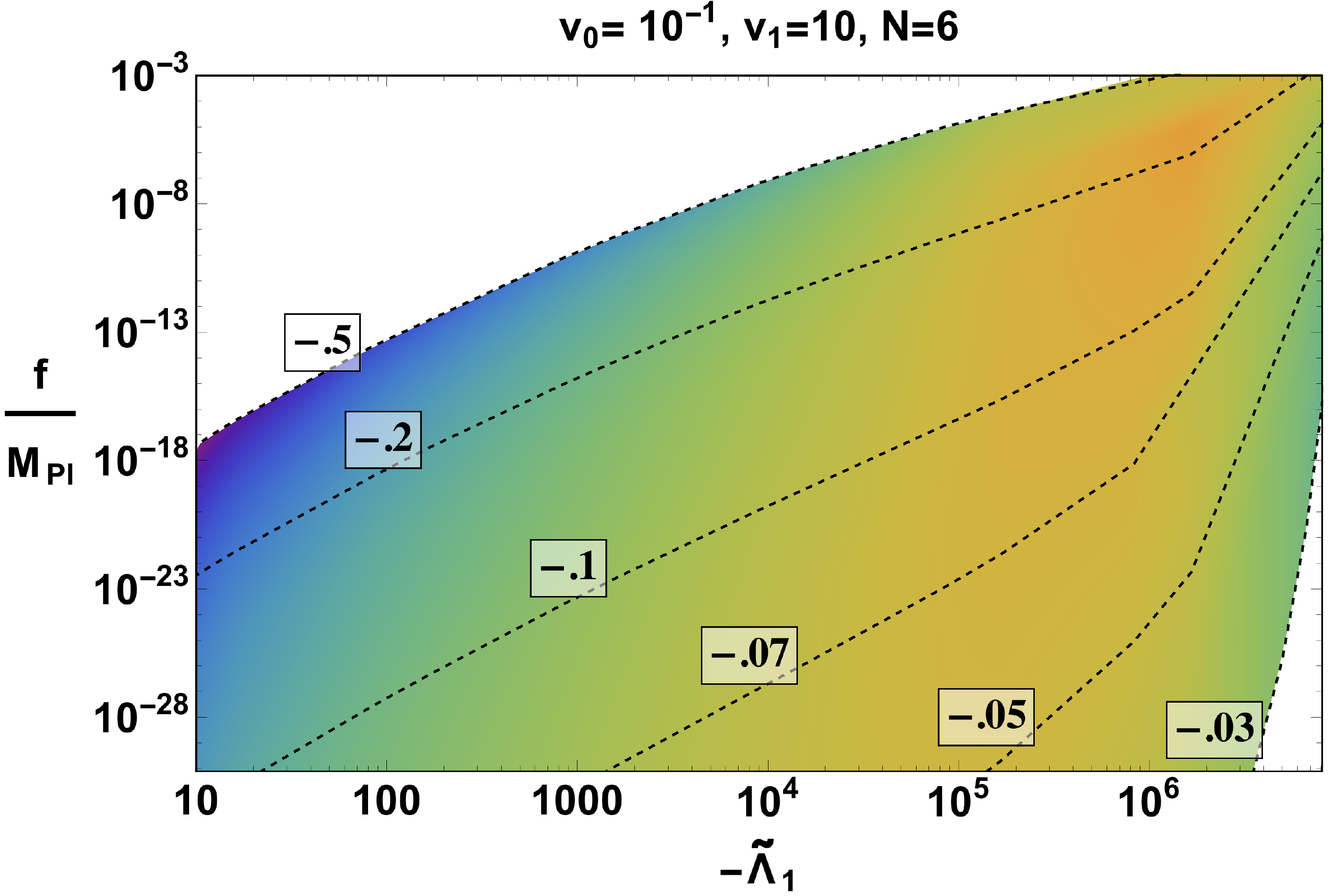}
\includegraphics[width=0.49\textwidth]{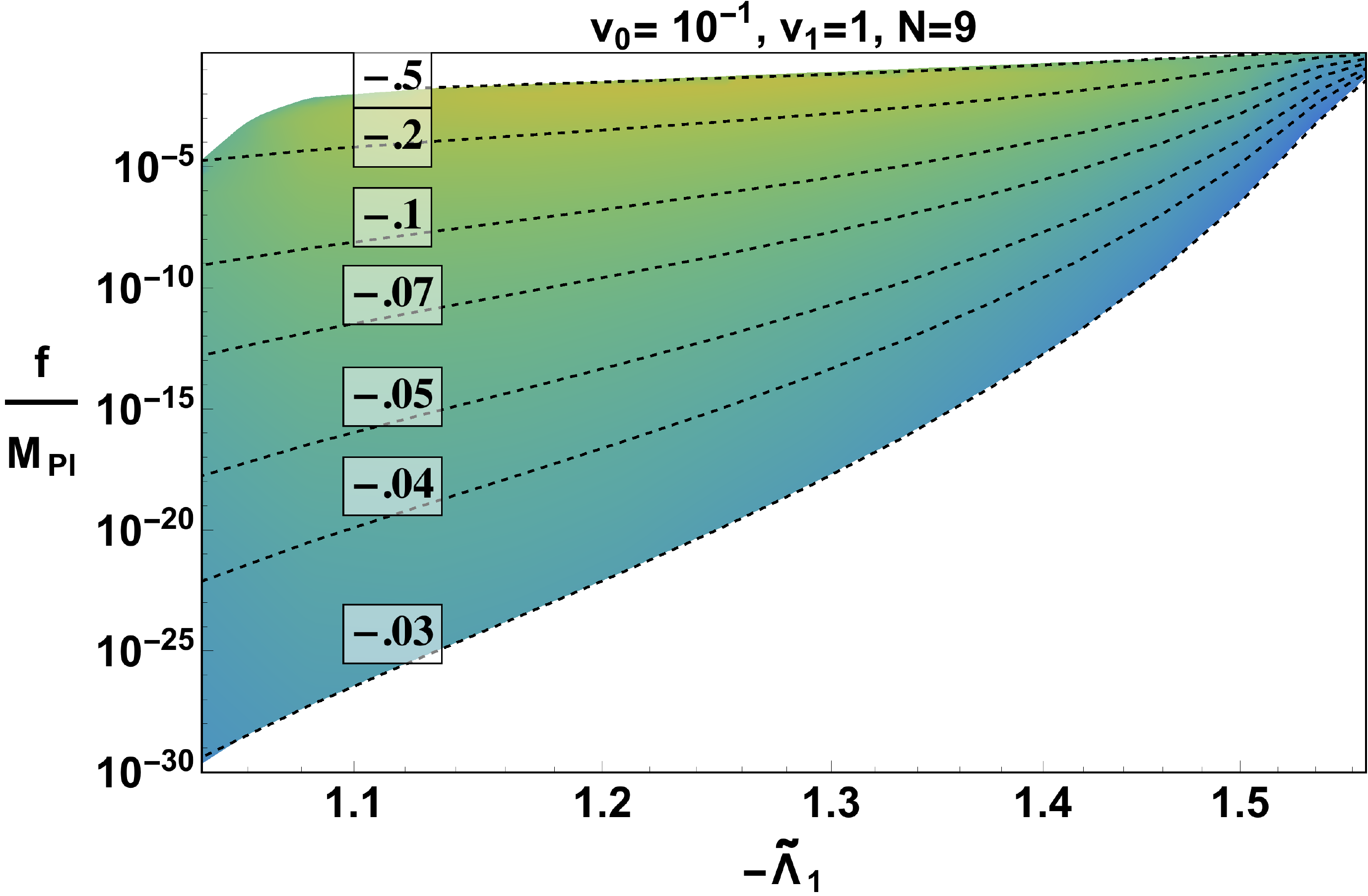}
\includegraphics[width=0.49\textwidth]{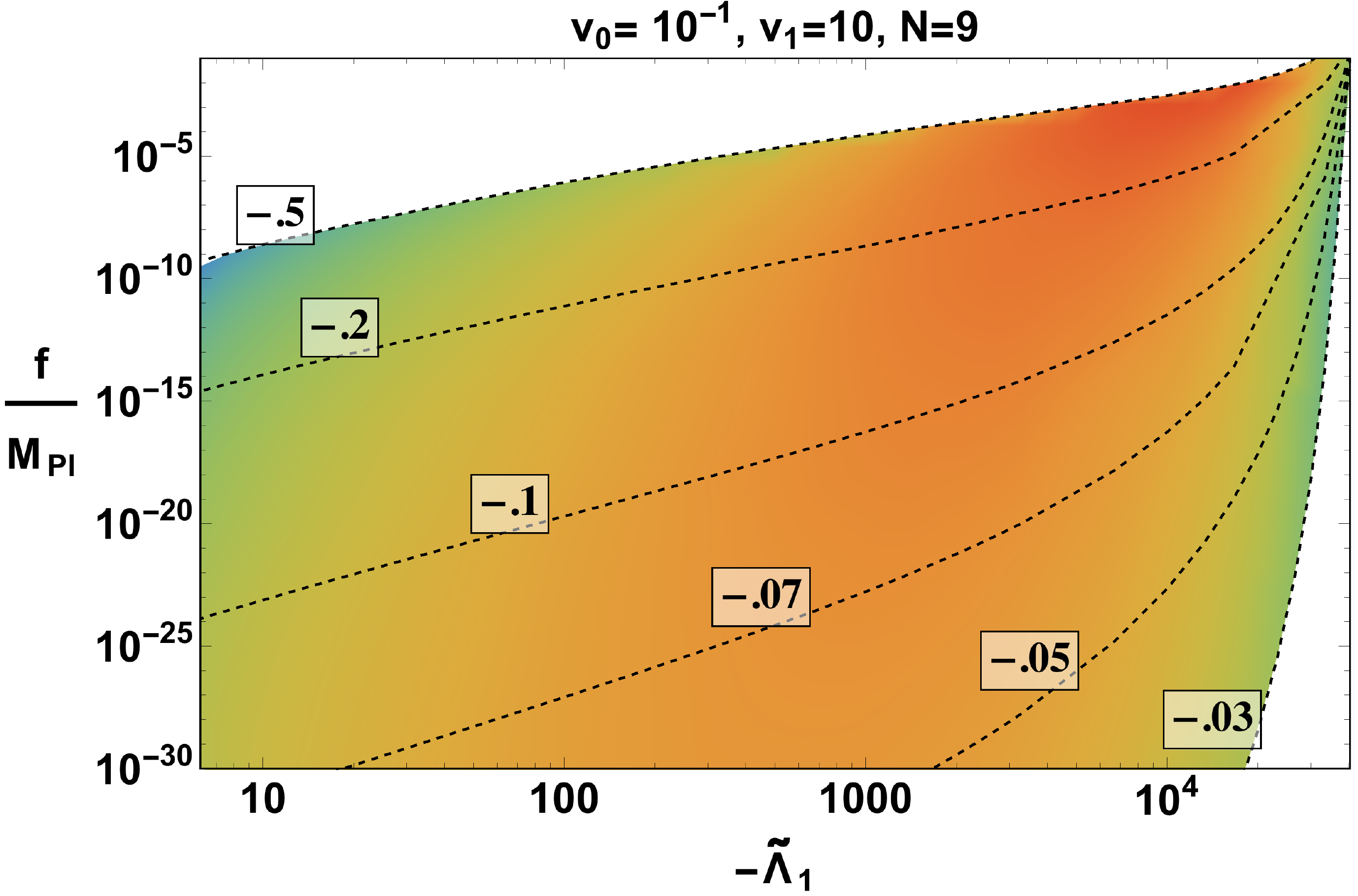}
\includegraphics[width=0.49\textwidth]{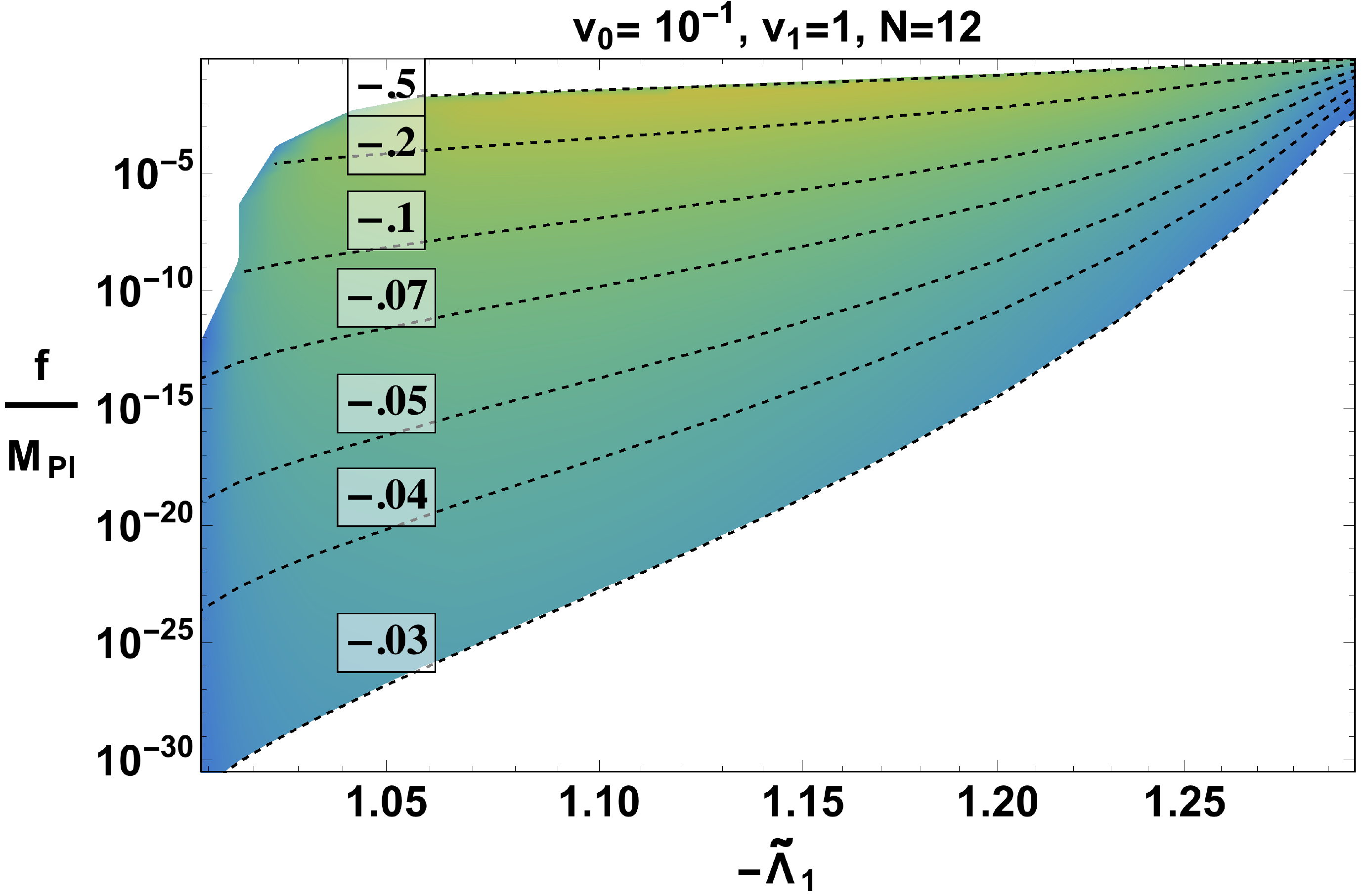}
\includegraphics[width=0.49\textwidth]{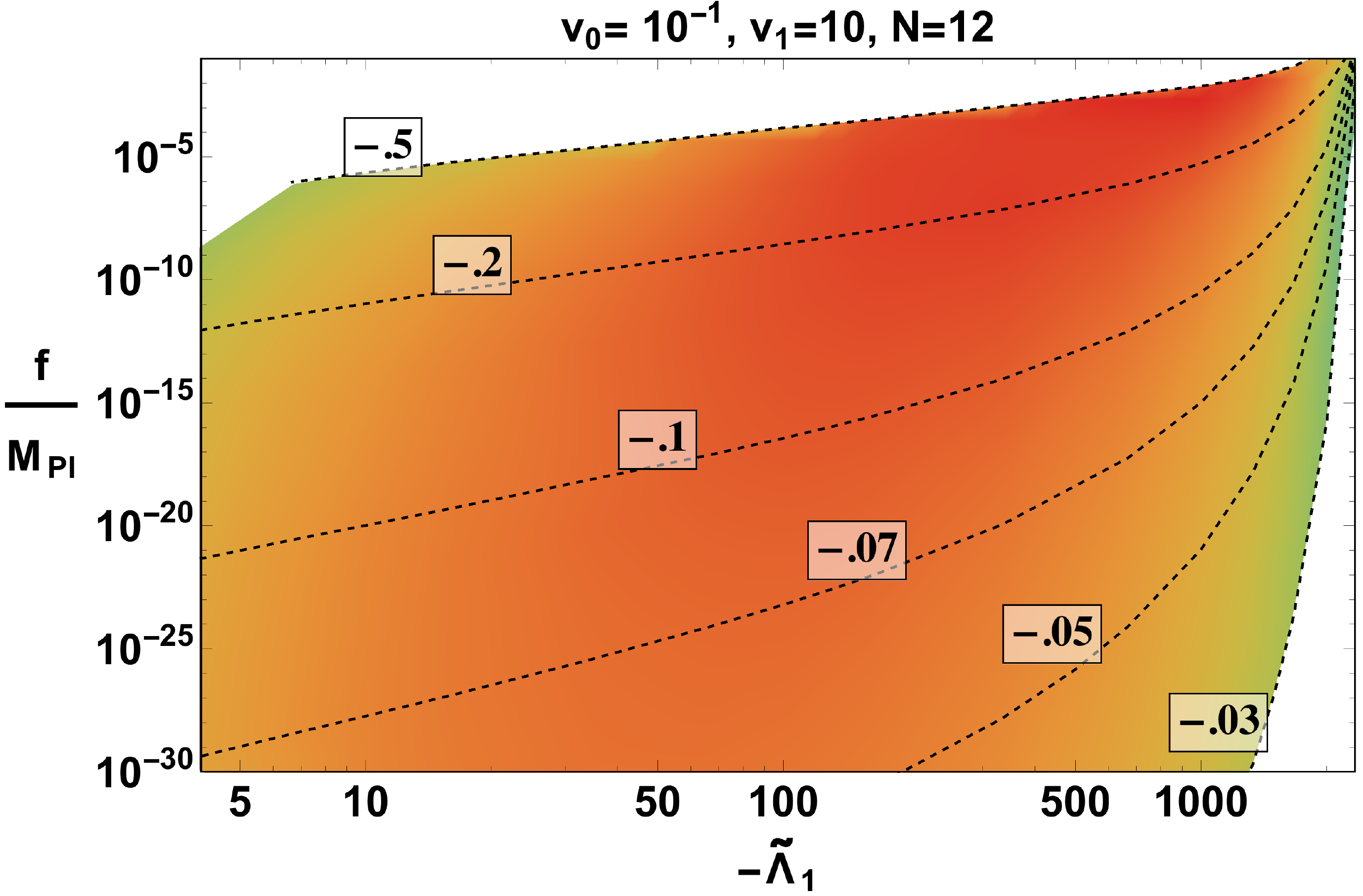}
\includegraphics[width=.9\textwidth]{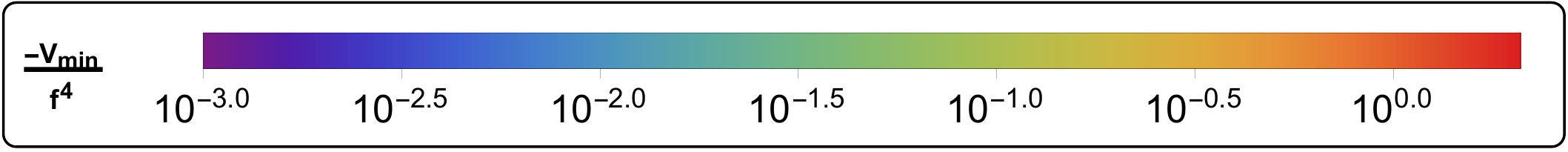}
\caption{These are plots of $f$ vs $\tilde{\Lambda}_1$ for different values of $N$ (the rows correspond to $N=3$, $6$, and $12$), different values of $v_1$ (left plots are $v_1 = 1$, right are $v_1 = 10$), and on each plot, various values of $\epsilon$ correspond to the various curves.  The color shading corresponds to values of $\frac{|V_\text{min}|}{f^4}$.}.
\label{fig:fvsLambda1}
\end{figure}

In Figure~\ref{fig:fvsLambda1} properties of the zero temperature dilaton potential are displayed for various values of the bulk mass term, $\epsilon$, and for various values of the IR brane tension.  The hierarchy between the 4D effective Planck scale and the dilaton vev, $f$, are shown.  In addition,  color shading indicates the value of the dilaton potential at the minimum (effectively the contribution to the cosmological constant from the dilaton vev).  We show in two columns the dependence on the IR brane value of the bulk scalar field.  On the left are plots that correspond to small back-reaction, which is the usual hard wall RS model where $v_1 = 1$.  On the right, we display results for the soft-wall model where $v_1 = 10$, and backreaction in the IR region is large.

The values of $\tilde{\Lambda}_1$ that accomplish breaking of conformal symmetry are much larger than in the hard-wall model for small values of $N$.  One might ask whether this is consistent with estimates arrived at using the tools of naive dimensional analysis.  One must be careful in applying these tools in the soft-wall:  Low mass graviton KK-modes do not penetrate the soft-wall to reach the IR brane, whereas higher mass KK modes have more significant overlap with the IR brane.  This means that the local cutoff associated with violation of unitarity in KK-graviton scattering is much higher as the self coupling of the gravitons is very small in this region.  It is not clear that the values for small $N$ are completely reasonable, but we are most interested anyway in larger values of $N$, corresponding to perturbative 5D gravity models.  As $N$ increases, the values of $\tilde{\Lambda}_1$ rapidly approach much smaller values that are easily consistent with NDA expectations.

%%%%%%%%%%%%%%%%%%%%%%%%%%%%%%%%%%%%
%%%%%%%%%%%%%%%%%%%%%%%%%%%%%%%%%%%%
%%%%%%%%%%%%%%%%%%%%%%%%%%%%%%%%%%%%
%%%%%%%%%%%%%%%%%%%%%%%%%%%%%%%%%%%%
\section{Finite Temperature}
\label{sec:finiteT}

The phase structure of near-conformal 4D theories is of interest both as a theoretical question and one of phenomenology.  If naturalness of the electroweak symmetry breaking sector is due to strong near-conformal dynamics, it is important to study the cosmology of such theories.   Studies of the RS1 phase transition indicate that it is strongly first order, with a critical temperature suppressed in comparison with the value of the condensate.  This is due to the presence of a near flat direction at the minimum of the dilaton potential.  It is this which allows for the light dilaton, and also for a suppression in the contribution of condensates to the effective IR value of the cosmological constant.  At finite temperature, such non-compact flat directions are lifted, sending the dilaton field value to the origin, thus evaporating the condensate.

In order to study the theory at finite temperature, the class of geometries we study is opened up to include the possibility of a horizon (or ``black brane") at some point $y = y_h$ in the 5D coordinate~\cite{Hawking:1982dh,Witten:1998zw,Hebecker:2001nv}.  In AdS space, the Hawking radiation from such a black hole allows the black hole to reach equilibrium with the thermal bath.  The action associated with the classical solution corresponds to the thermodynamical free energy of the system.  The geometry we study has metric function
\begin{equation}
ds^2 = e^{-2y} \left[ h(y) dt^2 + d\vec{x}^2 \right] + \frac{1}{h(y)} \frac{dy^2}{G(y)}.
\end{equation}
The presence of a horizon is associated with a zero in the horizon function $h(y)$ at position $y_h$.  As we are considering a thermal partition function, we work in Euclidean metric signature, with the time coordinate compactified on a circle: $t \in [ 0,1/T )$.  

The equations of motion for the metric functions $h$ and $G$, and for the scalar field $\phi$ are given by
%~\footnote{While it is not of immediately practical value in this work, it seems worth noting that combining the first and third of these equations yields a rather peculiar and suggestive relationship between $h$ and the bulk scalar potential: $\frac{d}{dy}\log V = 8 - \frac{d}{dy}\log h - 2 \frac{d}{dy}\log \frac{d}{dy}\log h+ \frac{d}{dy}\log \frac{d}{dy}\log \frac{d}{dy}\log h$.}
\begin{align}
 \frac{\ddot{h}}{\dot{h}} &= 4 -\frac{1}{2} \frac{\dot{G}}{G} \label{eq:hddot} \\
\frac{\dot{G}}{G} &= \frac{2 \kappa^2}{3} \dot{\phi}^2 \nonumber \\
G &= -\frac{ \frac{\kappa^2}{6} \frac{V(\phi)}{h}}{1 - \frac{1}{4} \frac{\dot{h}}{h} - \frac{\kappa^2}{12}\dot{\phi}^2} \\
\ddot{\phi} &= 4 \left( \dot{\phi} - \frac{3}{2\kappa^2} \frac{\partial \log V}{\partial \phi} \right) \left( 1-\frac{1}{4} \frac{\dot{h}}{h} -\frac{\kappa^2}{12} \dot{\phi}^2\right).
\end{align}
The effective potential is still given by a pure boundary term, although the singular terms due to orbifold boundary conditions at a putative black hole horizon require special treatment, as we discuss later in this section.  

The bulk contribution to the effective potential arises from using the equations of motion to express the bulk action as a total 5-derivative:
\begin{equation}
V_\text{bulk} = -\frac{2}{\kappa^2} \int d y \partial_5 \left[ e^{-4y} \sqrt{G} h \right] = \frac{2}{\kappa^2} \left [e^{-4 y_0} h(y_0) \sqrt{G(y_0)} - e^{-4 y_1} h(y_1) \sqrt{G(y_1)} h(y_1) \right].
\end{equation}
The curvature tensor has singularities at the orbifold fixed points that give additional contributions to the effective action.  Integrating the action over these singularities at the UV and IR branes gives the following contribution to the effective potential:
\begin{equation}
V_\text{sing} = - \frac{1}{\kappa^2} \left[  e^{-4 y_0} \sqrt{G(y_0)} \left(  8 h(y_0)  - \dot{h}(y_0) \right) - e^{-4 y_1} \sqrt{G(y_1)} \left(  8 h(y_1)  - \dot{h}(y_1) \right) \right]
\end{equation}
Note that the equation of motion for $h$ enforces an exact cancellation between the two $\dot{h}$ terms.

\begin{figure}[!htbp]
	\center
\includegraphics[angle=90,width=0.8\textwidth]{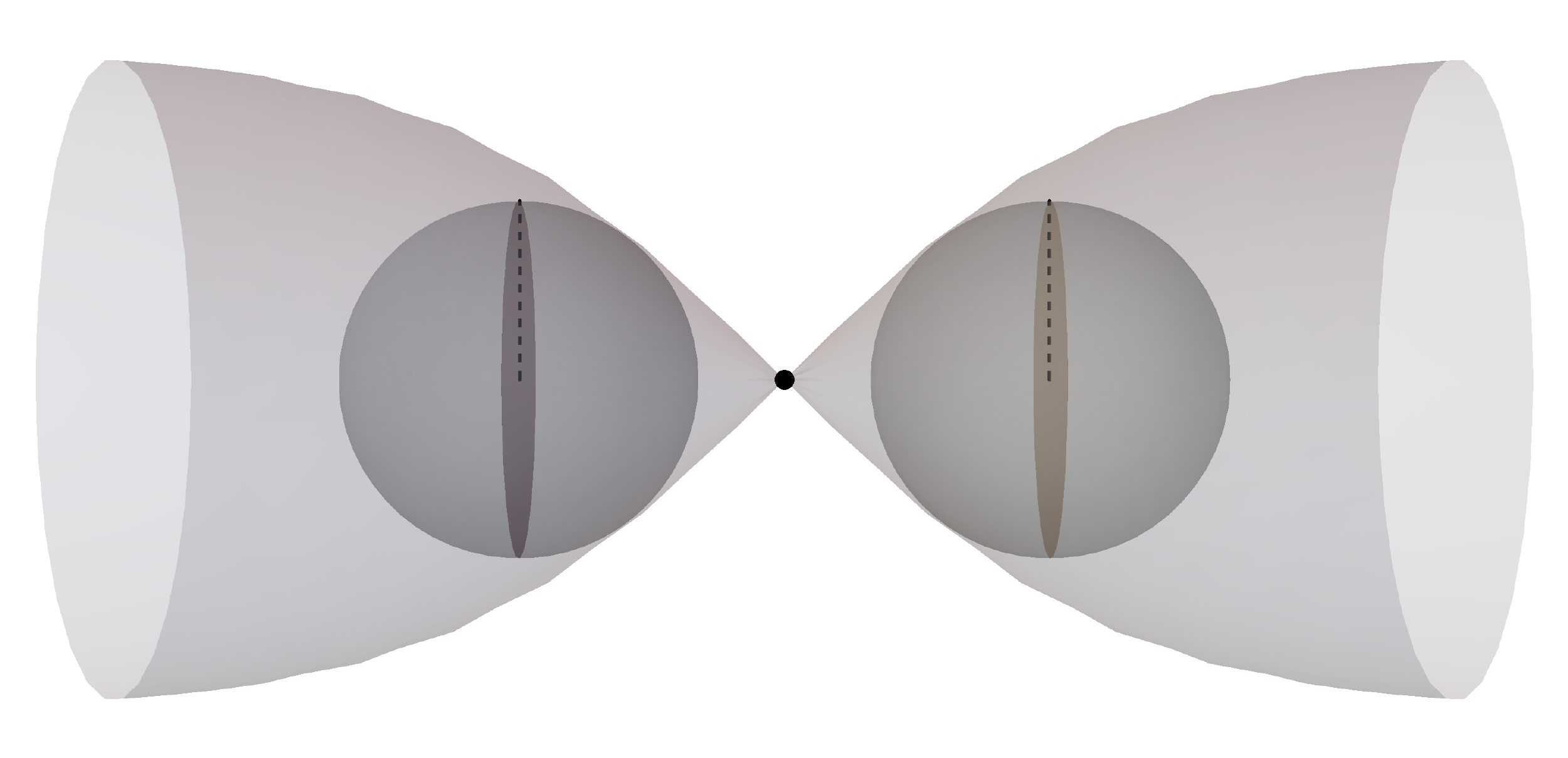}
\caption{This figure displays the conical singularity appearing in the t-y slice of the geometry. A spherical cap of small radius, r is put to regularize the singularity. }
\label{fig:conical}
\end{figure}

In summary, adding together the contributions to the potential when there is no black hole horizon, including the two brane localized potentials which each contribute $\sqrt{g_\text{ind}(y_i)} V_i$, are:
\begin{align}
V_\text{dilaton} &= e^{-4y_0} \left[ \sqrt{h(y_0)} V_0(\phi(y_0)) - \frac{6}{\kappa^2} h(y_0) \sqrt{G(y_0)} \right] \nonumber \\
&+e^{-4y_1}  \left[  \sqrt{h(y_1)} V_1(\phi(y_1)) + \frac{6}{\kappa^2}  h(y_1) \sqrt{G(y_1)} \right]
\label{eq:hotdilpot}
\end{align}
Our goal is to replace the IR brane at $y_1$ with a black hole horizon at $y_h$, such that $h(y_h)=0$ ~\cite{ArkaniHamed:2000ds}, however due to the structure of the manifold near the horizon, one cannot simply take $h(y_h) = 0$ in the above equation.  The reason for this is that the manifold near the horizon is typically singular, with a cone feature appearing in a given $t-y$ slice of the geometry, as shown in Figure~\ref{fig:conical}.

In order to study such a horizon for generic bulk scalar potential, we presume that the horizon function has a zero for some finite $y=y_h$.  We further presume that $h(y)$ is analytic, such that it has a Taylor expansion in the vicinity of the horizon.  In this case, we have $\frac{\dot{h}}{h} \approx \frac{1}{y-y_h}$, with the sign determined by the fact that $h$ is positive in the physical region $y < y_h$, and that it is passing through zero.

This behavior of the horizon function determines a boundary condition for $\phi$ that arises from taking the near-horizon limit of the scalar field equation of motion:
\begin{equation}
\left. \dot{\phi} \right|_{y_h} = \left. \frac{3}{2\kappa^2} \frac{\partial \log V}{\partial \phi} \right|_{y_h}.
\label{eq:horbc}
\end{equation}
This boundary condition enforces regularity of the solution for $\phi$ at the horizon - without this condition, $\phi$ diverges in the approach to the horizon~\cite{Gubser:2001vj}.

To compute the effective potential when the IR is screened by a black hole horizon, we need to pay closer attention to the treatment of singular terms at the orbifold fixed points at $y=y_0$ and $y=y_h$.  The scalar curvature is singular in both places.  In the UV, the singular terms can be treated as before, yielding a contribution to the effective potential that is given by 
\begin{equation}
V_\text{UV}^{\text{BH}} = -\frac{1}{\kappa^2}  e^{-4 y_0} \sqrt{G (y_0)} \left[ 8 h(y_0) - \dot{h} (y_0) \right].
\end{equation}
The IR contribution is calculated via a proper regularization of the 2D conical singularity.  There is a conically singular geometry near the black hole horizon corresponding to a system that is out-of-equilibrium.  Quantum effects will generally cause the singularity to emit radiation until it reaches equilibrium with the surrounding thermal bath, at the minimum of the free energy of the thermodynamical system.

If a theory admits solutions to the $h$ function which vanish at some finite value of $y_h$, then we can study such systems in the near-horizon limit.  Considering the near-horizon limit of the metric, where $h\approx \dot{h}(y_h) ( y-y_h )$, we have (displaying only the $dt$ and $dy$ components of the metric):
\begin{equation}
ds^2 \approx e^{-2y_h} \dot{h}(y_h) (y-y_h) dt^2 + \frac{dy^2}{\dot{h}(y_h) (y-y_h) G(y_h)}.
\end{equation}
We now go to ``good" coordinates, $( y-y_h ) = \frac{\tilde{y}^2}{4} \dot{h}(y_h) G(y_h) $,  $t = \frac{\theta}{2 \pi T}$ where the metric is manifestly that of a cone:
\begin{equation}
ds^2 = e^{-2 y_h} \dot{h}^2(y_h) G(y_h) \frac{\tilde{y}^2}{4(2 \pi T)^2} d\theta^2 + d\tilde{y}^2.
\end{equation}
The opening angle of the cone is given by
\begin{equation}
\sin \alpha = -e^{-y_h} \frac{\dot{h} (y_h) \sqrt{G(y_h)} }{4 \pi T},
\end{equation}
with the overall minus sign ensuring positivity of the angle since $\dot{h}$ is negative at the horizon.  By capping the cone with a sphere of radius $r$, which has constant curvature $2/r^2$, the contribution to the action is rendered finite and $r$ independent, allowing a sensible $r\rightarrow 0$ limit:
\begin{equation}
\Delta S_\text{IR} = \int d^3 x \frac{4 \pi}{\kappa^2} \left( 1- \sin \alpha \right) e^{-3y_h} = \int d^3 x\left[ \frac{4 \pi}{\kappa^2} e^{-3 y_h} +\frac{1}{T \kappa^2} e^{-4 y_h} \dot{h}(y_h) \sqrt{G(y_h)}  \right].
\end{equation} 
Note that a factor of two has been included as the integral is over the entire $S_1$ space in the $S_1/Z_2$ orbifold, leading to a double-copy of the spherical cap, one on each side of the orbifold fixed point. The singular IR contribution to the 4D effective potential energy is then given by 
\begin{equation}
V_\text{sing}^\text{IR} = -\left[  \frac{1}{ \kappa^2} e^{-4 y_h} \dot{h}(y_h) \sqrt{G(y_h)}  +\frac{4 \pi}{\kappa^2} e^{-3 y_h} T \right].
\end{equation}
The first term cancels exactly the corresponding UV term, and we can write the complete effective potential in the presence of the black hole horizon as
\begin{equation}
F = e^{-4 y_0} \left[ \sqrt{h(y_0)} V_0 (\phi(y_0)) - \frac{6}{\kappa^2} h(y_0) \sqrt{G(y_0)} \right] - \frac{4 \pi T}{\kappa^2} e^{-3 y_h}.
\label{eq:ftpot}
\end{equation}
This expression for the free energy, $F = U - TS$ separates into an energetic component $U$ that is completely localized on the UV brane and an entropic component $-TS$ arising from the Bekenstein-Hawking entropy of the black hole.

  The value of $y_h$ that minimizes the free energy as a function of the horizon location is obtained by inverting the following relation:
\begin{equation}
T = -\frac{\kappa^2}{12 \pi} \frac{dU}{d y_h} e^{3 y_h}.
\end{equation}
The right hand side of this equation for arbitrary $y_h$ is interpreted as the temperature of the black hole.  The value of the free energy at the minimum is
\begin{equation}
V_\text{min} = U+ \frac{1}{3} \frac{dU}{d y_h}.
\end{equation}

Up to terms that violate conformal invariance due to the introduction of the Planck brane or the Goldberger-Wise potential, the equilibrium temperature that minimizes the effective potential as a function of $y_h$ is associated with the value of $y_h$ that removes the conical singularity.
We can use the $h$ equation of motion to express this equilibrium temperature in terms of the near AdS-Schwarzchild UV behavior of $h$ and $G$: $\dot{h}(y_0) \approx -4 e^{4 (y_0-\tilde{y}_h)}$ and $G(y_0) \approx k^2$.  Note that $\tilde{y}_h$ is the position where the horizon would be if there were no deformation of the geometry due to the varying $\phi$ field.  In the absence of scalar backreaction, $\tilde{y}_h = y_h$.  From the equations of motion one finds that the presence of the back-reaction delays the onset of the horizon, establishing the inequality $\tilde{y}_h \le y_h$.  

\begin{equation}
T_\text{eq} = T_h = \frac{k}{\pi} e^{-y_h} e^{4 (y_h -\tilde{y}_h)}.
\label{eq:Th}
\end{equation}
As the position of the horizon, $y_h$, is greater than $\tilde{y}_h$, the equilibrium temperature is larger than it would be in the absence of scalar backreaction.  This is potentially problematic, as this would mean that the temperature is not necessarily a monatonic function of the position of the horizon.  The temperature would in fact grow when the backreaction becomes sizable, causing a deviation between $y_h$ and $\tilde{y}_h$.  We see that the temperature grows with increasing $y_h$ when $\frac{d y_h}{d \tilde{y}_h} > 4/3$.  Note however, that the entropy $S = \frac{4\pi}{\kappa^2} e^{-3y_h}$ is monotonically decreasing with increasing $y_h$. These high temperature solutions with low entropy are disfavored relative to those of equal temperature but small $y_h$ and thus larger entropy.

%%%%%%%%%%%%%%%%%%%%%%%%%%%%%%%%%%%%
%%%%%%%%%%%%%%%%%%%%%%%%%%%%%%%%%%%%
%%%%%%%%%%%%%%%%%%%%%%%%%%%%%%%%%%%%
%%%%%%%%%%%%%%%%%%%%%%%%%%%%%%%%%%%%
%%%%%%%%%%%%%%%%%%%%%%%%%%%%%%%%%%%%
\subsection{Constant Bulk Potential at Finite Temperature}
In the case of $V(\phi) = -\frac{6k^2}{\kappa^2}$, with no dependence on $\phi$, the scalar field equation of motion has a significantly simplified relationship to $h$:
\begin{equation}
\frac{d}{dy} \log \dot{\phi} = \frac{d}{dy} \log \frac{\dot{h}}{h}
\end{equation}
This scalar field equation of motion is integrable, and we find that the solution is given by:
\begin{equation}
\phi = \phi_0 + C_l \log h,
\end{equation}
where $C_l$ is an integration constant.  We note that this equation immediately excludes the case of constant bulk potential as a candidate for a spontaneously broken CFT at finite temperature, or where $h= 0$ for some finite $y$ in a non-trivial scalar field configuration.  Clearly, if $h$ is vanishing, but $C_l$ is finite then $\phi$ must be divergent at the position of the horizon, and the horizon boundary condition Eq.~(\ref{eq:horbc}) cannot be satisfied.  

The equations can be satisfied for one particular value: $C_l=0$, which corresponds to $\phi = $ constant.  Solving the Einstein equations for this case yields $h = 1 - e^{4(y_h-y)}$ and $G(y)=k^2$, corresponding to the AdS-Schwarzchild geometry.  This configuration is dual to an unbroken exact CFT at finite temperature.

%%%%%%%%%%%%%%%%%%%%%%%%%%%%%%%%%%%%
%%%%%%%%%%%%%%%%%%%%%%%%%%%%%%%%%%%%
%%%%%%%%%%%%%%%%%%%%%%%%%%%%%%%%%%%%
%%%%%%%%%%%%%%%%%%%%%%%%%%%%%%%%%%%%
%%%%%%%%%%%%%%%%%%%%%%%%%%%%%%%%%%%%
\subsection{Generic Potential at Finite Temperature}
We now calculate the results for the free energy when a non-trivial bulk potential is considered. Using the potential explored in the zero-temperature analysis, Eq.~(\ref{eq:bulkpot}), we numerically solve the coupled scalar and Einstein equations for a range of temperature and the free parameters of the model.
At high temperatures the theory is in a quasi-AdS-Schwarzchild geometry, with a free energy given by Eq.~(\ref{eq:ftpot}), while at low temperatures, the theory transitions to the zero temperature geometry studied in Section 2. 

\begin{figure}[htbp!]
	\centering
	\includegraphics[width=.485\hsize]{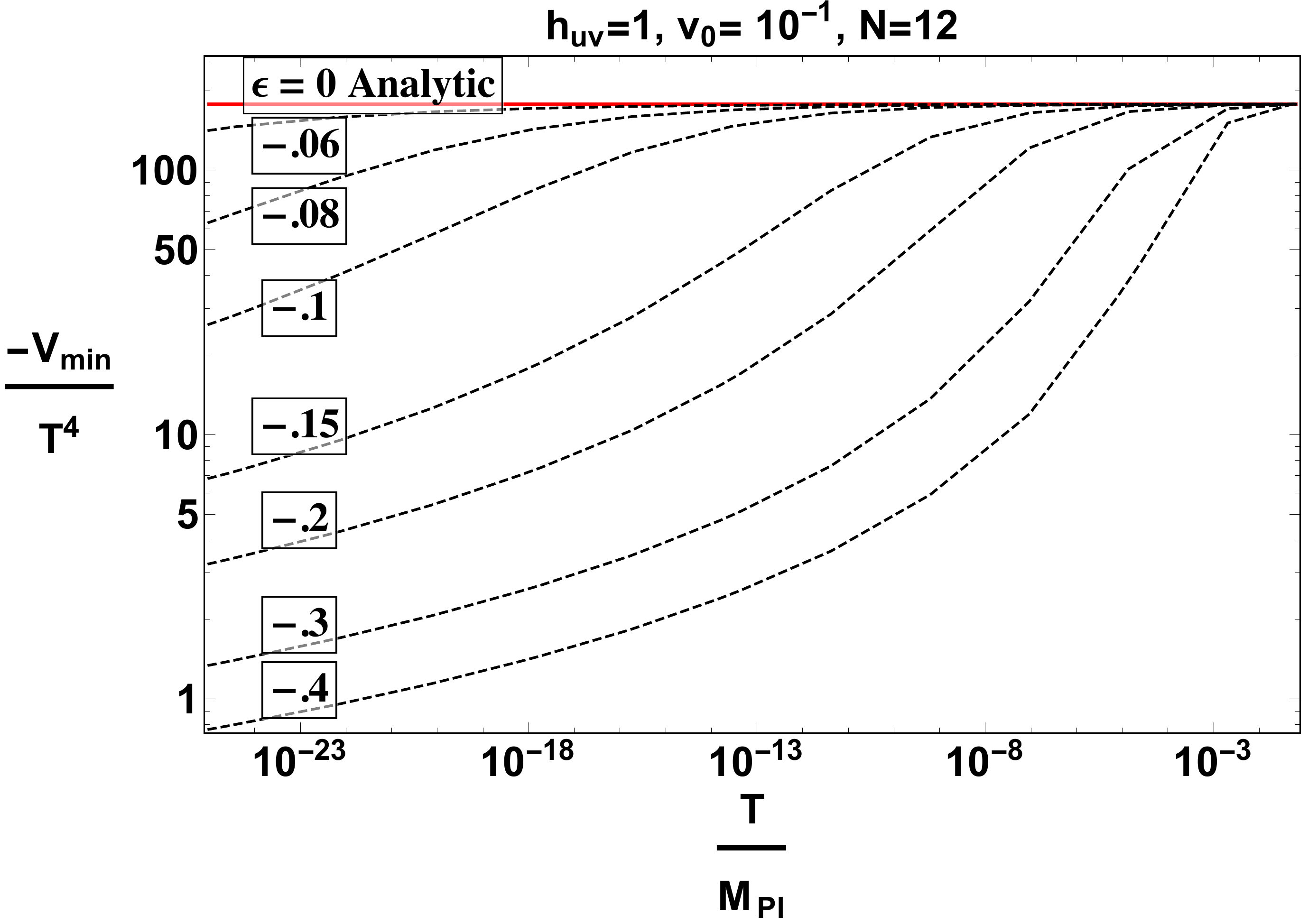}
	\includegraphics[width=.505\hsize]{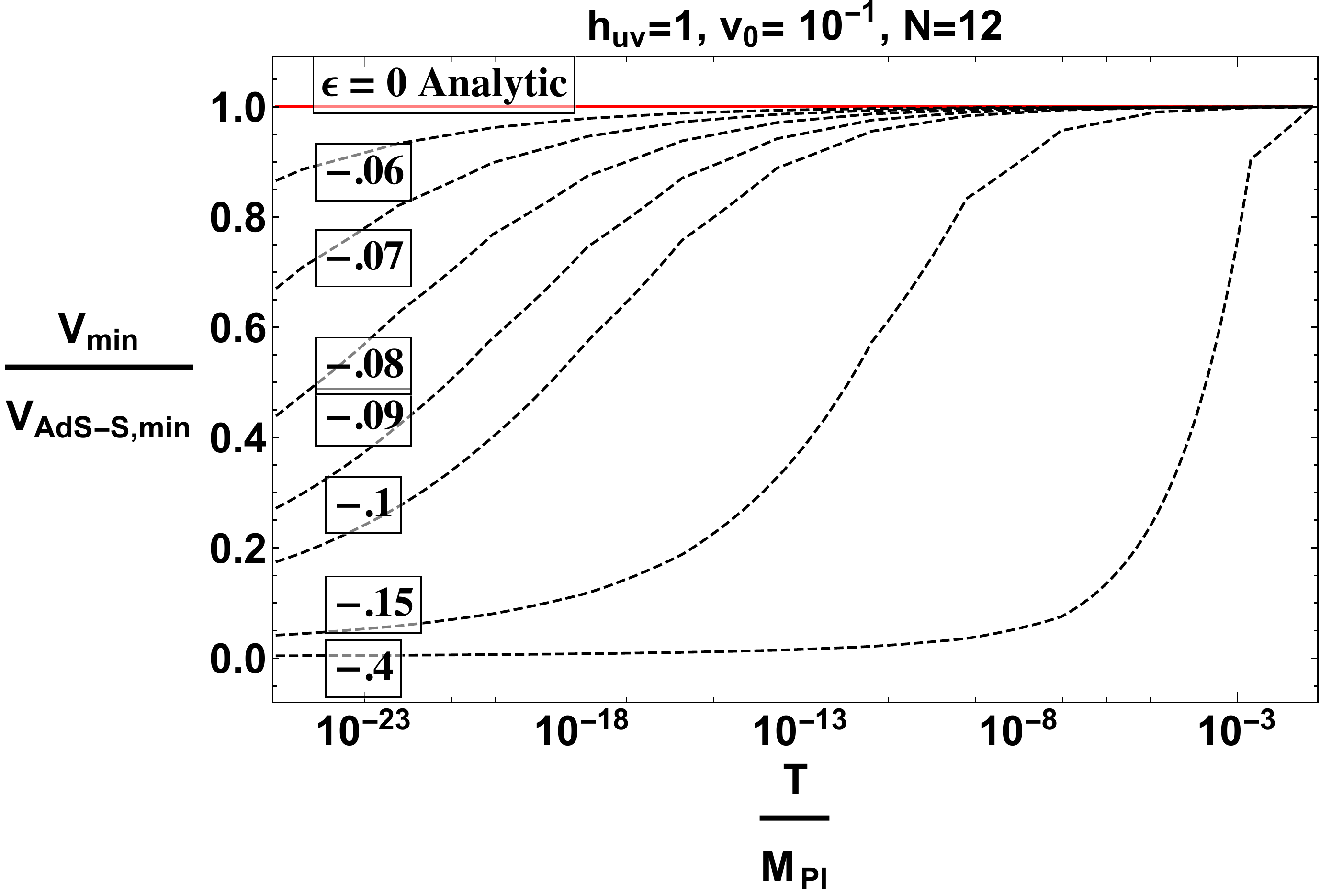} 
	\caption{In this figure, we show the minimum of the free energy function rescaled by the fourth power of the temperature.  The temperature is set by the inverse compactification radius of the time coordinate.  On the right hand side, for comparison, we give the minimum of the free energy divided by the result obtained while neglecting the back-reaction on the geometry, with very large deviations clear at lower values of the temperature.  Results are presented in terms of the temperature divided by the 4D effective Planck scale.}
	\label{fig:finiteTfig}
\end{figure}

Geometries that minimize the free energy can be found for a large range of temperatures for each configuration of parameters.  The analysis reveals that there are striking differences between the free energy as a function of the temperature when one includes or does not include the effects of backreaction on the metric.  In Figure~\ref{fig:finiteTfig}, we show the results of the numerical analysis in terms of the value of the free energy at the minimum, having extremized over the position of the black hole horizon.  The curve corresponding to $\epsilon = 0$ gives the free energy for the AdS-Schwarzchild solution, or equivalently, the free energy where backreaction is neglected.  The remaining curves have non-trivial scalar field profiles due to non-vanishing values of the bulk scalar mass term, and affect the free energy.  Backreaction effects generally increase the value of the free energy at the minimum for a given temperature, meaning that it will be easier to make the transition:  the critical temperature when the value of the free energy is equal to the minimum of the zero temperature effective dilaton potential is higher.

%%%%%%%%%%%%%%%%%%%%%%%%%%%%%%
%%%%%%%%%%%%%%%%%%%%%%%%%%%%%%
%%%%%%%%%%%%%%%%%%%%%%%%%%%%%%
%%%%%%%%%%%%%%%%%%%%%%%%%%%%%%

\section{Phase Transitions}
\label{sec:cosmoPT}
The phase transition connecting the finite temperature approximate CFT to the zero temperature soft-wall description dual to spontaneously broken near-conformal symmetry is first order, and thus proceeds via bubble nucleation~\cite{Cline:2000xn,Randall:2006py,Nardini:2007me}.  In the early universe, bubble nucleation competes with Hubble expansion, with the phase transition proceeding to completion only if bubble creation outpaces Hubble dilution~\cite{Guth:1983pn,Hawking:1982ga}.

This requirement can be phrased as the necessity of having one or more bubbles in one unit of Hubble 4-volume.  If there is a rate of bubble nucleation per unit volume given by $\Gamma/V$, then model parameters that satisfy
\begin{equation}
\frac{\Gamma}{V} \gtrsim H^4
\end{equation}
at some time period in the early universe will correspond to a successful phase transition.  In the hot conformal phase, the universe is radiation dominated, and we have 
\begin{equation}
H^2 =\frac{ 8 \pi G \rho }{3} \sim \frac{\pi^3 G N^2 T^4}{3}.
\end{equation}
The decay rate is proportional to $\Gamma/V \propto e^{-S_E}$, where $S_E$ is the euclidean action associated with the fields evolving from their initial to their final values at nucleation.  The coefficient of proportionality is difficult to calculate in general, involving an ``obdurate" functional determinant, but dimensional analysis says that this factor should be of order $f^4$.  Similarly, the nucleation temperature is not expected to be drastically different from $f$, and later analysis in this section confirms this, with results shown in Figure~\ref{fig:nucleation}.  Putting this together, the criterion for bubble nucleation is roughly
\begin{equation}
S_E \lesssim 4 \log \left( \frac{M_\text{Pl}}{f} \right)
\label{eq:Scrit}
\end{equation}
Many terms have been left out that are subdominant for large $M_\text{Pl}/f$ hierarchies.  The large gap between $f$ and $M_\text{Pl}$ along with the exponential sensitivity to $S_E$ justifies using simple dimensional analysis on the coefficient of the rate.

The Euclidean action is associated with the path in field space from the initial state far away from the bubble to the final state associated with its interior.  At finite temperature, the geometry is compactified along the time direction, with compactification radius given by the inverse temperature: $t \in ( 0, 1/T ]$.  Two types of bubbles are possible - those with $O(3)$ symmetry whose radius is large in comparison with $1/T$, and those with small radius, where bubbles exhibit invariance under the full $O(4)$.  The one with lower action is the one that will determine the decay rate.

At low temperatures, or at the interior of a bubble in the cooling universe, there exists a warped extra dimensional solution with time compactified and no black hole horizon.   At high temperatures, or far away from the bubble, a near-AdS-Schwarzchild solution (AdS-S) with horizon is seen, as discussed in the previous section.  To determine the rate at which the phase transition proceeds, one requires the action associated with moving from one phase to the other.

Comparison of the minimum total free energy for each geometry identifies the preferred vacuum at various temperatures, and the structure of the potential that interpolates between the AdS-S minimum and the conformal breaking minimum specifies the dynamics that interpolates between the two phases.  Our analysis parallels earlier work on conformal phase transitions~\cite{Creminelli:2001th, Randall:2006py}, but with emphasis on the particulars of the soft-wall light dilaton construction.

%%%%%%%%%%%%%%%%%%%%%%%%%%%%%%
%%%%%%%%%%%%%%%%%%%%%%%%%%%%%%
%%%%%%%%%%%%%%%%%%%%%%%%%%%%%%
%%%%%%%%%%%%%%%%%%%%%%%%%%%%%%

\subsection{Nucleation of soft-wall bubbles}

The dynamics of the phase transition interpolate between the black hole solution and the zero-temperature soft-wall dilaton geometry.  We assume that Hubble expansion adiabatically cools the finite temperature solution, with the black hole horizon position tracking the minimum of the free energy.  Eventually, when the temperature is such that the criteria for tunneling are satisfied, bubbles in the black hole horizon form, with the interior of the bubbles containing the brane that cuts off the zero temperature geometry in the infrared region.

The criteria for tunneling are two-fold.  First, the transition must be energetically favorable, with the minimum of the black hole solution free energy being greater than the minimum of the zero-temperature effective dilaton potential.  This defines the critical temperature, $T_c$.  However, the rate of nucleation may not yet be high enough to overcome Hubble dilution, which is the second criterion.  It is only when the bubble action reaches the critical value in Eq.~(\ref{eq:Scrit}) that bubbles begin to nucleate.  The temperature associated with the critical action crossing is the nucleation temperature, $T_n$, but the action is minimum not for $f$ at the bottom of the effective potential, but rather for \emph{smaller} values of $f$.  We denote the value of $f$ inside a bubble as the nucleation scale, $f_n$.  After the phase transition is completed, the dilaton will oscillate and decay down to the true minimum.

The full action interpolating between the black hole and stabilized dilaton solutions is not accessible in this calculation without a UV completion, as the black hole solution at large $y_h$ and the zero temperature small $f$ regions both involve 
trans-planckian excursions of the bulk curvature and scalar field, and also it is not clear how to properly normalize fluctuations in the position of the black hole horizon.  However one can estimate the bubble action in several hypothesis scenarios that depend on the size of the bubble radius relative to the inverse temperature, and the maximum size of the gradient of the fields as one moves from the interior to the exterior of a bubble.  We presume that the contribution to the bubble action from evolution on the black hole side of the transition is small, and the finite temperature calculation serves only to give the proper nucleation temperature.  The size of the bubble determines whether the bubble has $O(3)$ or $O(4)$ symmetry in the Euclidean action, and the steepness of the bubble wall determines whether a thick~\cite{Anderson:1991zb} or thin ~\cite{ColemanVacDecay} wall approximation is a better estimate for the minimum action.

Thick walls typically dominate when the latent heat associated with the phase transition is comparable to the barrier height separating the minima.  While the barrier height cannot be calculated due to loss of control of the theory in the small $T_h$, small $f$ regions, it appears in numerical simulation that the trend is to maintain a shallow potential.  We have also checked the action for thin wall bubbles, and indeed the thick wall solutions have values of the action that are typically of order $1/10$ that of thin wall bubbles.  Estimation of $O(4)$ and $O(3)$ thick wall bubble actions show that for some regions of parameter space, $O(3)$ bubbles have smaller action, and dominate the phase transition, and in other regions, it is the $O(4)$ symmetric bubbles that have smaller action.

The Euclidean action associated with a bubble during a finite temperature phase transition is calculated on a geometry where the time coordinate is compactified on a circle of radius $1/T$.  Small bubbles with $R < 1/T$ exhibit $O(4)$ symmetry, and the action above reduces to a radial integral along which $f$ varies from its nucleation value out to zero at the boundary of the bubble:
\begin{equation}
S_E^{O(4)} = S_4 = 2 \pi^2 \int r^3 \left[\frac{{\mathcal N}}{2} \left(\vec{\nabla} f \right)^2 + V\left(f, T\right)\right] dr
\end{equation}
For the larger $O(3)$ symmetric bubbles which wrap the time direction, one obtains
\begin{equation}
S_E^{O(3)}=S_3/T = \frac{4 \pi}{T} \int r^2 \left[\frac{{\mathcal N}}{2} \left(\vec{\nabla} f \right)^2 + V\left(f, T\right)\right] dr.
\end{equation}
${\mathcal N}$ is a normalization factor associated with canonically normalizing the fluctuations of the soft wall dilaton. 

The bubble action can be approximated by~\cite{Anderson:1991zb}
\begin{align}
S_{3} &= 4 \pi R^{2} \left[ {\mathcal N} \frac{f^2}{\Delta R^2} \Delta R + \frac{R}{3} \bar{V} \right] \nonumber \\
S_4 & = 2 \pi^2 R^3  \left[ {\mathcal N} \frac{f^2}{\Delta R^2} \Delta R + \frac{R}{4} \bar{V} \right] 
\end{align}
where $\Delta R$ is the region where $f$ is changing significantly, and  $\bar{V}$ is the volume averaged value of the potential inside the bubble.  
For the case of the dilaton potential, which is typically shallow for small values of $\epsilon$, the thick wall action is smaller than the thin wall, and so the phase transition is driven by thick wall bubbles with $f$ varying throughout.  Thus, we can take $\Delta R$ to be the same as $R$, and minimize the bubble action over $R$.  This yields:
\begin{align}
S_3/T \text{(min)} &= \frac{4 \pi}{3} \frac{ {\mathcal N}^{3/2} f^3}{\sqrt{2 |\bar{V}|}} \nonumber \\
S_4 \text{(min)} &= \pi^2 \frac{ {\mathcal N}^2 f^4 }{2 | \bar{V} |}
\end{align}
The normalization factor $\mathcal N$ we take to be $\mathcal N = 3 N^2/2 \pi^2$.  This is likely \emph{larger} than the actual normalization, which is affected by backreaction, and thus we expect our values of the action to be conservatively large.  For the average value of the potential, we use the difference between the finite temperature potential minimum and the value of the soft-wall potential at the nucleation value of the dilaton, $f_n$:
\begin{equation}
|\bar{V}| \approx F_\text{min} (T) - V_\text{dilaton} (f).
\end{equation} 

\begin{figure}[!htbp]
	\center
	\includegraphics[width=.48\textwidth]{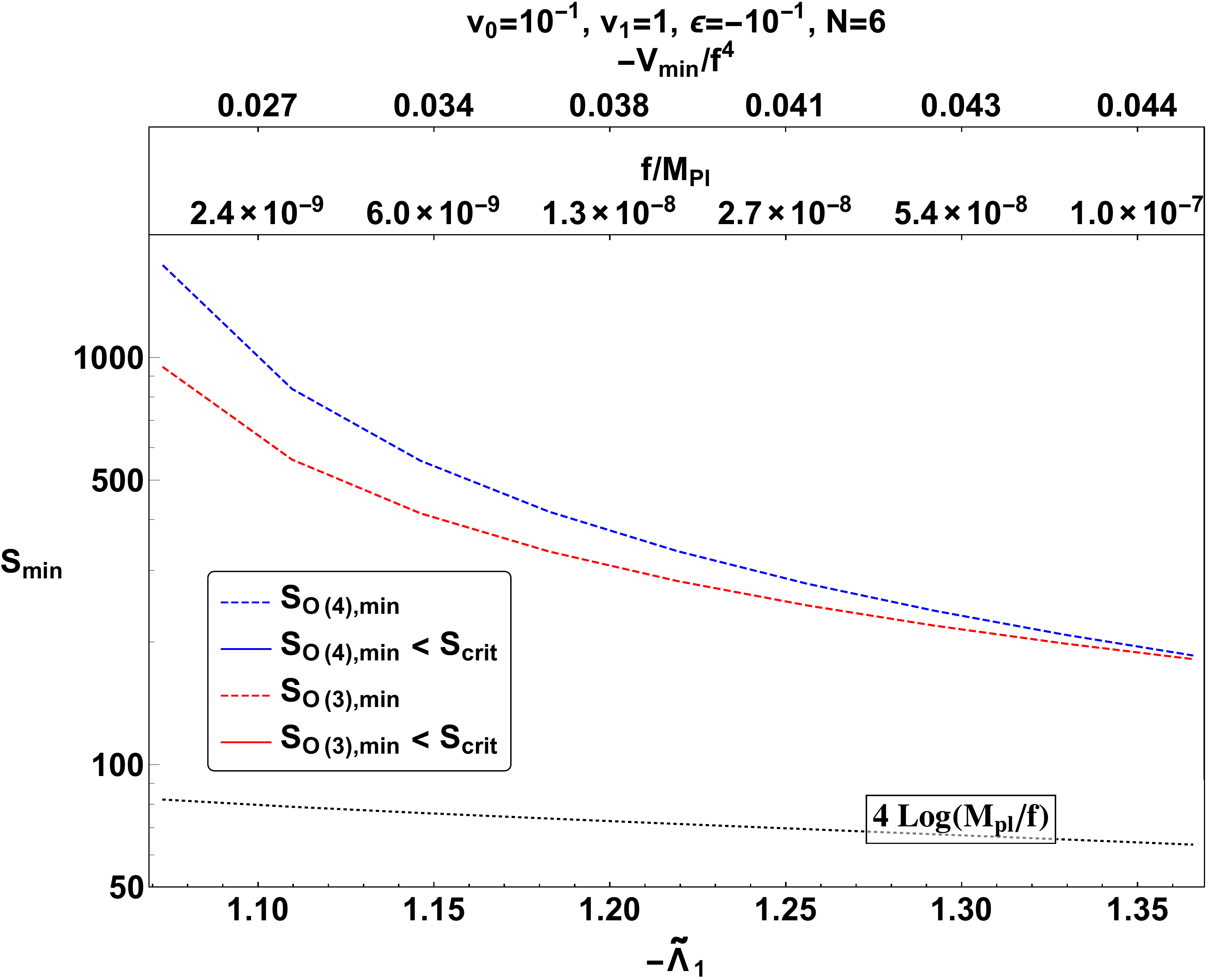}
	\includegraphics[width=.48\textwidth]{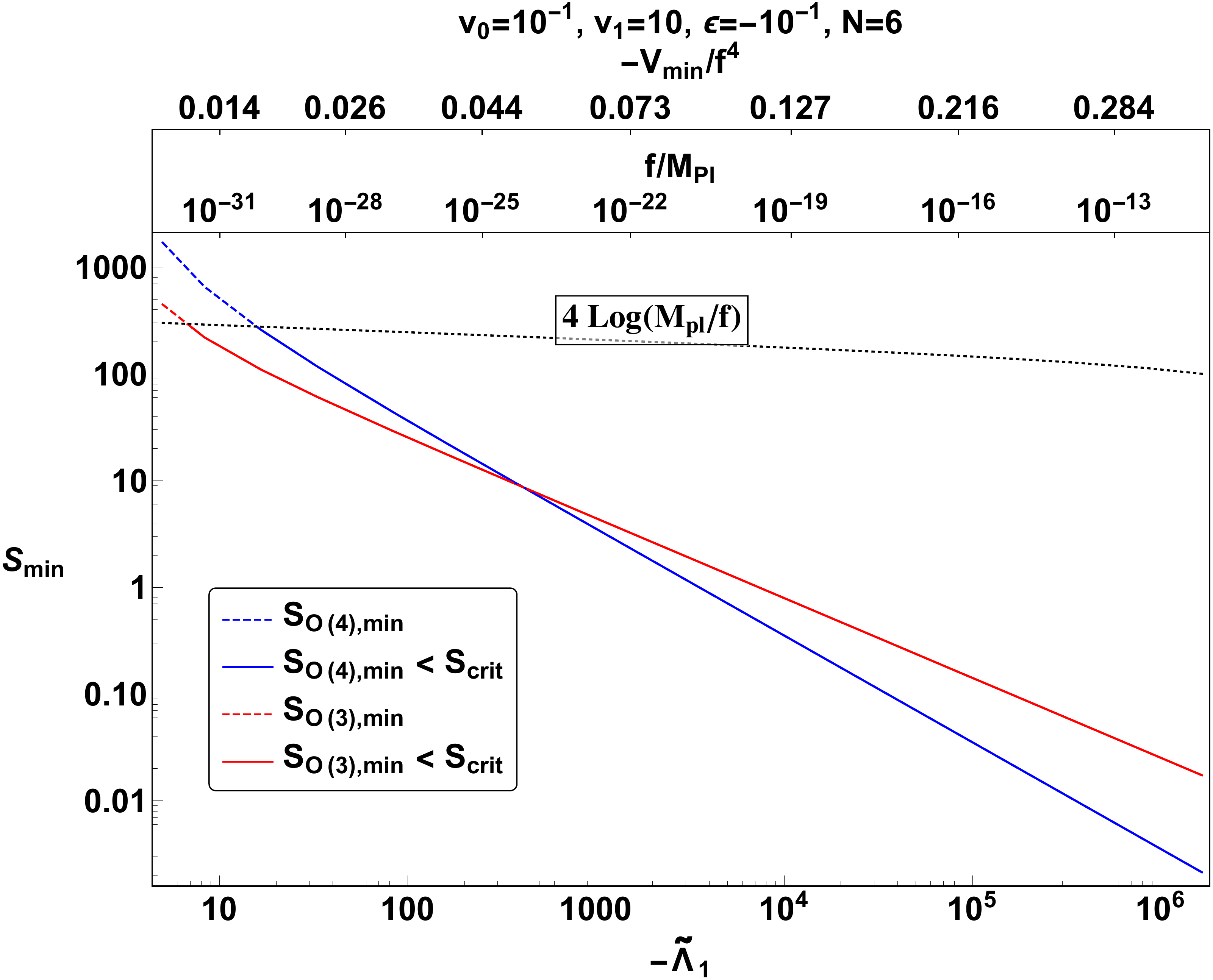}
	\includegraphics[width=.48\textwidth]{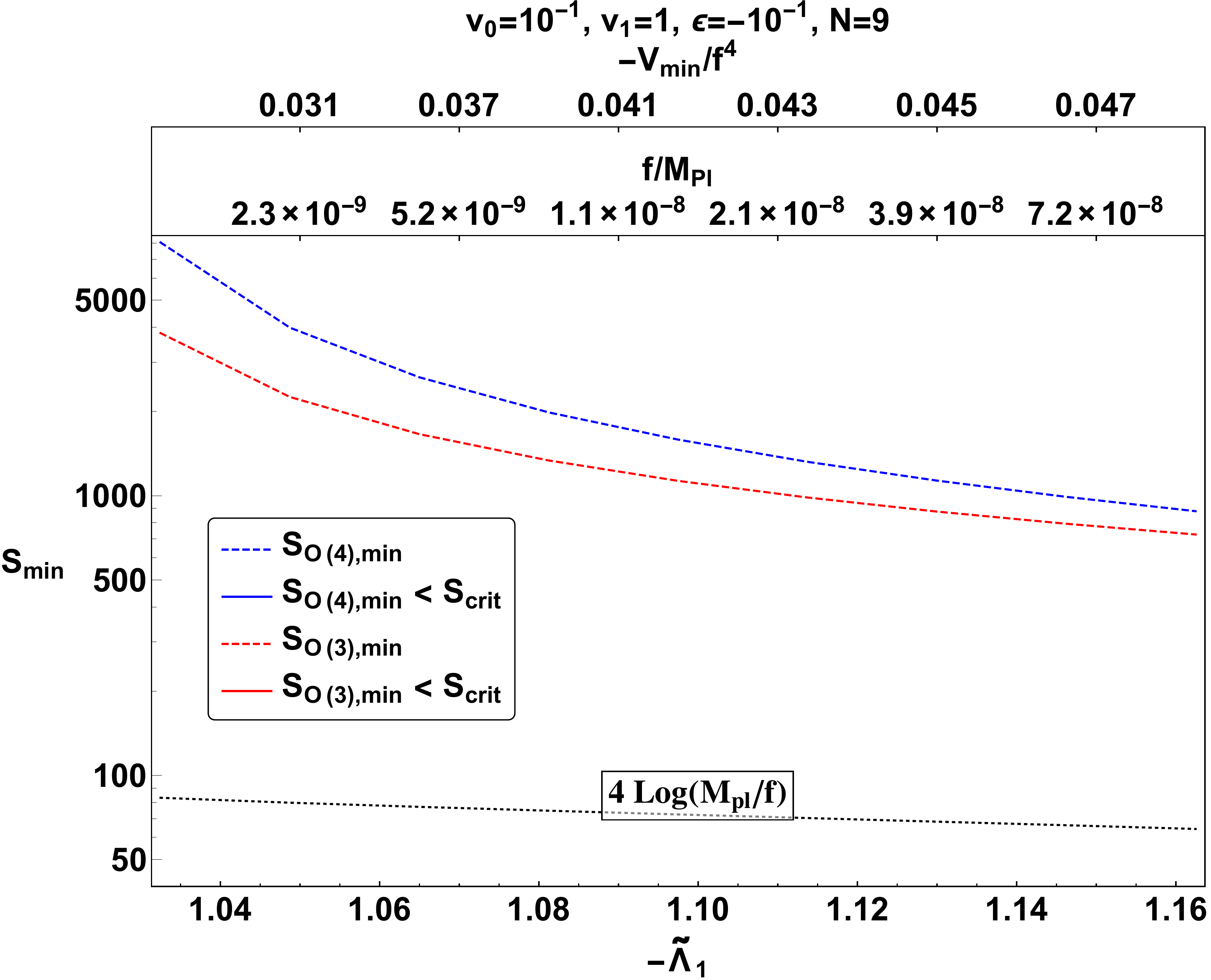}
	\includegraphics[width=.48\textwidth]{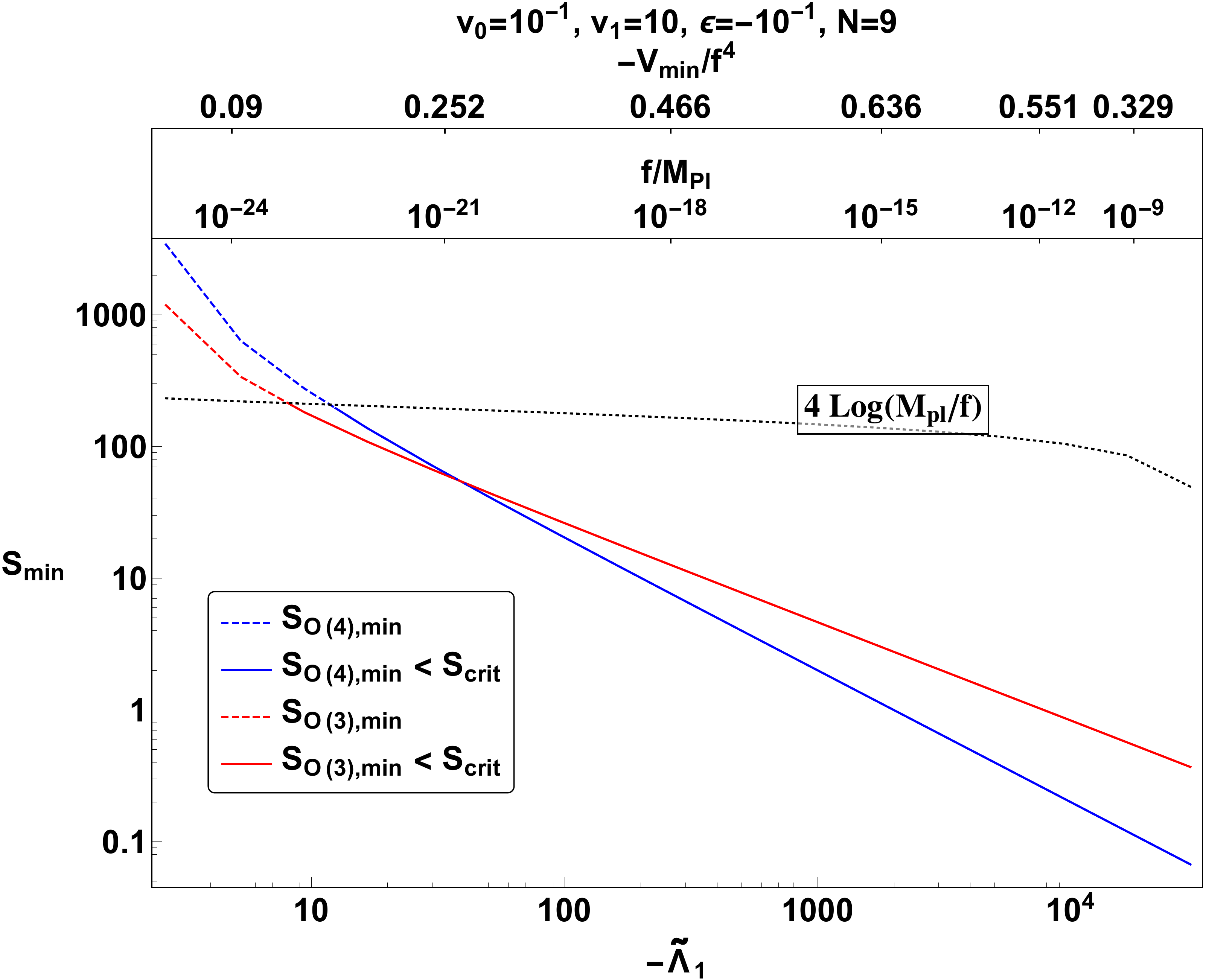}
	\includegraphics[width=.48\textwidth]{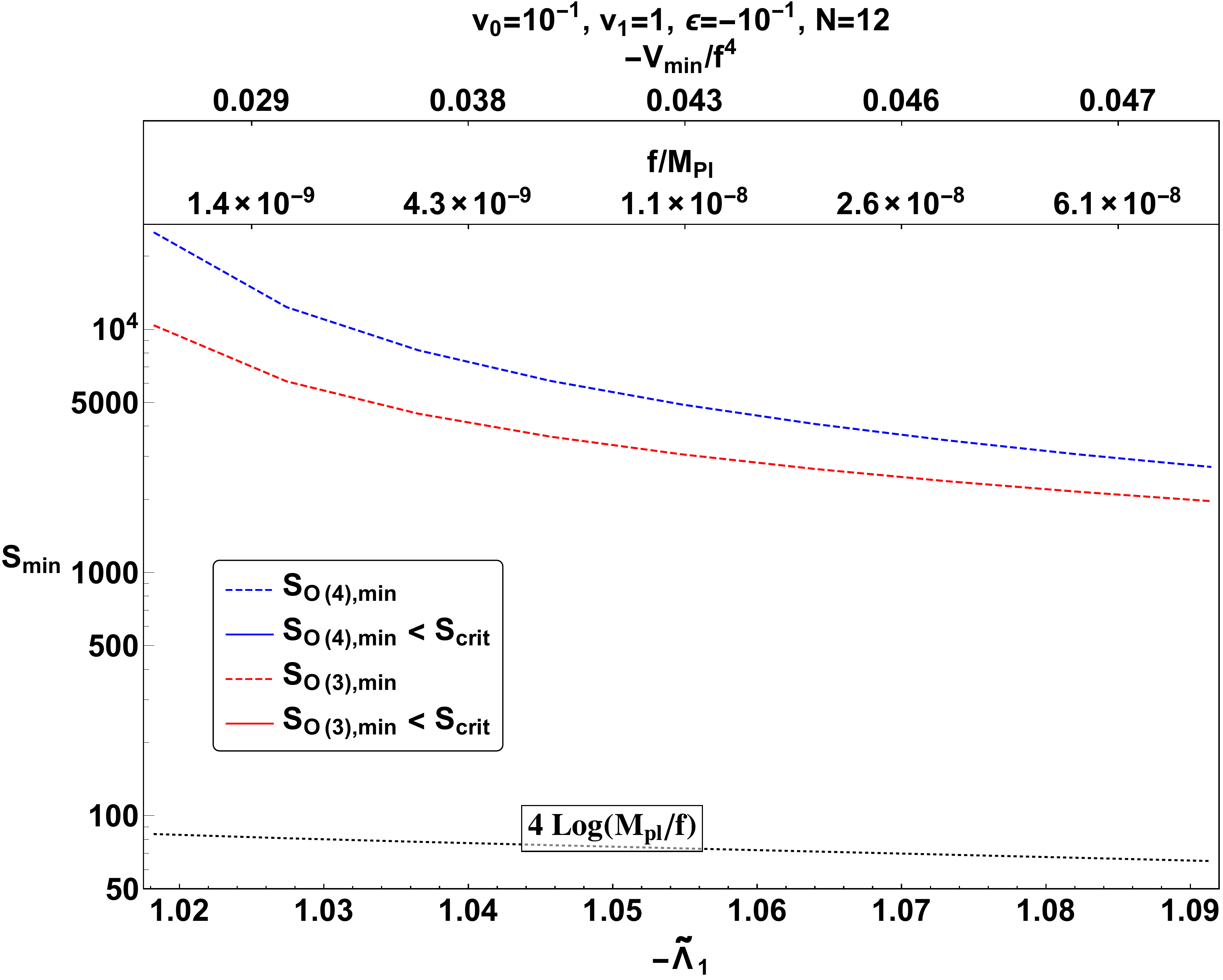}
	\includegraphics[width=.48\textwidth]{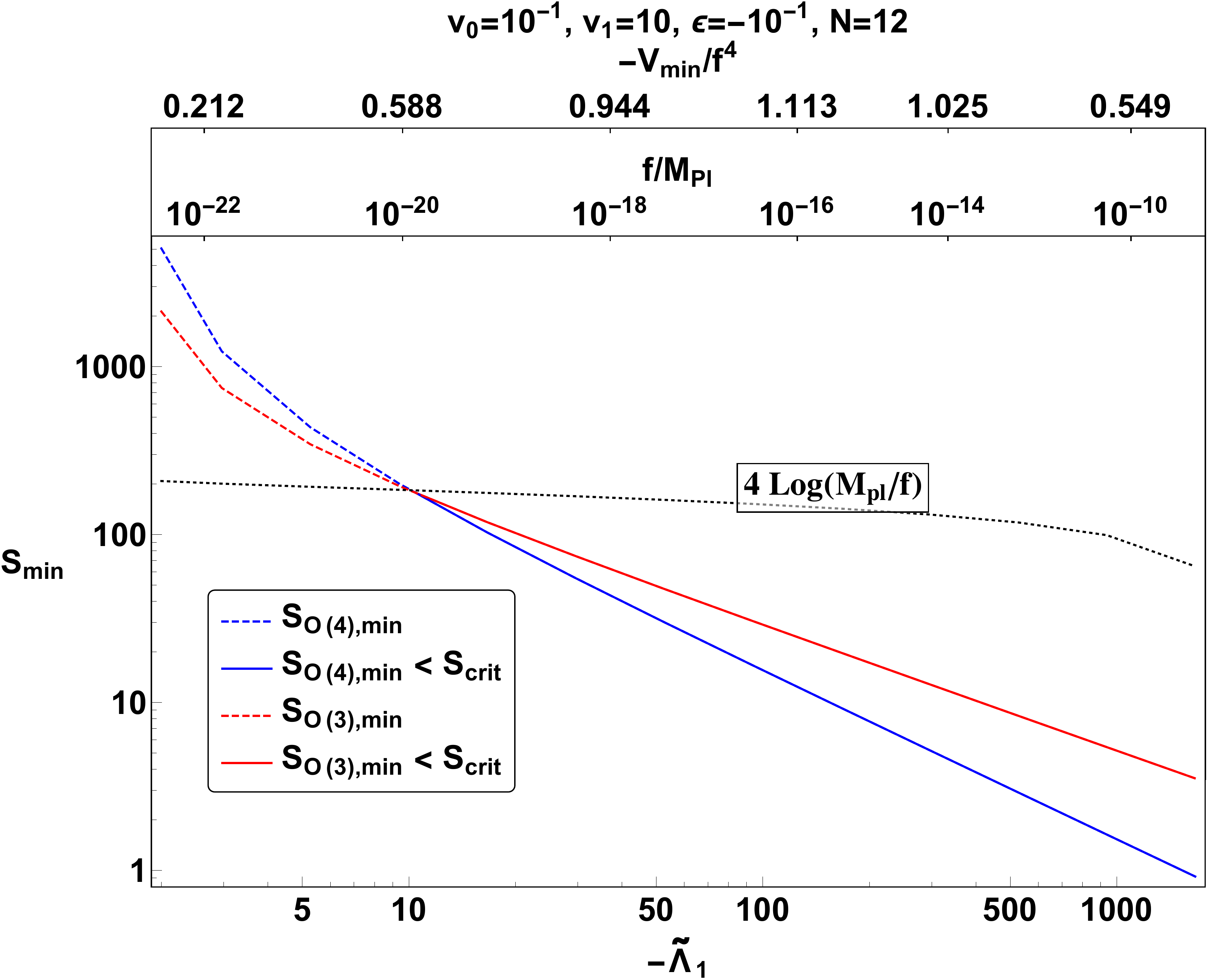}
	\caption{Each panel includes plots of the values for the $O(3)$ and $O(4)$ symmetric bubble actions as a function of the dimensionless IR brane tension $\tilde{\Lambda}_1$.  The critical value for the action as a function of $f/M_\text{Pl}$ is shown.  Plots on the left side correspond to $v_1=1$, while plots on the right correspond to $v_1=10$.  All plots take $\epsilon = -0.1$.
	}
	\label{fig:action}
\end{figure}

The properties of the finite temperature near-conformal phase transition can be calculated as a function of the input parameters in the 5D model.  In Figure~\ref{fig:action},we display in each panel the value of the minimum bubble action (both for $O(3)$ and $O(4)$ bubbles) as a function of the dimensionless IR brane tension $\tilde{\Lambda}_1$.  Each point on the $\tilde{\Lambda}_1$ axis has an associated value of $f$ at the minimum of the zero temperature potential, and a value of the potential itself at the minimum.  On additional horizontal axes on the top of each plot, we display the hierarchy between $f$ and the 4D Planck scale as well as $| V_\text{min}| /f^4$ for reference.  In the figure, we show both ``hard" and soft-wall values of $\phi_1$, with the hard wall calculation displaying excellent agreement with analytical calculations in the literature~\cite{Randall:2006py}.  As $N$ increases, the bubble action increases, making it more difficult for the phase transition to complete for a fixed ratio of $f/M_\text{Planck}$.

We note that the phase transition completes over a much wider range of parameter space, which is a strong success of the soft-wall models.  In the hard wall description, smaller values of $N$ are typically necessary, and perturbativity of the 5D gravity theory is not guaranteed.  

\begin{figure}[!htbp]
	\center
	\includegraphics[width=.48\textwidth]{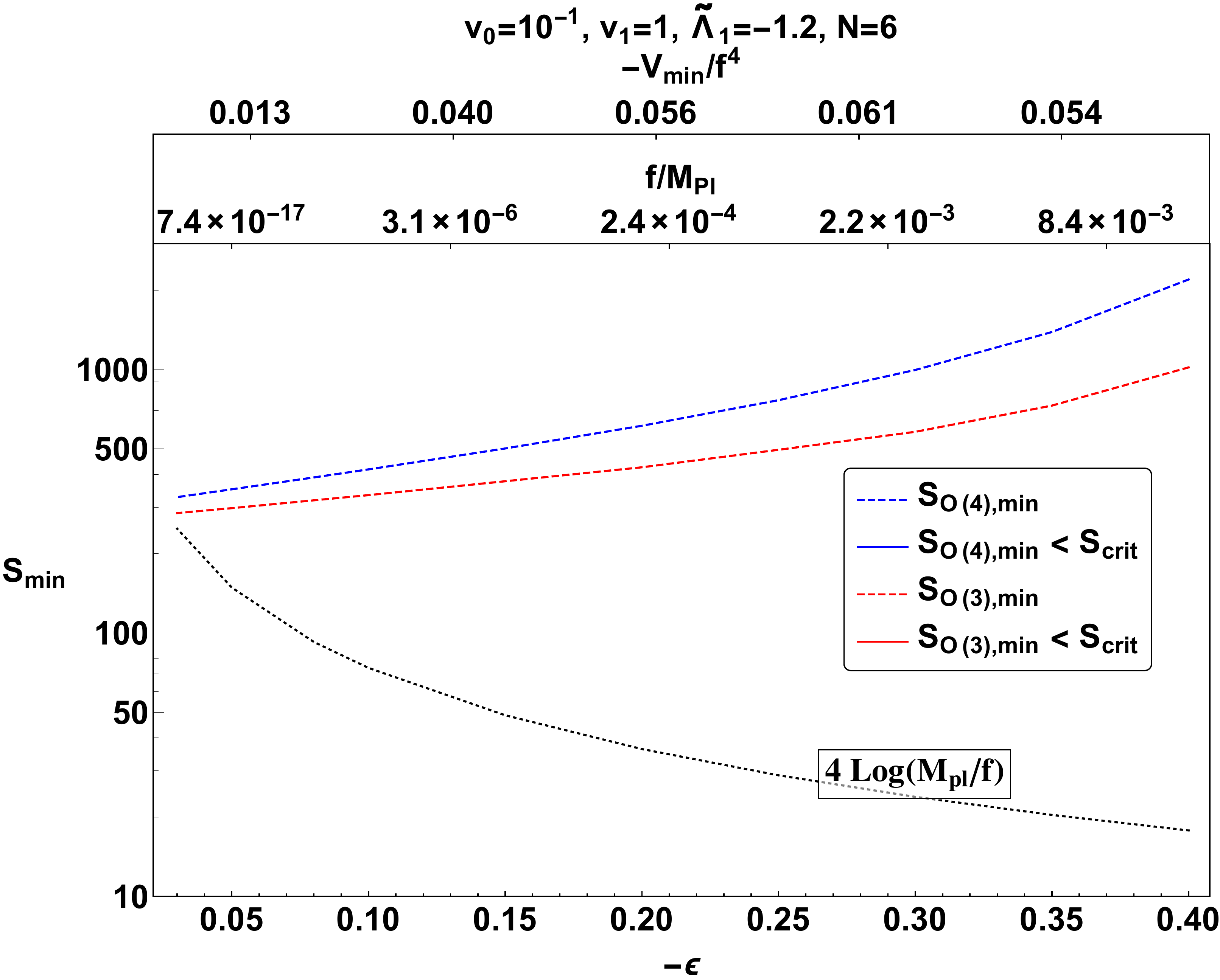}
	\includegraphics[width=.48\textwidth]{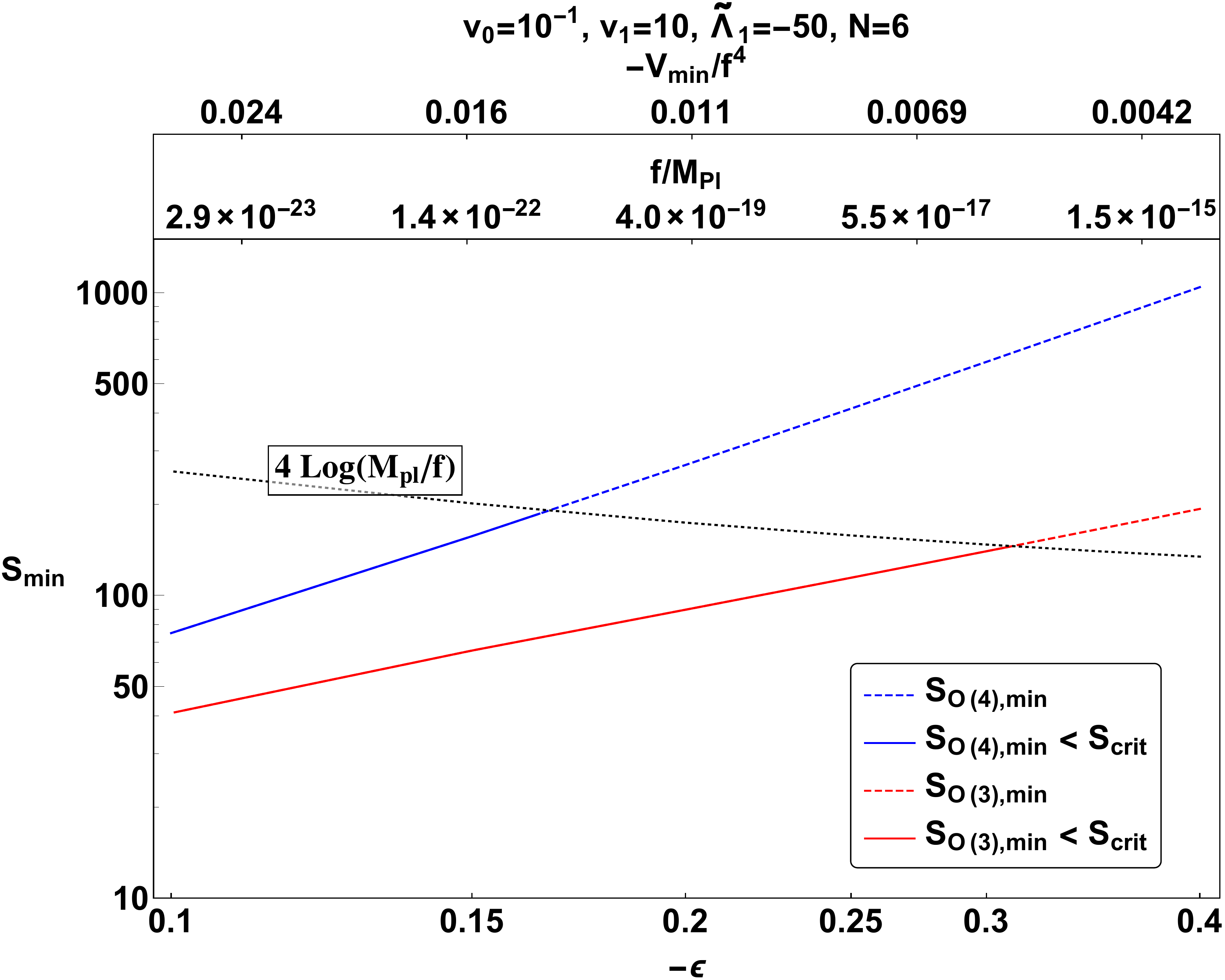}
	\includegraphics[width=.48\textwidth]{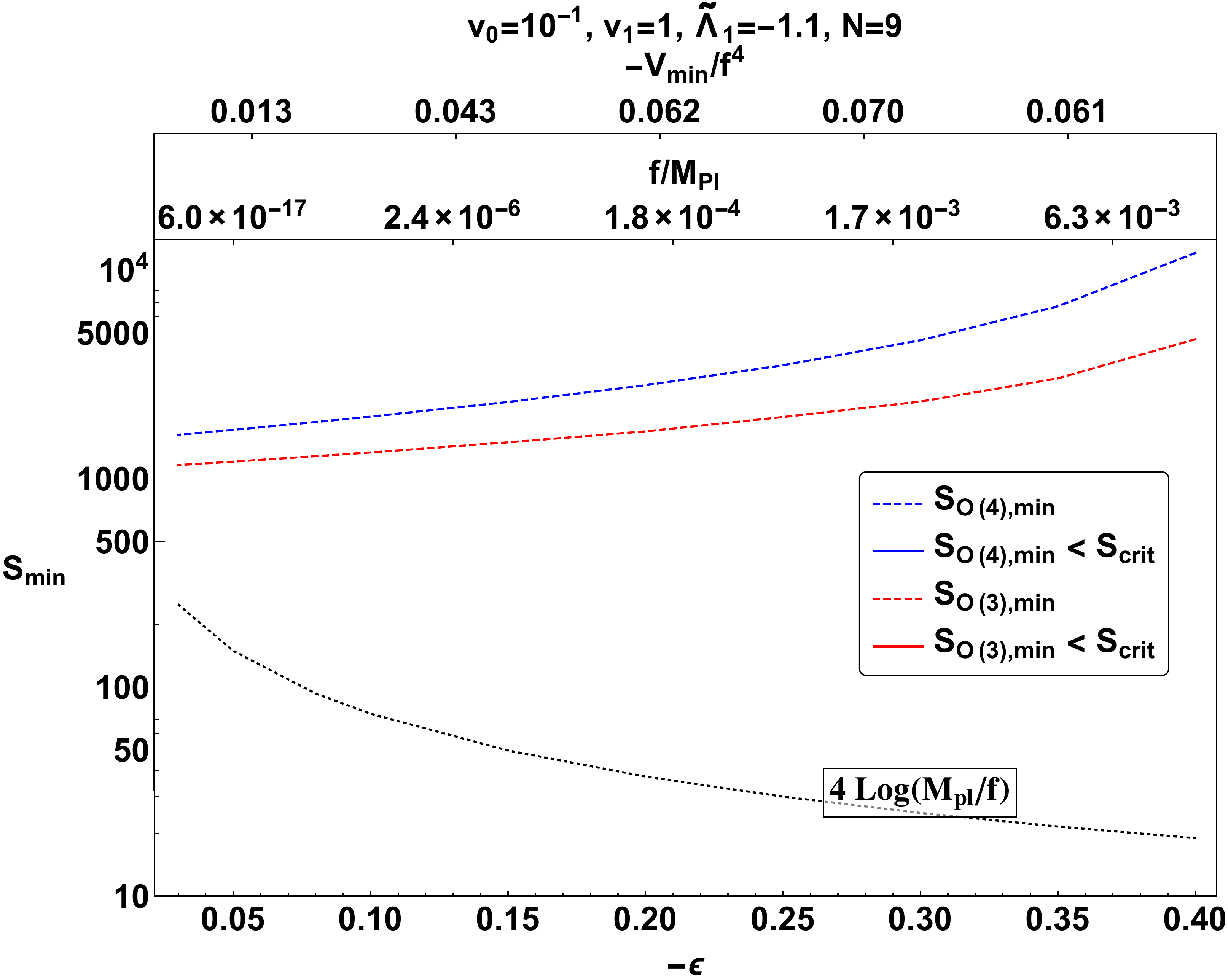}
	\includegraphics[width=.48\textwidth]{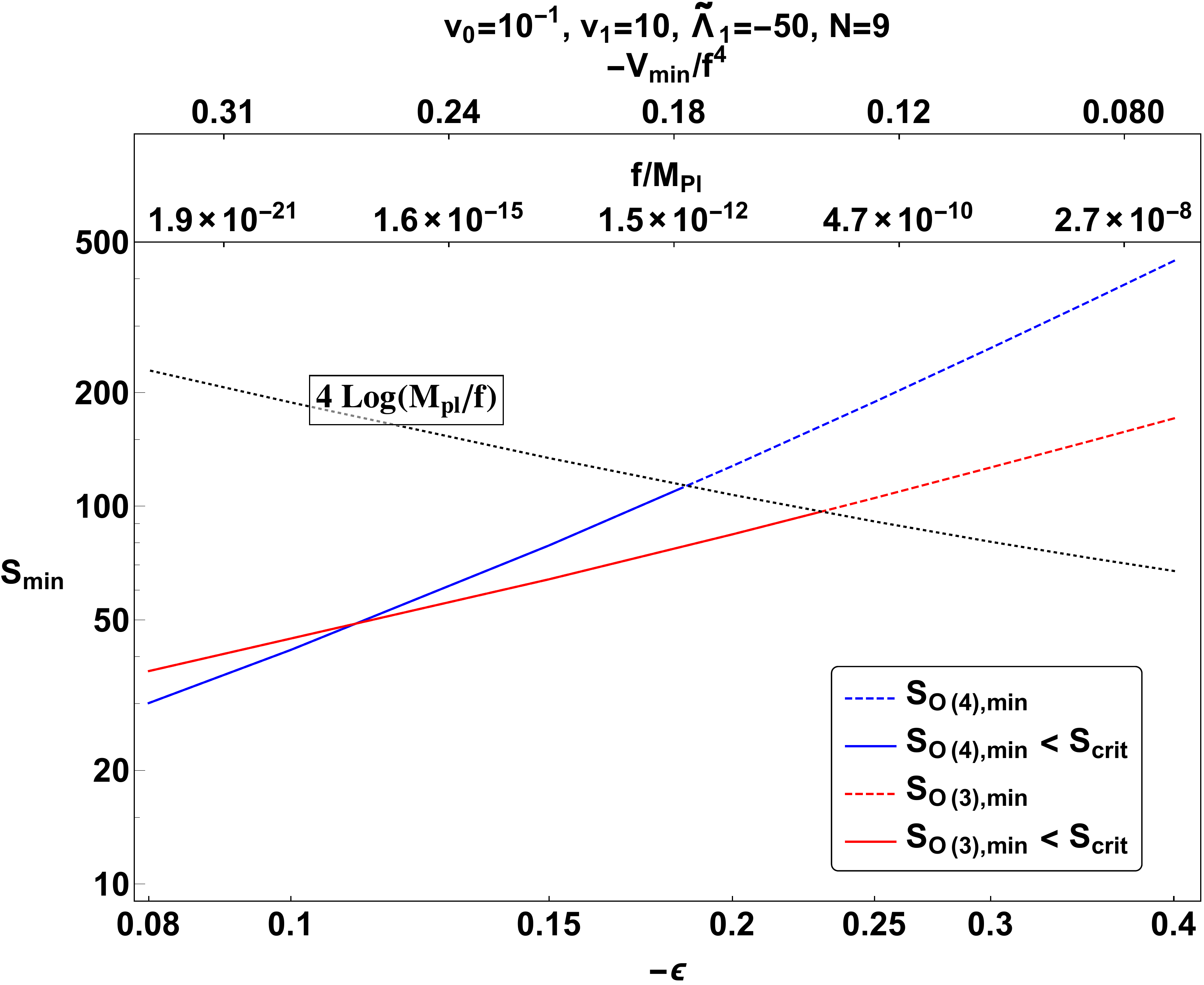}
	\includegraphics[width=.48\textwidth]{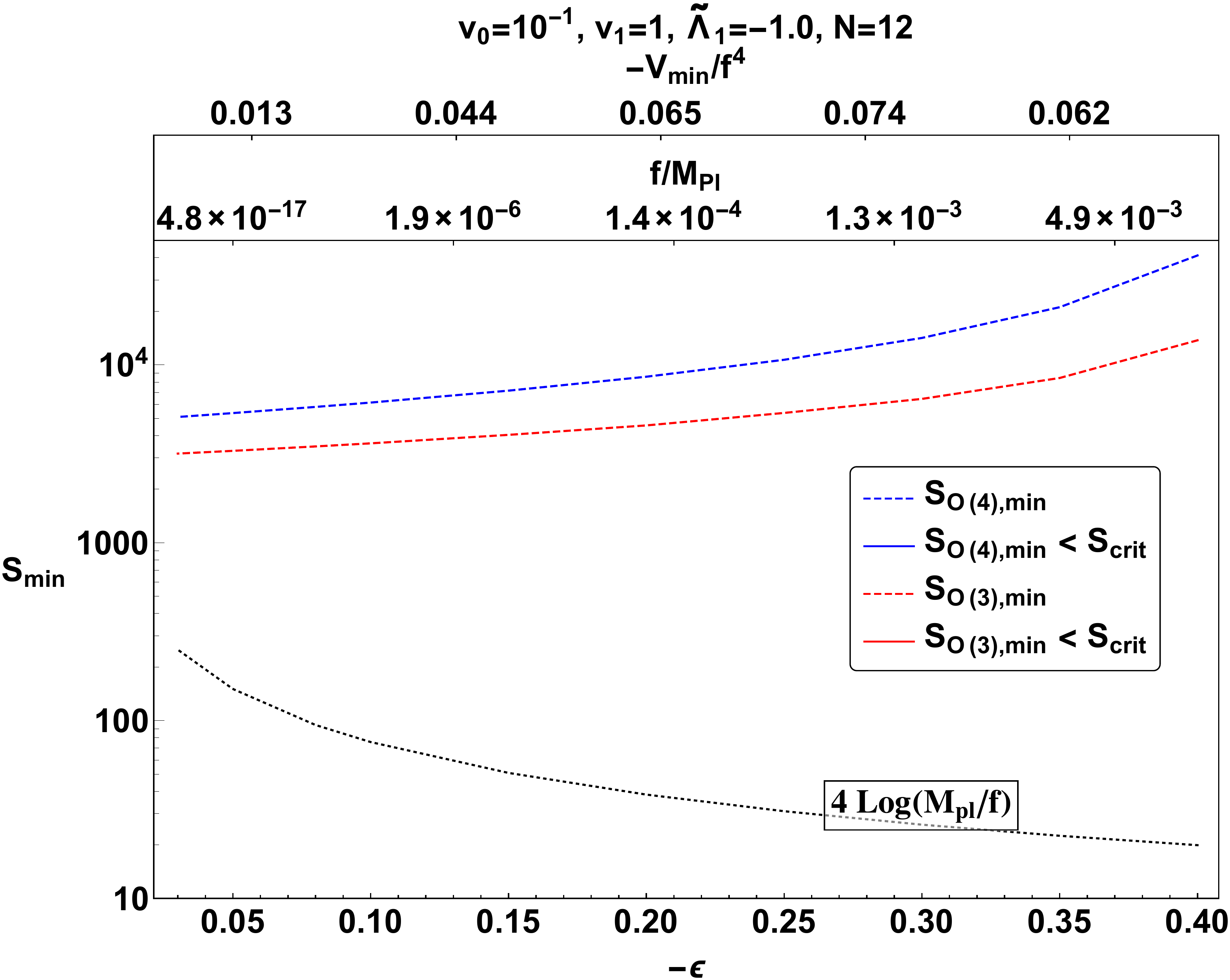}
	\includegraphics[width=.48\textwidth]{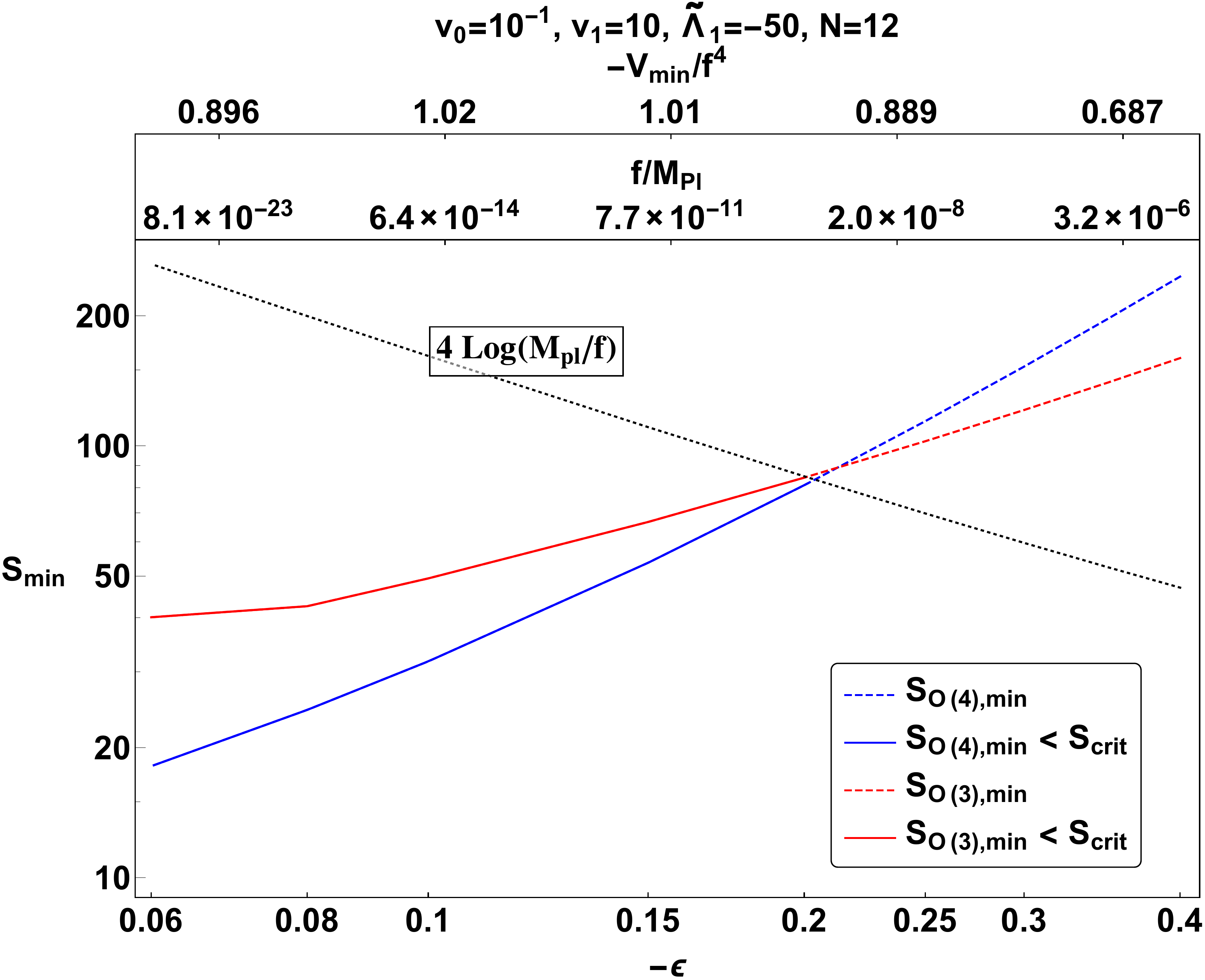}
	\caption{In this figure, we show the value of the $O(3)$ (red) and $O(4)$ (blue) bubble actions for various values of the bulk scalar mass parameter, $\epsilon$.  Again, a comparison is made with a smaller value of $v_1=1$ (plots on the left side) vs larger values $v_1=10$ (plots on the right side).}
\label{fig:epsilonaction}
\end{figure}

In Figure~\ref{fig:epsilonaction}, we show the values for the bubble action as a function of the bulk scalar mass, $\epsilon$.  Values of $\tilde{\Lambda}_1$ are chosen so as to center the plots with $\epsilon = 0.1$ corresponding to a  ratio $f/M_\text{Planck}$ of order the TeV-Planck hierarchy.  

\begin{figure}[!htbp]
	\centering
	\includegraphics[width=.48\hsize]{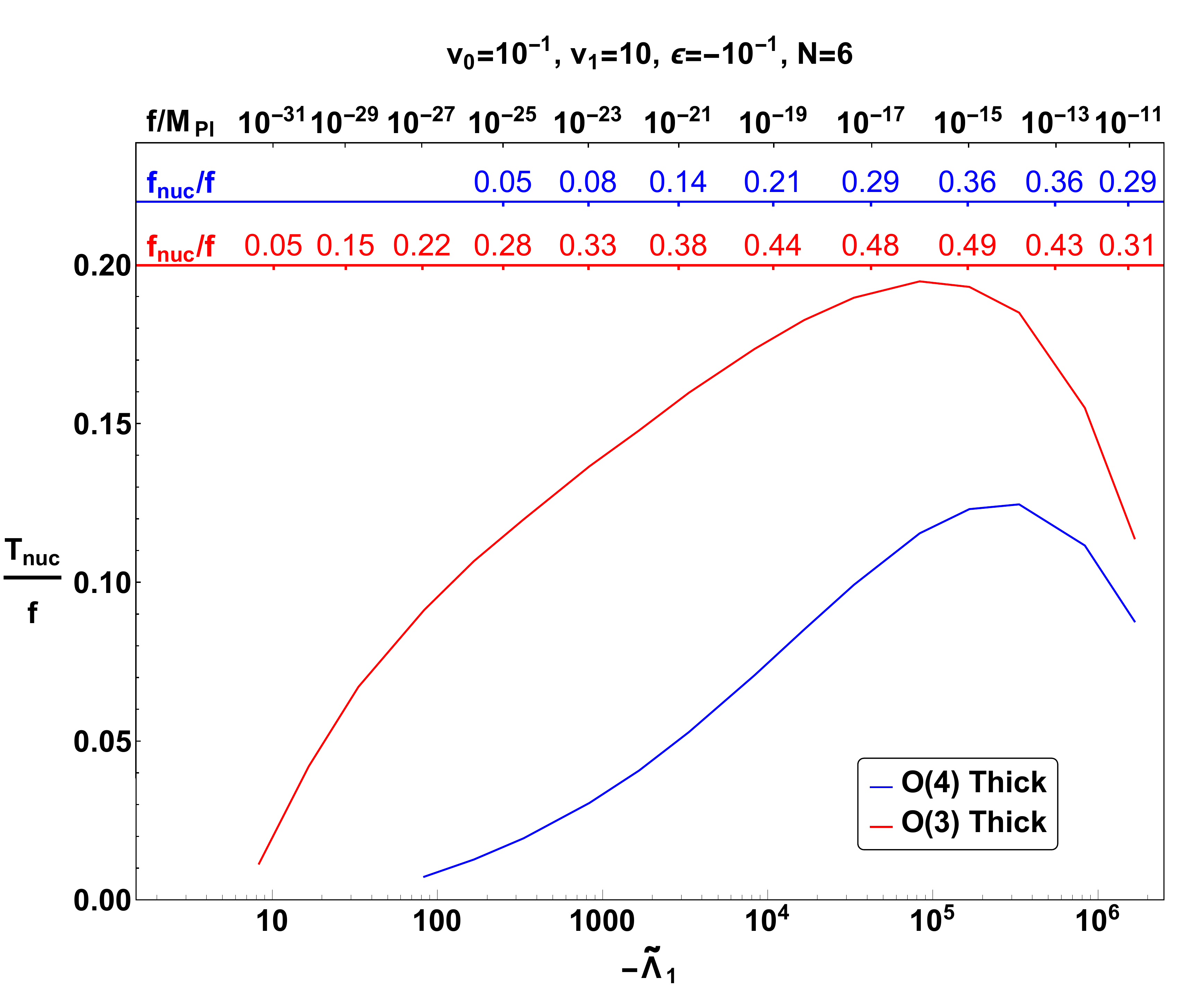}
	\includegraphics[width=.48\hsize]{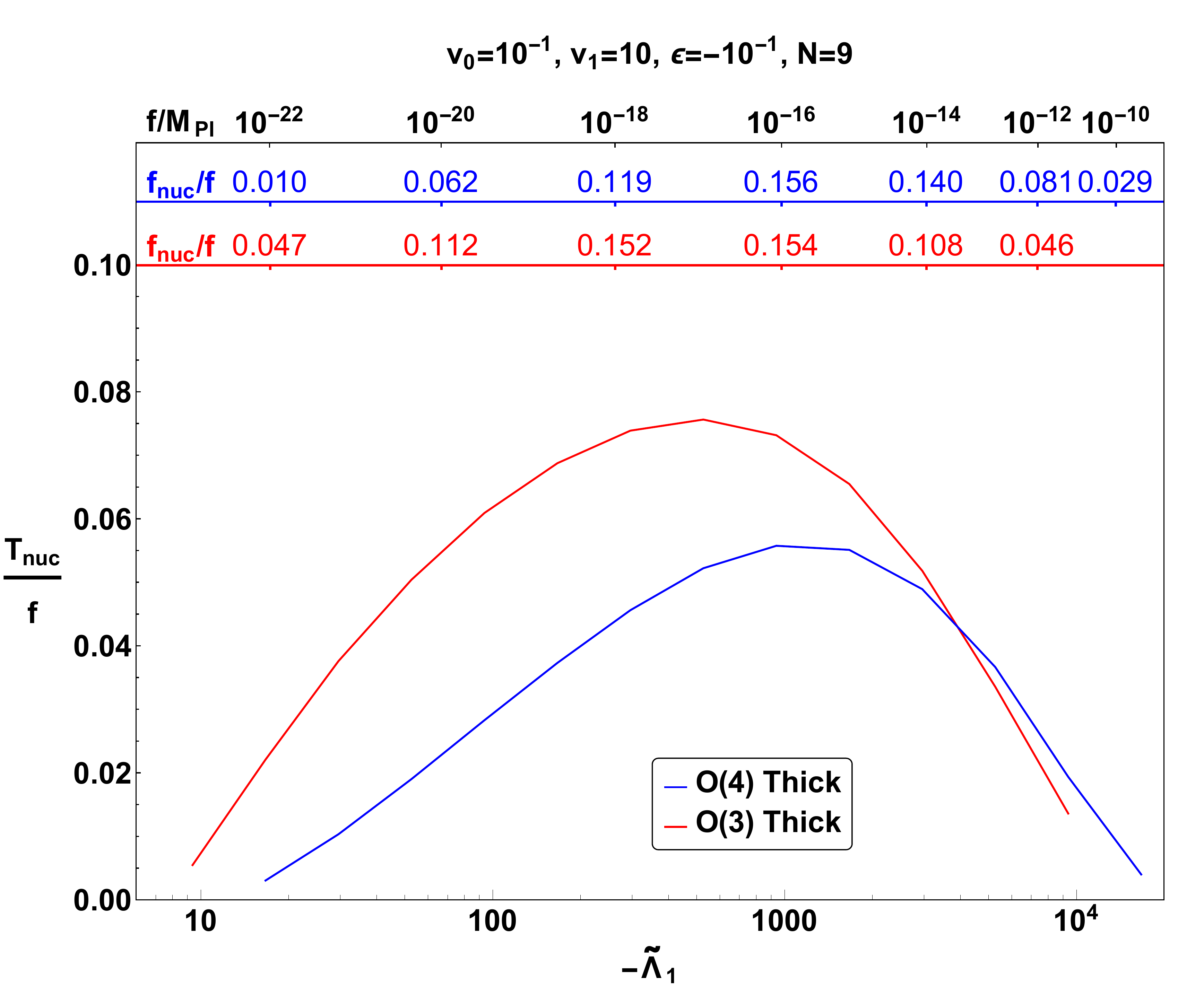}
	\includegraphics[width=.48\hsize]{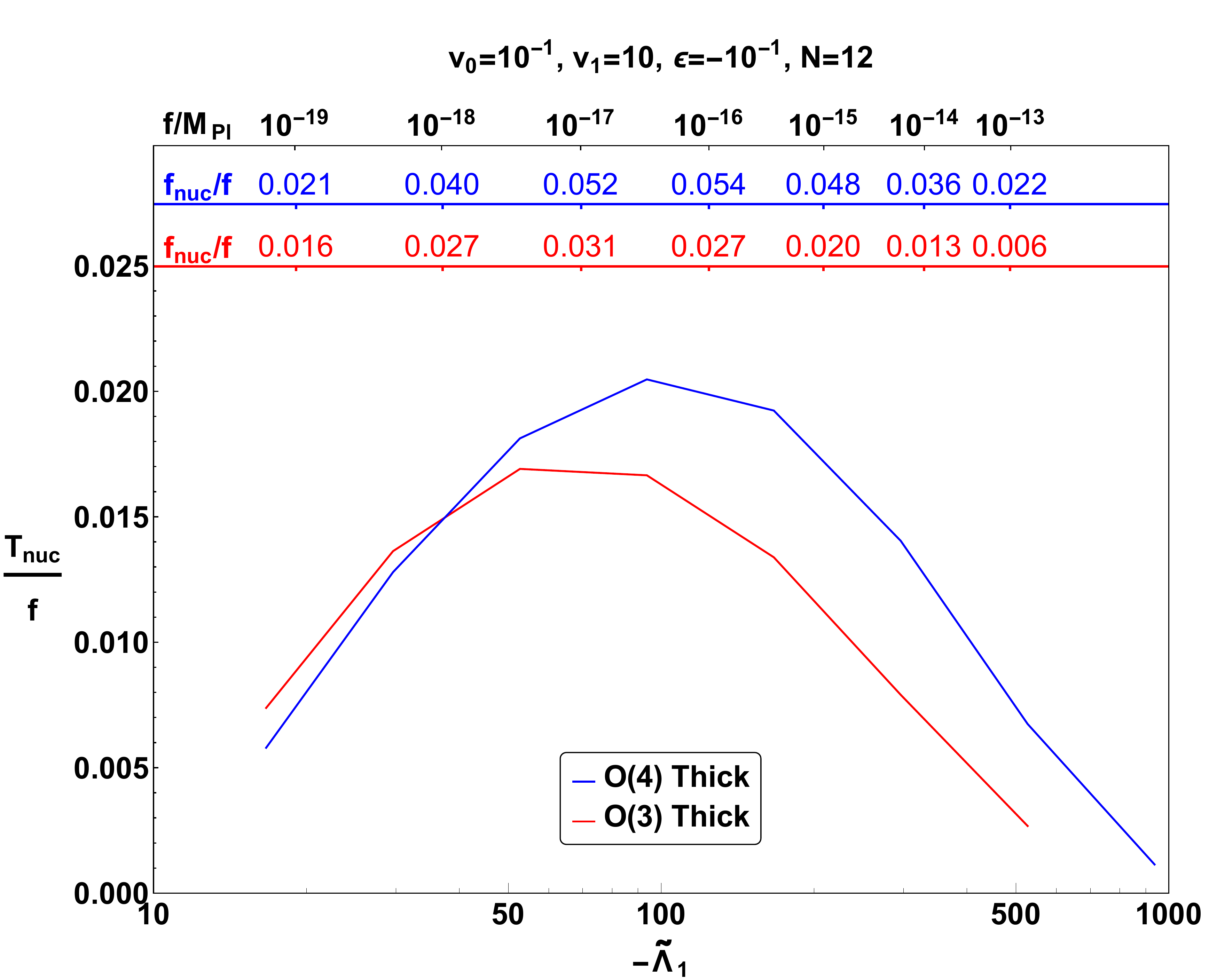}
\caption{These plot shows the nucleation properties for the soft wall with $v_1 = 10,$ and various values of $N$.  Both the temperature at which nucleation occurs and the value of $f_\text{nuc}$ tunneled to are displayed, along with the corresponding values of $f$ at the minimum of the zero temperature potential in units of the physical 4D Planck scale.}
\label{fig:nucleation}
\end{figure}

\begin{figure}[!htbp]
	\centering
	\includegraphics[width=.48\hsize]{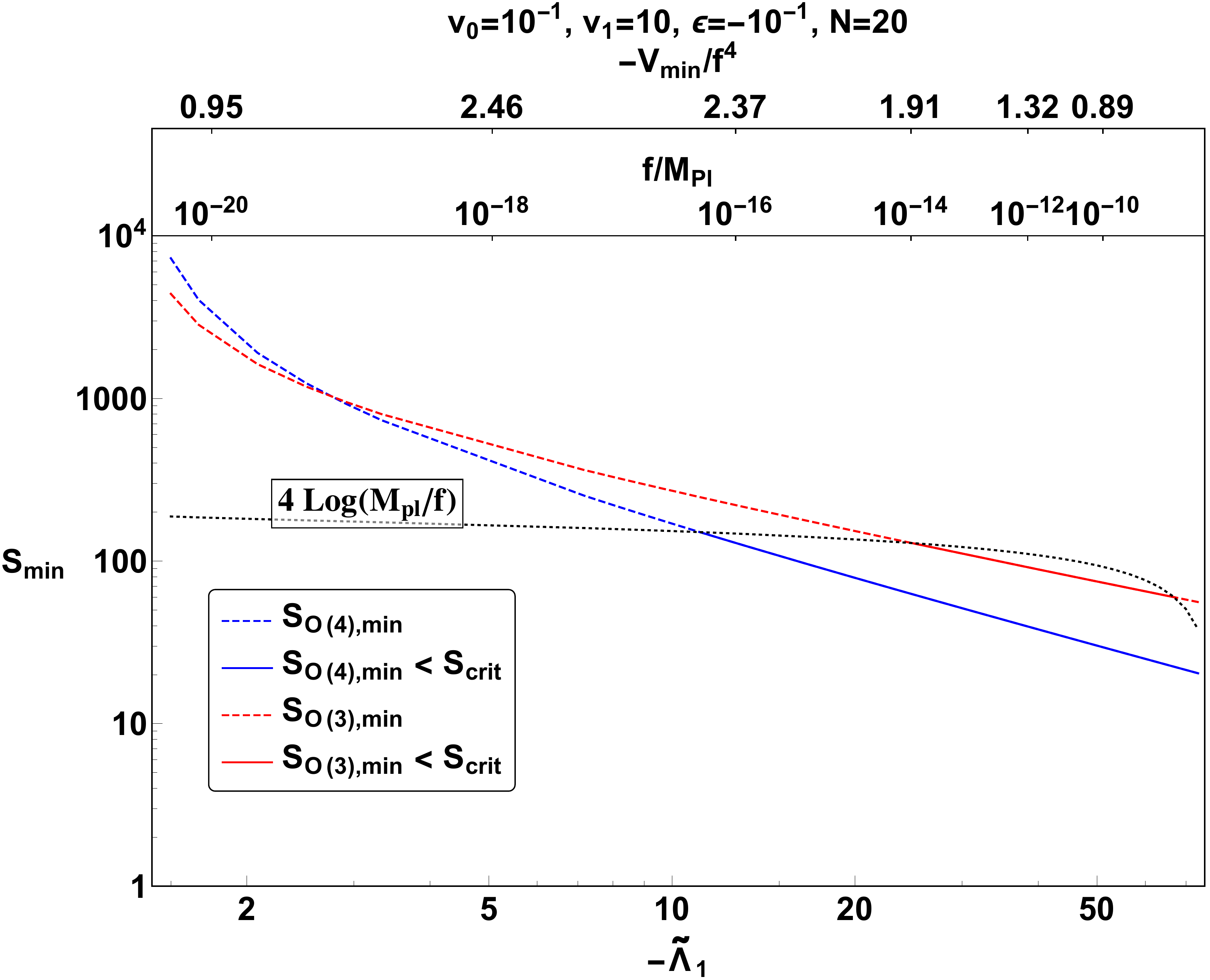}
	\includegraphics[width=.48\hsize]{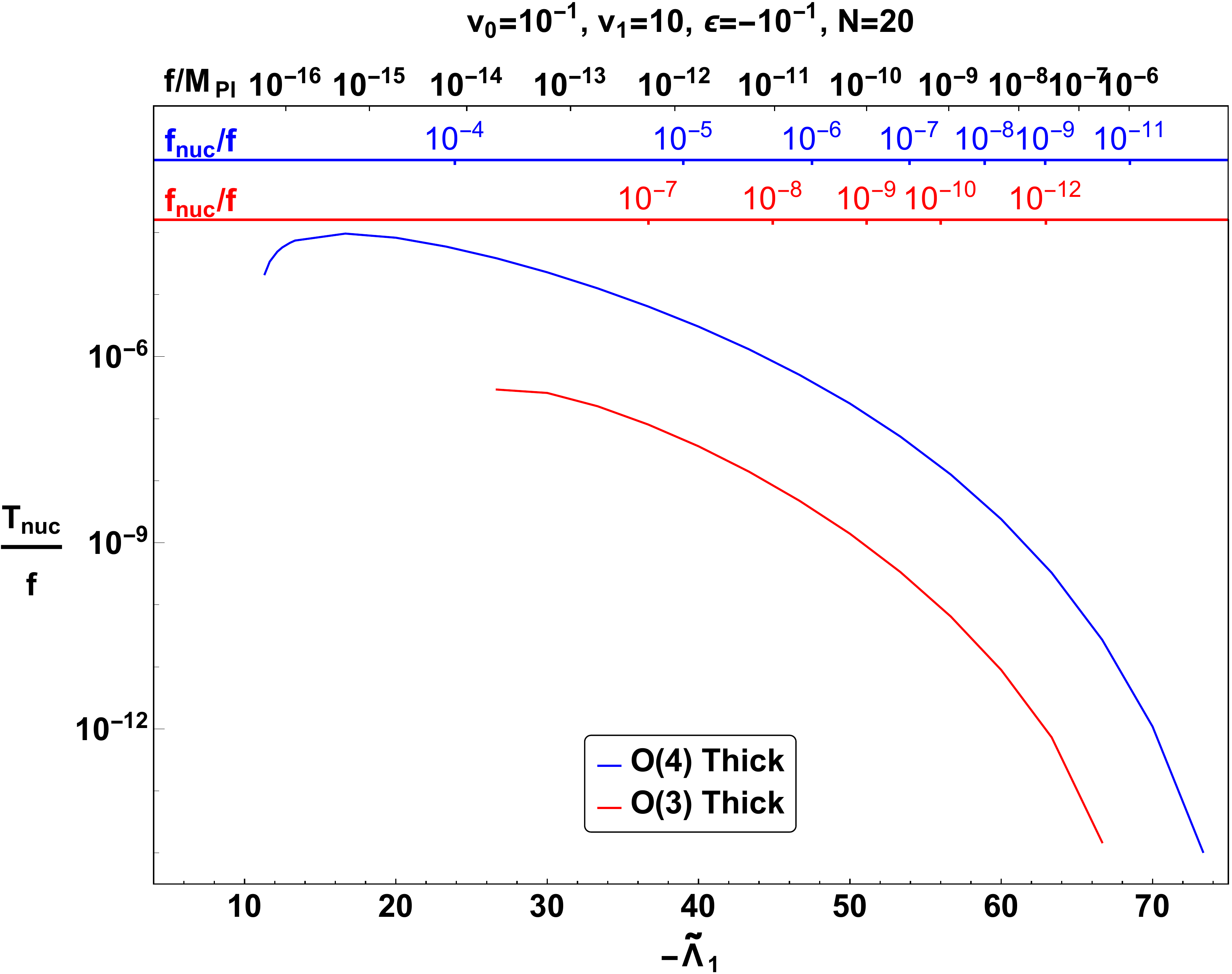}
	\caption{In this figure, we display the values of the $O(3)$ and $O(4)$ bubble actions for the special case of $N=20$.  For this large value of $N$, the phase transition is on the boundary of completing, and so the nucleation temperature, $T_n$ and the value of the nucleation condensate $f_n$ are very suppressed in relation to the value of $f$ at the minimum of the potential.}
\label{fig:N20}
\end{figure}

In Figure~\ref{fig:nucleation}, for the softwall scenario where $v_1 = 10$, we display the values of the temperature at which nucleation begins to occur for each scenario, as well as the value of $f_n$ that the system initially tunnels to.  Both $O(3)$ and $O(4)$ bubbles are shown.  Of note is the behavior of the nucleation temperature and nucleation value of the condensate.  Relative to the value of $f$ at the minimum, both of these quantities become small as the value of $N$ is increased.  
Finally, in Figure~\ref{fig:N20}, we display the bubble action and nucleation properties for the largest value of $N$ found for which a hierarchy of TeV-Planck will complete for $\epsilon = -0.1$.  For this large value of $N$ the values of $T_n/f$ and $f_n/f$ are quite small, $10^{-4}$ and $10^{-3}$ respectively.

%%%%%%%%%%%%%%%%%%%%%%%%%%%%%%
%%%%%%%%%%%%%%%%%%%%%%%%%%%%%%
%%%%%%%%%%%%%%%%%%%%%%%%%%%%%%
%%%%%%%%%%%%%%%%%%%%%%%%%%%%%%

\section{Gravitational Waves}

During a first order phase transition that proceeds by bubble nucleation, gravitational waves are sourced both by the collisions of the bubbles themselves, which break the spherical symmetry of the solutions discussed above, and also by turbulence in the finite temperature plasma as the bubbles move through it.

The stochastic spectrum is determined primarily by only coarse features of the phase transition.  The latent heat difference determines a parameter $\alpha$, which is the ratio of the latent heat compared with the energy density in the finite temperature false vacuum phase:
\begin{equation}
\alpha = \frac{ V_{T=0} (f_n) }{V_T (T = T_n)} - 1
\end{equation}

Secondly, there is a parameter which describes the rate of variation of the bubble nucleation rate.  This parameter, $\beta$, can be derived in terms of the variation of the bubble action with respect to temperature at the bubble nucleation temperature:
\begin{equation}
\frac{\beta}{H^*} = T^* \left. \frac{dS}{dT} \right|_{T^*}
\end{equation}

\begin{figure}[!htbp]
	\centering
	\includegraphics[width=.4\hsize]{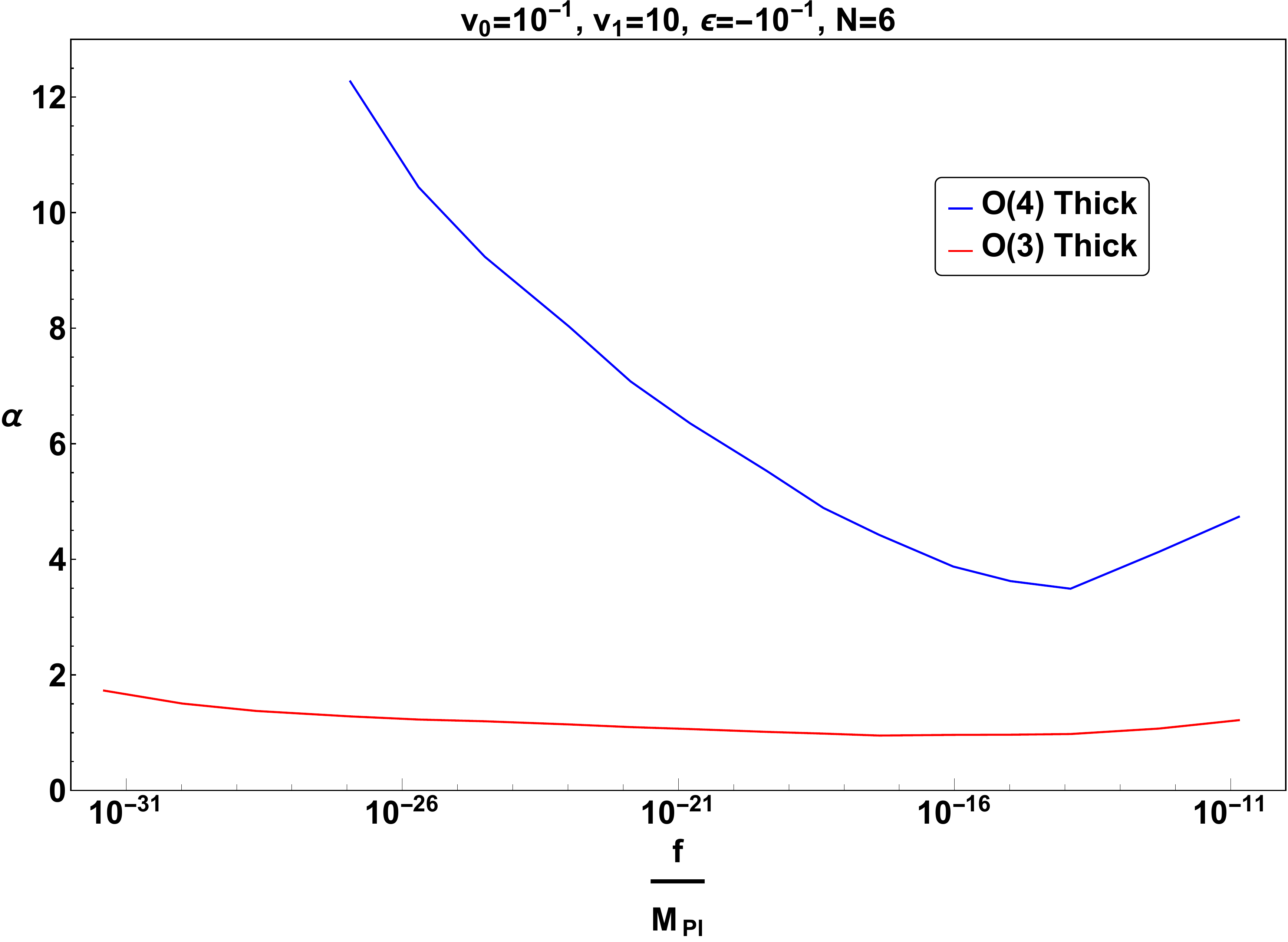}
	\includegraphics[width=.4\hsize]{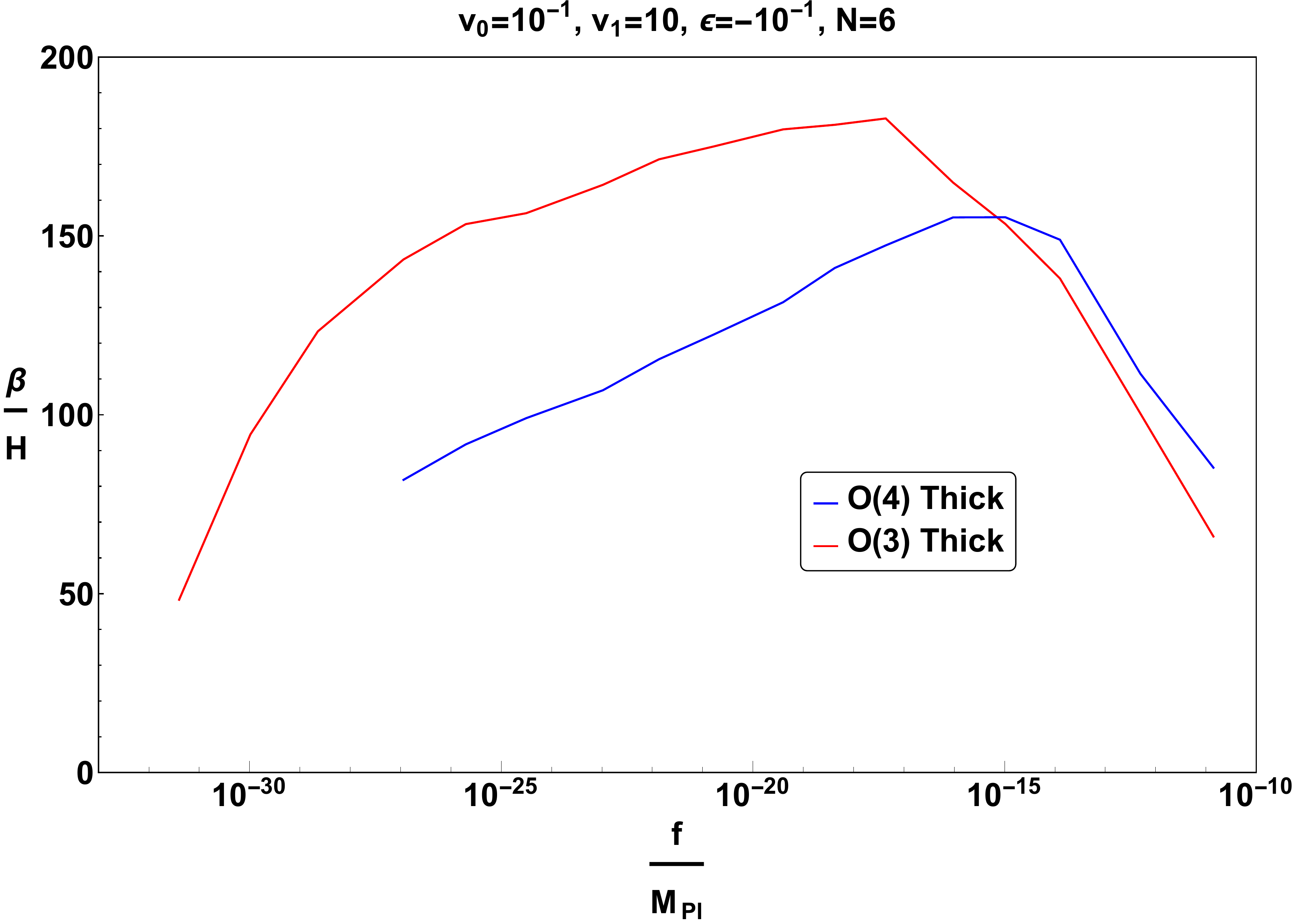}
	\includegraphics[width=.4\hsize]{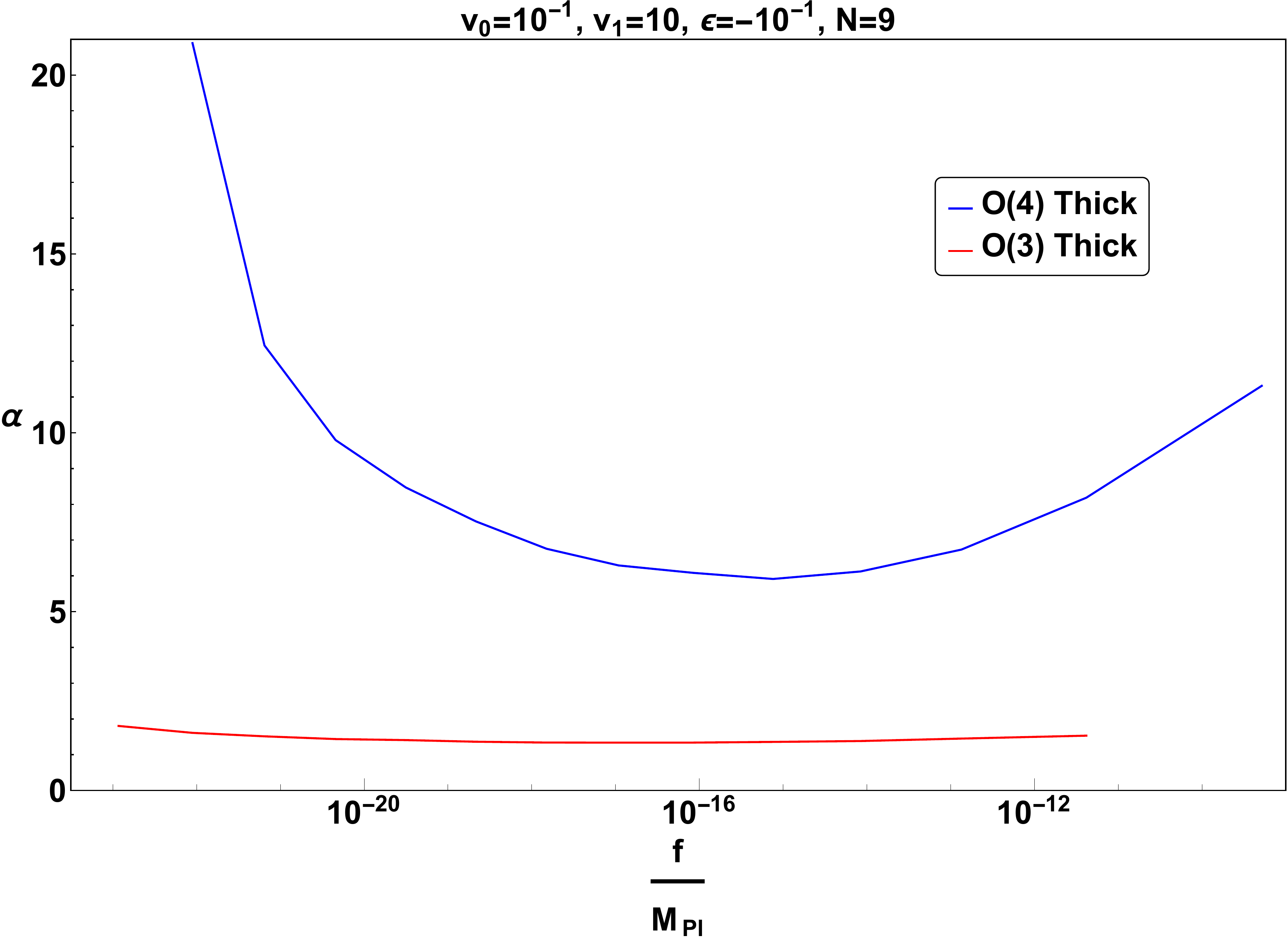}
	\includegraphics[width=.4\hsize]{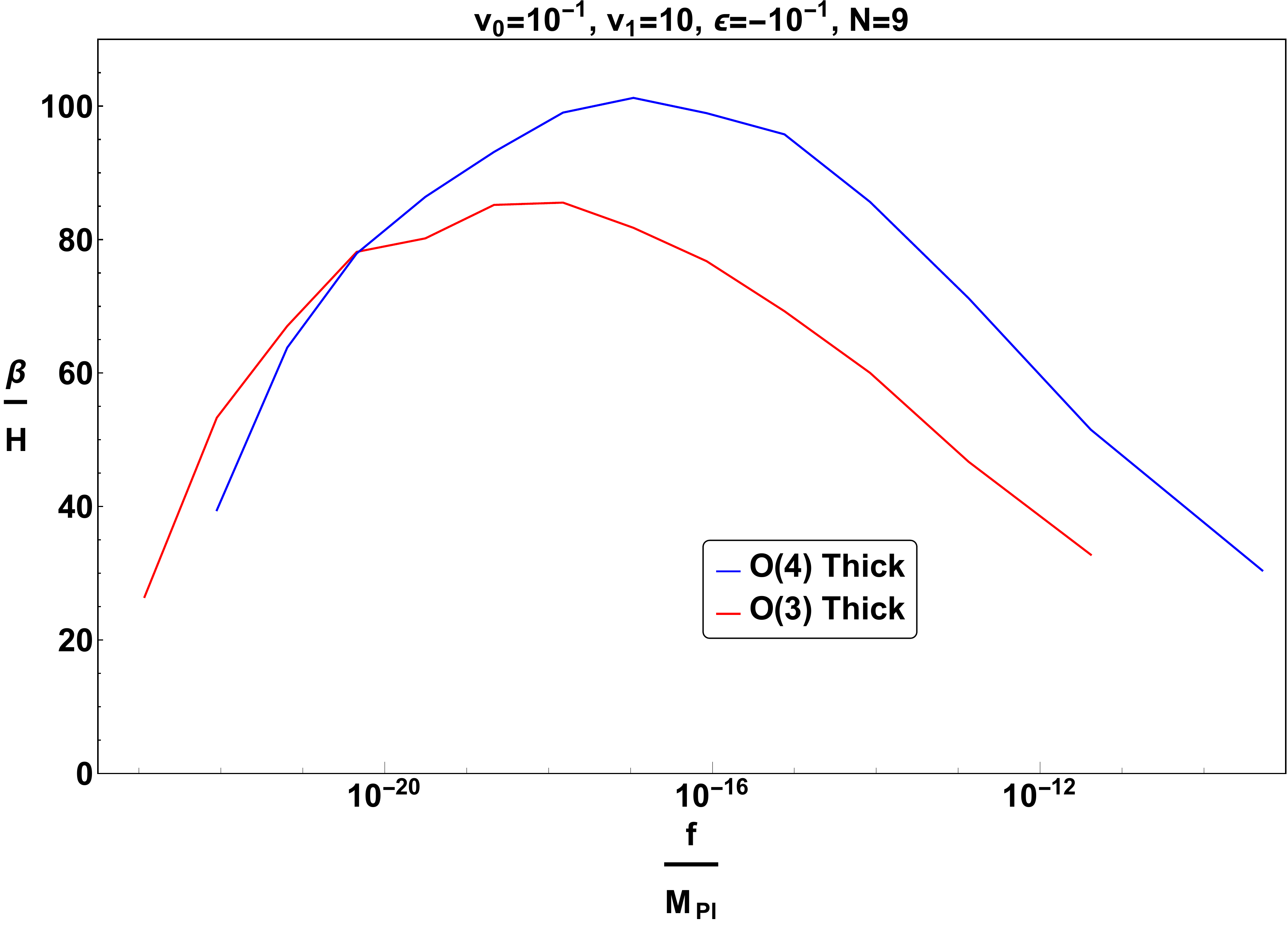}
	\includegraphics[width=.4\hsize]{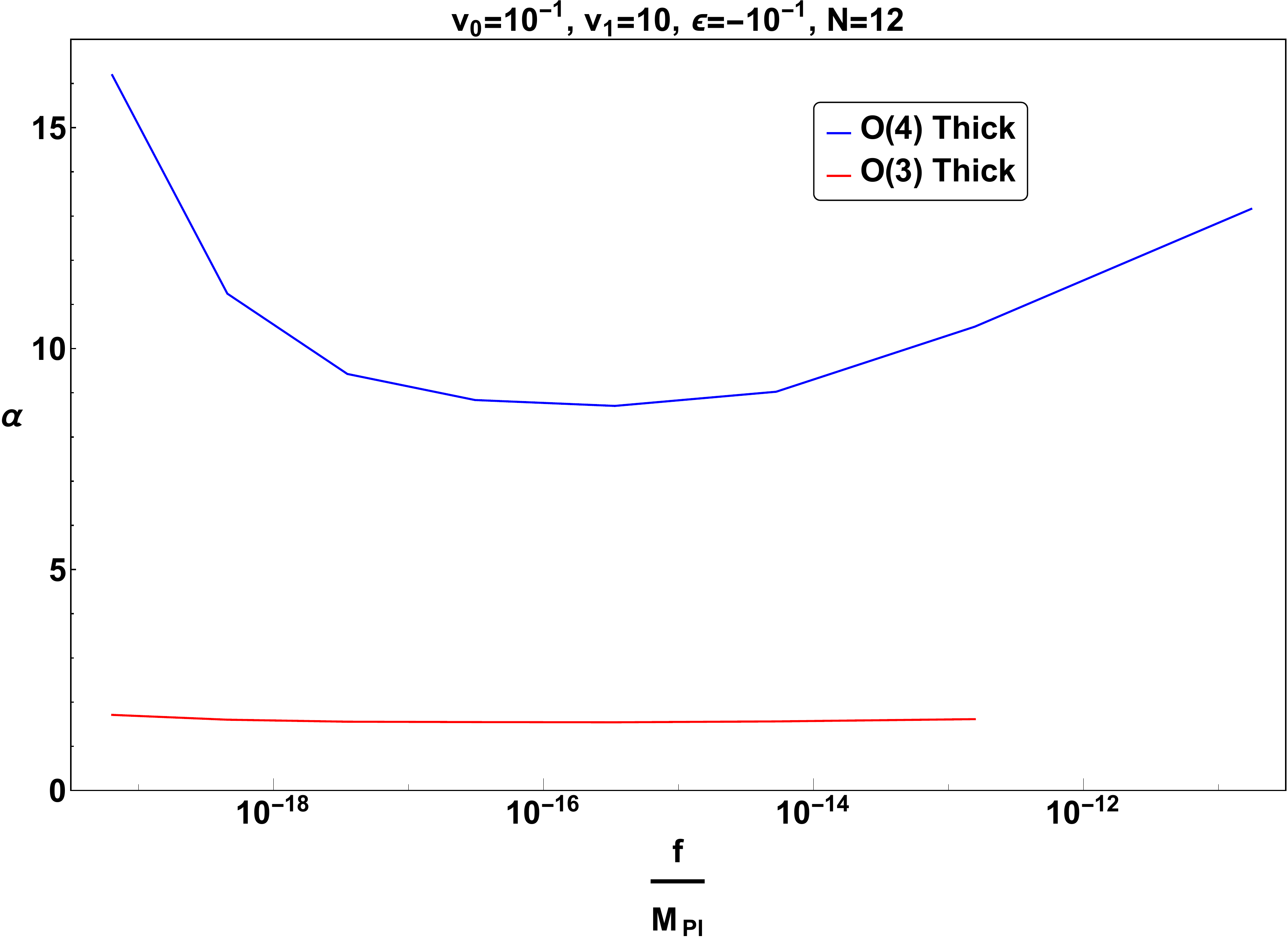}
	\includegraphics[width=.4\hsize]{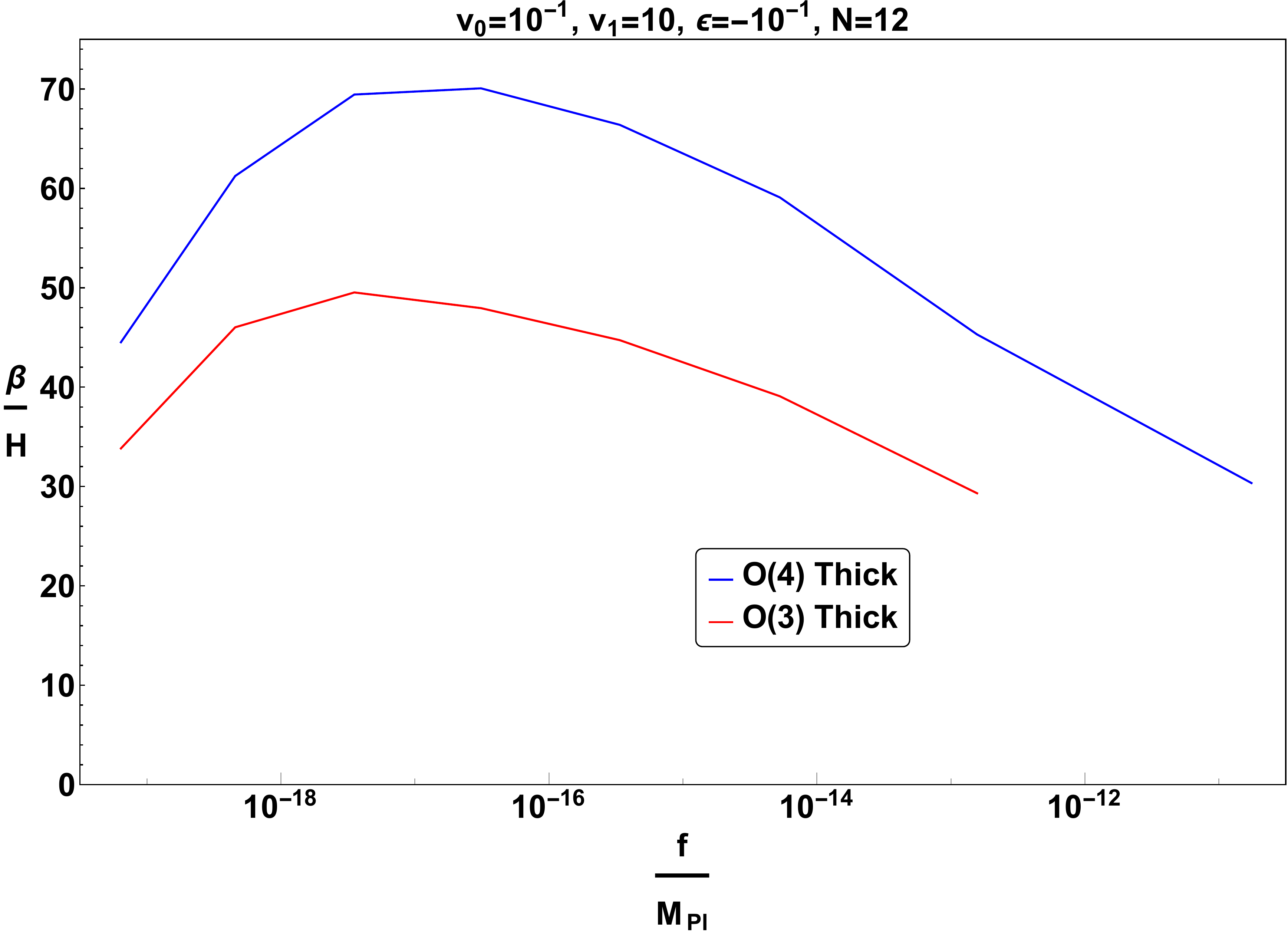}
	\includegraphics[width=.4\hsize]{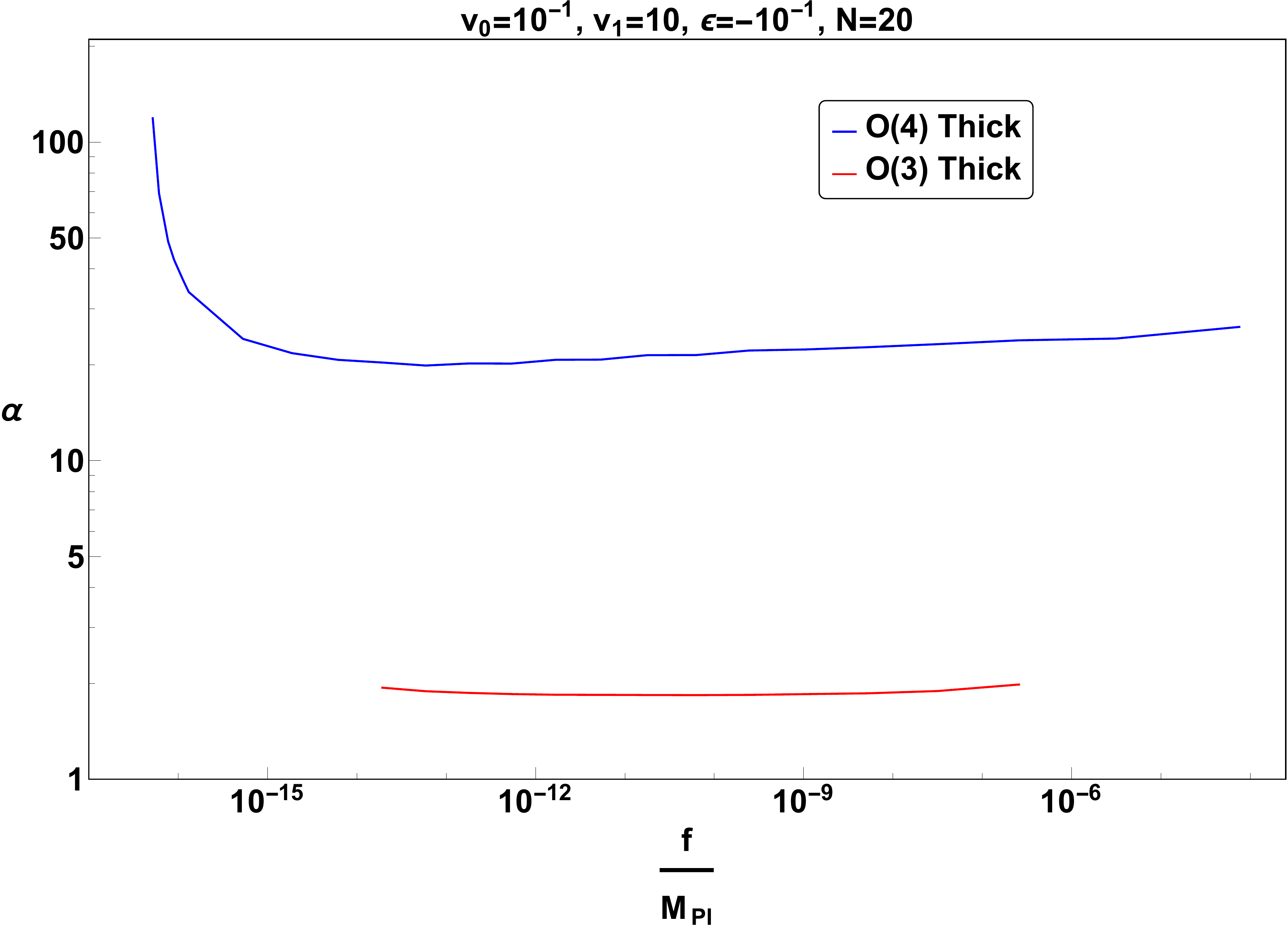}
	\includegraphics[width=.4\hsize]{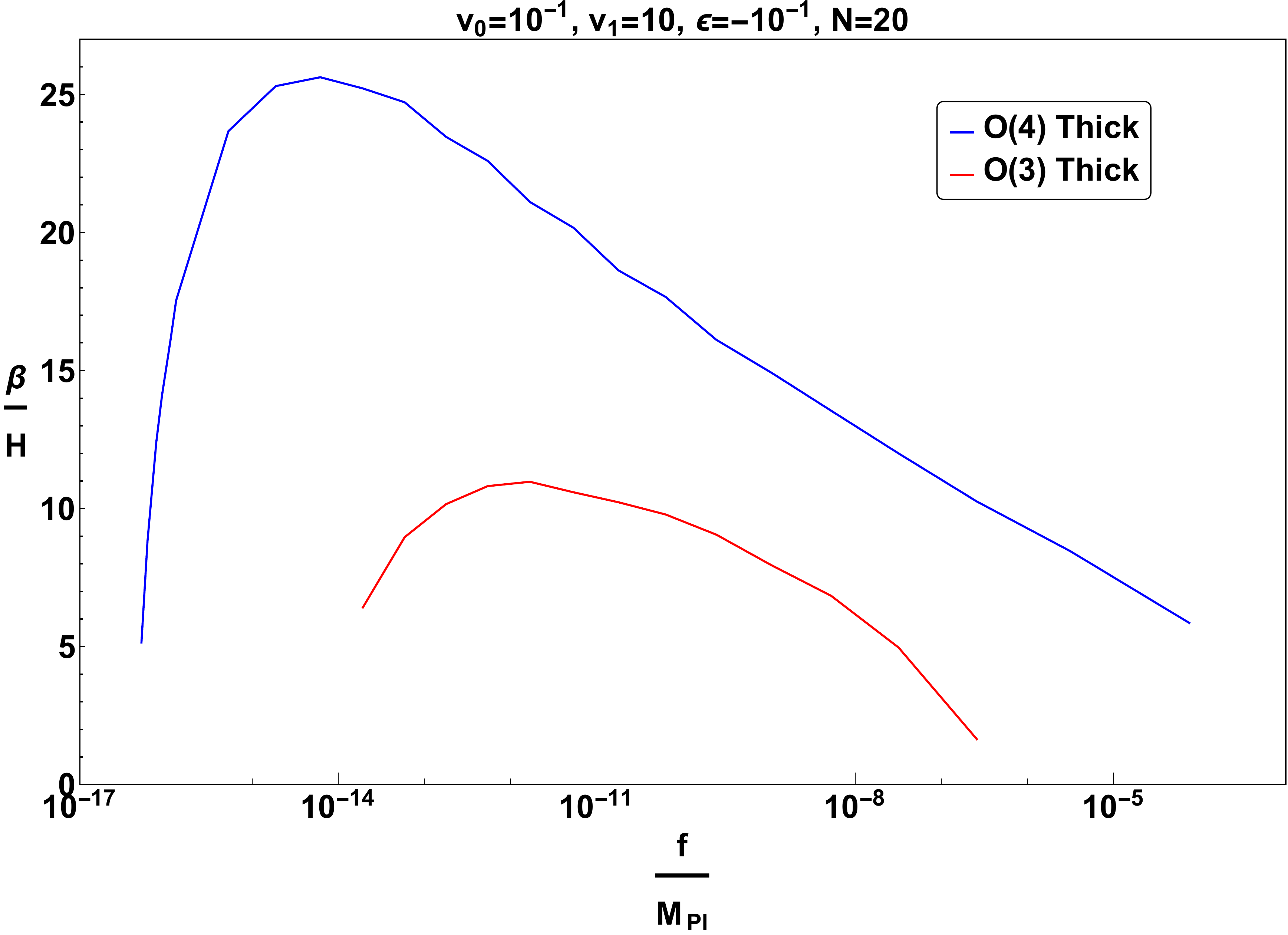}
\caption{These plots display the parameters $\alpha$ and $\beta$ for a range of $N$ as a function of the hierarchy between $f$ and $M_\text{Planck}$ at the minimum of the  zero temperature potential after the cosmological phase transition completes.  Both $O(3)$ and $O(4)$ bubbles are shown.}
\label{fig:alha}
\end{figure}

\begin{figure}[!htbp]
	\centering
	\includegraphics[width=.75\hsize]{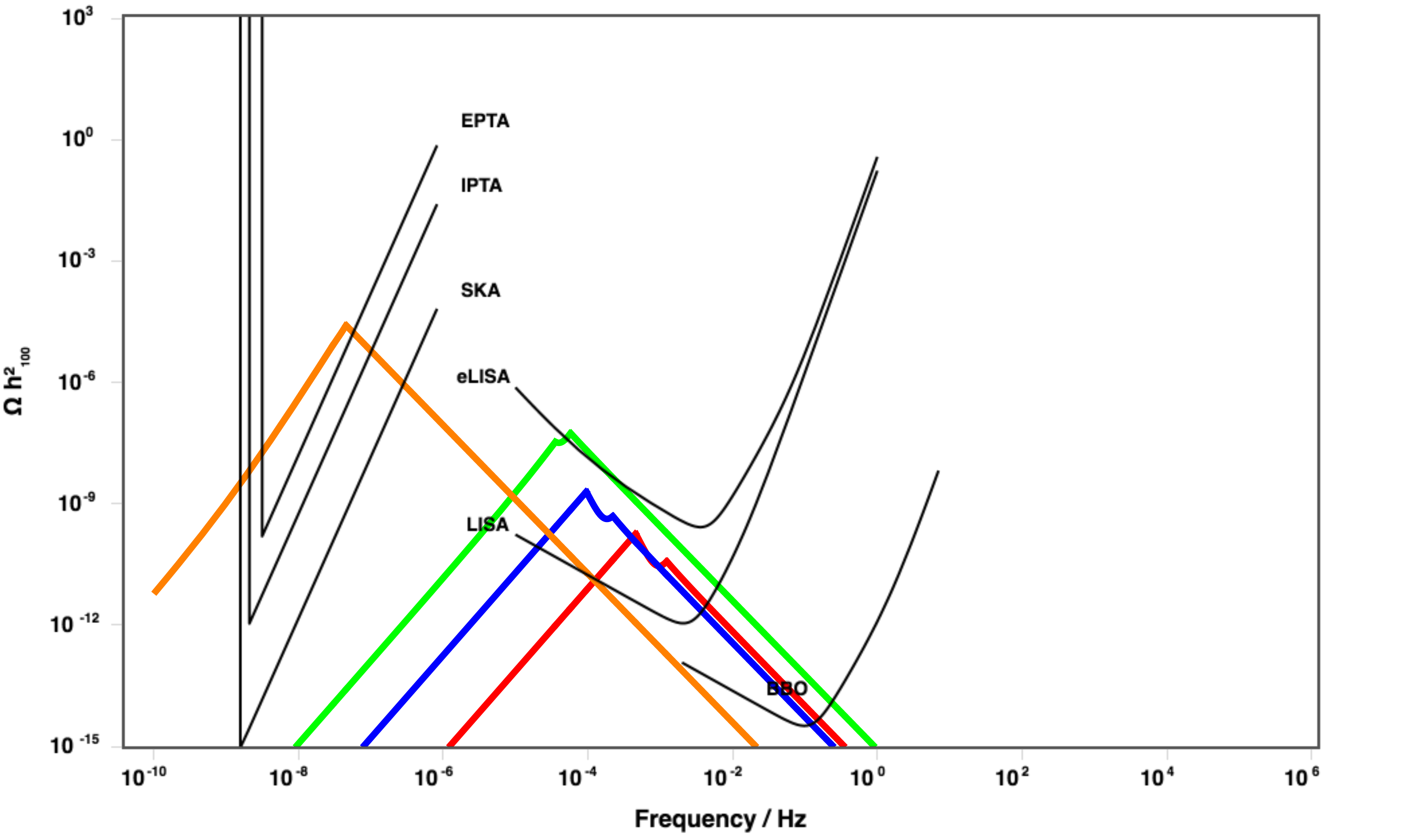}
\caption{This is the expected signal strength, expressed in terms of the gravitational wave contribution to the the energy spectrum as a function of frequency.  The parameters used in the plots are:  $\epsilon = 0.1$, $v_0 = 0.1$, $v_1 = 10$, and $f = 1$ TeV and the three different colors correspond to $N = 6$ (red), $9$ (blue),  $12$ (green), and $20$ (orange).  In this plot the signal data calculated in this work is overlaid on sensitivity curves derived from~\cite{Moore:2014lga}, and available at \url{http://rhcole.com/apps/GWplotter/}}
\label{fig:gwsig}
\end{figure}

The two types of sources of gravitational waves generate spectra with different values of the frequency at the peak of the signal, and different power laws for the fall-off on the tails of the signal.  The results of~\cite{Grojean:2006bp} are used for the purpose of calculating the characteristics of the signal.  In Figure~\ref{fig:gwsig}, we display the expected density spectrum of gravitational waves generated from the phase transition.  The signal strength increases quickly with $N$, and as the nucleation temperature decreases with increasing $N$ for fixed $f$, the higher values of $N$ have a peak at lower frequency.  Smaller values of $N$ For $N = 6$ and $N=9$ may be visible at LISA, and perhaps eLISA if our approximations are overly conservative.  For $N=20$, the frequency looks too small to be detectable at LISA, however proposed pulsar timing array experiments would probe this region with sufficient sensitivity.

Of course, the type of transition we discuss is not restricted to be at the TeV scale, although we would not be solving the electroweak hierarchy in this case.  The scale $f$ could instead be associated with other higher energy scales such as the GUT scale, or perhaps a Peccei-Quinn scale.  In this case, the signal peak would be at higher frequency.

%%%%%%%%%%%%%%%%%%%%%%%%%%%%%%
%%%%%%%%%%%%%%%%%%%%%%%%%%%%%%
%%%%%%%%%%%%%%%%%%%%%%%%%%%%%%
%%%%%%%%%%%%%%%%%%%%%%%%%%%%%%

\section{Conclusions}
This work explored solutions to 5D Einstein-Scalar theories with the goal of studying the cosmological phase transition of models with naturally large hierarchies of scale that are induced by geometric warping.  In particular, we have examined scenarios where gravitational backreaction is large, yet the theory remains approximately conformally invariant.  A numerical analysis of the finite and zero temperature potentials was performed in order to calculate the properties of the phase transition.

In early constructions of the Randall-Sundrum model, stabilized by a Goldberger-Wise mechanism with small gravitational backreaction, small values of $N$ were required to accomplish a successful cosmological phase transition.  Perturbativity of 5D gravity in these models with such small values of $N$ is in doubt.  In this work it was demonstrated that in the soft-wall construction where gravitational backreaction is taken into account, the bubble nucleation rate is faster for a given $N$, and that far larger values of $N$ can be accommodated in the sense that there is a successful early universe first order phase transition.  For the parameters we have considered, $N=20$ is near the threshold beyond which bubbles will not nucleate.

The gravitational wave signals associated with a first order transition near the weak scale, generated by collisions and turbulance in the plasma, are potentially strong enough to be visible at next generation gravitational wave observatories such as LISA (and perhaps eLISA for larger $N$) if the nucleation temperature corresponds to a peak in the spectrum not far from the targeted frequency range of these experiments.  As the nucleation temperature can be far from the value of the condensate, $f$, the peak in the spectrum can vary significantly while holding the condensate value fixed, with this separation growing with increasing $N$.

\section*{Acknowledgements}
The authors thank Alex Pomarol, Tony Gherghetta, Peter Cox, Geraldine Servant, Oriol Pujolas, Sungwoo Hong, Csaba Cs\'aki, John Terning, Javi Serra, and Brando Bellazzini for useful conversations during the preparation of this manuscript.  JH thanks the Aspen Center for physics for hospitality during the earliest stages of this work, and Cornell University for hospitality throughout.  DB thanks Hamilton College for research support during a large portion of the work on this manuscript.  JH acknowledges support from the US DOE, grant award number DE-SC0009998.  
%%%%%%%%%%%%%%%%%%%%%%%%%%%%
%%%%%%%%%%%%%%%%%%%%%%%%%%%%
%%%%%%%%%%%%%%%%%%%%%%%%%%%%
\setcounter{equation}{0}
\setcounter{footnote}{0}
%%%%%%%%%%%%%%%%%%%%%%%%%%%%%%%%%%%%%%%%%%%%%%%%%%%%%
%%%%%%%%%%%%%%%%%%%%%%%%%%%%
%%%%
\appendix
%%%%%%%%%%%%%%%%%%%%%%%%%%%%%%%%%%%%%%%%%%%%%%%%%%%%%
%%%%%%%%%%%%%%%%%%%%%%%%%%%%
%%%%

\section{Appendix:  Superpotential Method} 
\setcounter{equation}{0}
%%%%%%%%%%%%%%%%%%%%%%%%%%%%%%%%%%%%%%%%%%%%%%%%%%%%%
%%%%%%%%%%%%%%%%%%%%%%%%%%%%
%%%%
\label{app:SuperPot}

Much of the literature on the Goldberger-Wise stabilization mechanism employs a formalism referred to as the superpotential method~\cite{DeWolfe:1999cp,Csaki:2000zn}.  In this Appendix, we present an approach to the equations of motion that is based on this familiar framework.  

In the superpotential method, a function $W(\phi)$, the superpotential, that solves the first order equation
\begin{equation}
V(\phi) = \frac{1}{8} \left(\frac{\partial W}{\partial \phi} \right)^2 - \frac{\kappa^2}{6} W^2 
\end{equation}
can be used to generate solutions for $\phi$ according to the following relationship:
\begin{equation}
\dot{\phi} = \frac{3}{\kappa^2} \frac{1}{W} \frac{\partial W}{\partial \phi}.
\end{equation}
Also, the superpotential, $W$ is related to $G$:
\begin{equation}
\sqrt{G} = \frac{\kappa^2}{6} W(\phi)
\end{equation}
A superpotential that solves the above equation can be re-expressed as:
\begin{equation}
W(\phi)  = \sqrt{\frac{6}{\kappa^2}} \sqrt{-V(\phi)} \cosh\left[ \sqrt{\frac{4 \kappa^2}{3}} \left( \phi-\sigma(\phi) \right) \right],
\label{eq:Weq}
\end{equation}
and if we then have
\begin{equation}
\frac{\partial W}{\partial \phi} = \sqrt{8} \sqrt{-V(\phi)} \sinh\left[ \sqrt{\frac{4 \kappa^2}{3}} \left( \phi-\sigma(\phi) \right) \right],
\label{eq:Wpeq}
\end{equation}
then the superpotential equation is solved.  Consistency of these two equations then gives an equation of motion for $\sigma(\phi)$:
\begin{equation}
\frac{\partial \sigma}{\partial \phi} = \frac{1}{4} \frac{\partial \log V(\phi)}{\partial \phi} \sqrt{\frac{3}{\kappa^2}} \coth \left[ \sqrt{\frac{4 \kappa^2}{3}} \left( \phi-\sigma \right) \right].
\end{equation}
The solution for $\sigma$ is a trivial constant if the bulk potential is constant, and for a small deformation, $\sigma$ evolves slowly as a function of $\phi$.

In terms of the superpotential, the dilaton effective potential has the following form 
\begin{equation}
V_\text{dil} = e^{-4 y_0} \left[ V_0 (\phi_0) - W(\phi_0) \right] + e^{-4 y_1} \left[ V_1 (\phi_1) + W(\phi_1) \right] 
\end{equation}
where we can use the relation between $\phi$ and $W$ to extract the hierarchy associated with the brane separation:
\begin{equation}
y_1 -y_0= \frac{\kappa^2}{3} \int_{\phi_0}^{\phi_1} \frac{W}{\left( \frac{ \partial W}{\partial \phi} \right)} d\phi =\sqrt{\frac{\kappa^2}{12}} \int_{\phi_0}^{\phi_1} \coth \left[ \sqrt{\frac{4 \kappa^2}{3}} \left( \phi- \sigma (\phi) \right) \right] d\phi
\end{equation}
The value of the condensate is then given by
\begin{equation}
f^{-1}  =2 \int_{\phi_0}^{\phi_1} \frac{ \exp\left[ y(\phi) \right]}{\partial W/\partial\phi} d \phi
\end{equation}
where $y(\phi)$ is given by the integral equation above.

%%%%%%%%%%%%%%% BRIEF ANALYTICS FOR DRAFT  %%%%%%%%%%%%%%%%%%%%%%%%%%%%%%%%%%%%%%%%%%%%%%%%%%%%%%
%%%%%%%%%%%%%%%%%%%%%%%%%%%%%%%%%%%%%%%%%%%%%%%%%%%%%%%%%%%%%%%%%%%%%%%%%%%%%%
%%%%%%%%%%%%%%%%%%%%%%%%%%%%%%%%%%%%%%%%%%%%%%%%%%%%%%%%%%%%%%%%%%%%%%%%%%%%%%

\section{Appendix: Approximate Analytic Results}

While a full numerical analysis was needed to draw accurate conclusions regarding the phase structure of the  models under consideration, there are analytical expressions that give insight into the results.  In this appendix, we present these approximate analytic solutions and visually compare them to full numerical results, and, for comparison, to solutions used in the literature for models that have only small backreaction.

To begin, we observe that for the case of a constant bulk potential, the equation of motion for $ \tilde{\phi}  $ reduces to 
\beq
\ddot{\tilde{\phi}} = 4 \dot{\tilde{\phi}} \left(1 - \frac{1}{12}  \dot{\tilde{\phi}}^2 \right).
\eeq
It can be shown that starting in the UV and moving to the IR, this equation has asymptotic behavior where $\dot{\phi}\approx 0$ in the UV, and $\dot{\phi} \approx \sqrt{12}$ in the IR.   

For the case of a nontrivial bulk potential the equation of motion is 
 \beq
\ddot{\tilde{\phi}} = 4\left( \dot{\tilde{\phi}} - \frac{3}{2 } \frac{\partial \log V (\tilde{\phi})}{\partial \tilde{\phi}} \right) \left(1 - \frac{\kappa^2}{12}  \dot{\tilde{\phi}}^2 \right)
\eeq
For a bulk potential which depends only mildly on $\tilde{\phi}$ and is polynomial in $\tilde{\phi}$, the potential term is always suppressed.  For the case of a quadratic potential, $V(\tilde{\phi})= \Lambda(1+\frac{\epsilon}{3} \tilde{\phi}^2)$, when the backreaction and the value of $\phi$ are both small, the second derivative term goes like $\epsilon^2$, and can be ignored in a leading approximation.  When $\phi$ is large, the term is again suppressed as the term goes like $1/\tilde{\phi}$ in this limit, and $\ddot{\tilde{\phi}}$ is again especially small.  The equation is then relatively simple in cases where the potential term is small:
$\dot{\tilde{\phi}} - \frac{3}{2 } \frac{\partial \log V (\tilde{\phi})}{\partial \tilde{\phi}} \approx 0$.

\begin{figure}[h!]
	\center
	\includegraphics[width=.474\textwidth]{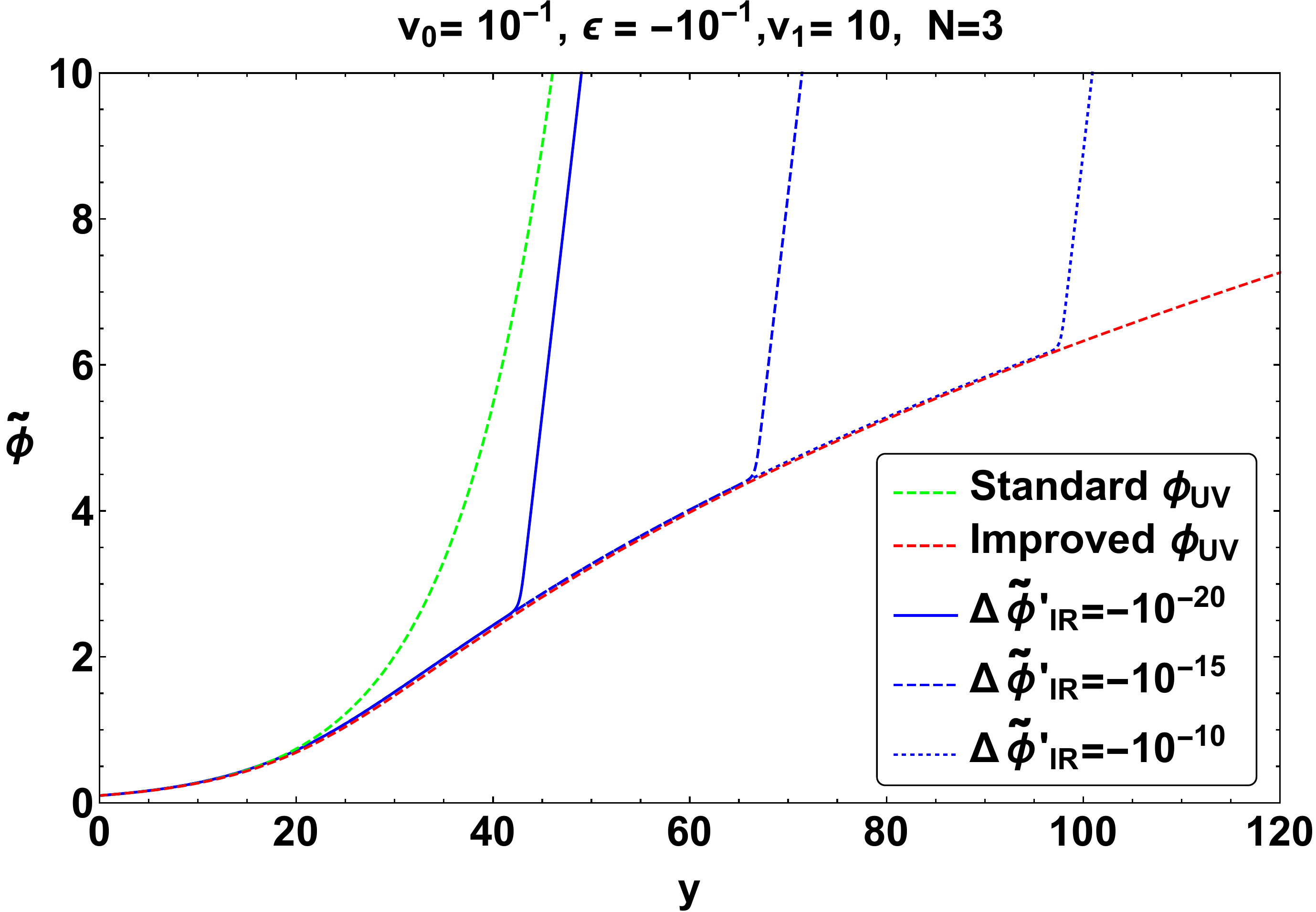}
	\includegraphics[width=.506\textwidth]{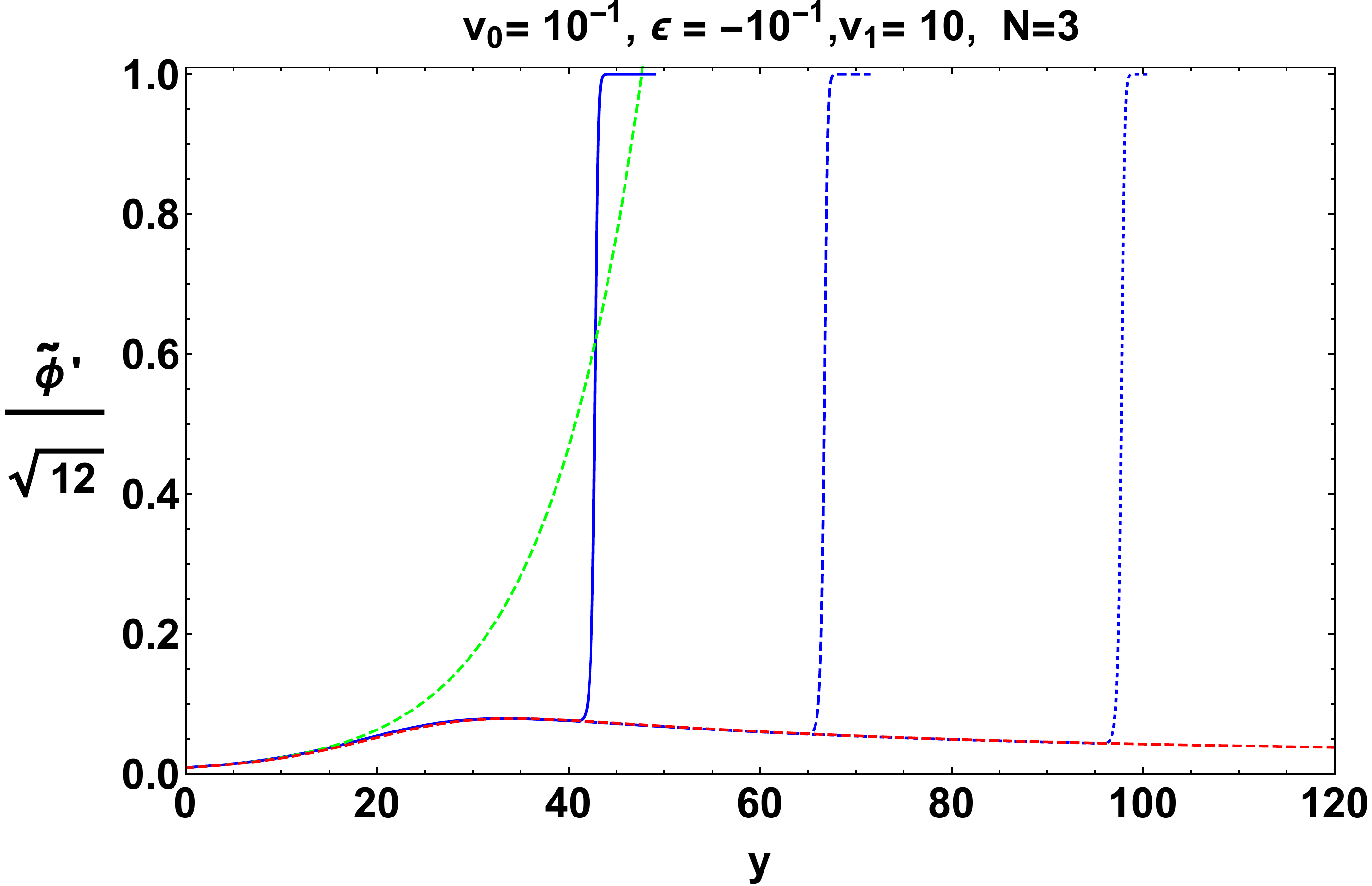}
	\caption{Plots for $\tilde{\phi}$  and its derivative. Green is the lowest order UV solution, $\tilde{\phi}_0 e^{\epsilon y} $, red is the analytic UV solution we are presenting, and blue are various full numerical results for the solution parameterized by $\Delta \tilde{\phi}'_{IR}  \equiv \sqrt{12}-  \tilde{\phi}'(y_{IR}) $. The agreement in the deep UV is good for all cases, as expected, but the lowest order UV solution disagrees severely.  Our approximate UV solution outlined above continues to track the unstable UV fixed point up until the IR condensate region develops. }
	\label{fig:phi_plot}
\end{figure}

For the case of the quadratic potential, we have $\dot{\tilde{\phi}} -  \frac{ \epsilon \tilde{\phi} }{ 1+ \frac{\epsilon}{3} \tilde{\phi}^2   } \approx 0$ which has an exact solution:
\beq
\tilde{\phi} (y)_\text{UV} = \sqrt{ \frac{3}{\epsilon}   PL \left[ \frac{\epsilon  \tilde{\phi}_\text{UV}^2 }{3} \exp  \left(  \epsilon( 6(y-y_0)+ \tilde{\phi}_\text{UV}^2 ) \right)       \right]   }
\eeq
where we have imposed the boundary condition $\tilde{\phi}(y_0)_\text{UV} \equiv \tilde{\phi}_{UV}$ and PL is the product log function. 
This UV solution can be contrasted with the lowest order UV solution by expanding the UV fixed point equation as  $\dot{\tilde{\phi}} -   \epsilon \tilde{\phi} = 0$ and then solving to obtain 

\beq
\tilde{\phi} (y)_\text{UV} = \tilde{\phi}_\text{UV} \exp(\epsilon y ).
\eeq
One could also obtain this solution from expanding the correct leading-order result to lowest order in $  \epsilon \tilde{\phi}_\text{UV}^2 $. We will refer to this as the lowest order UV solution for $\tilde{\phi}$, which is the solution for the Golberger-Wise field in the UV commonly found in the literature. While this solution is sufficient for the hardwall scenario as $\tilde{\phi} $ is always $\mathcal{O}(1)$, it is insufficient for a UV region that allows moderately larger values of $\tilde{\phi}$ since in that case $\epsilon \tilde{\phi}_{UV}^2 $ is not a small dimensionless quantity, and starts to dominate over the UV value of the 5D cosmological constant term.  A comparison between the lowest order solution and our UV solution, along with some exact numerical results are displayed in Figure~\ref{fig:phi_plot}. 

\begin{figure}[h!]
	\center
	\includegraphics[width=.474\textwidth]{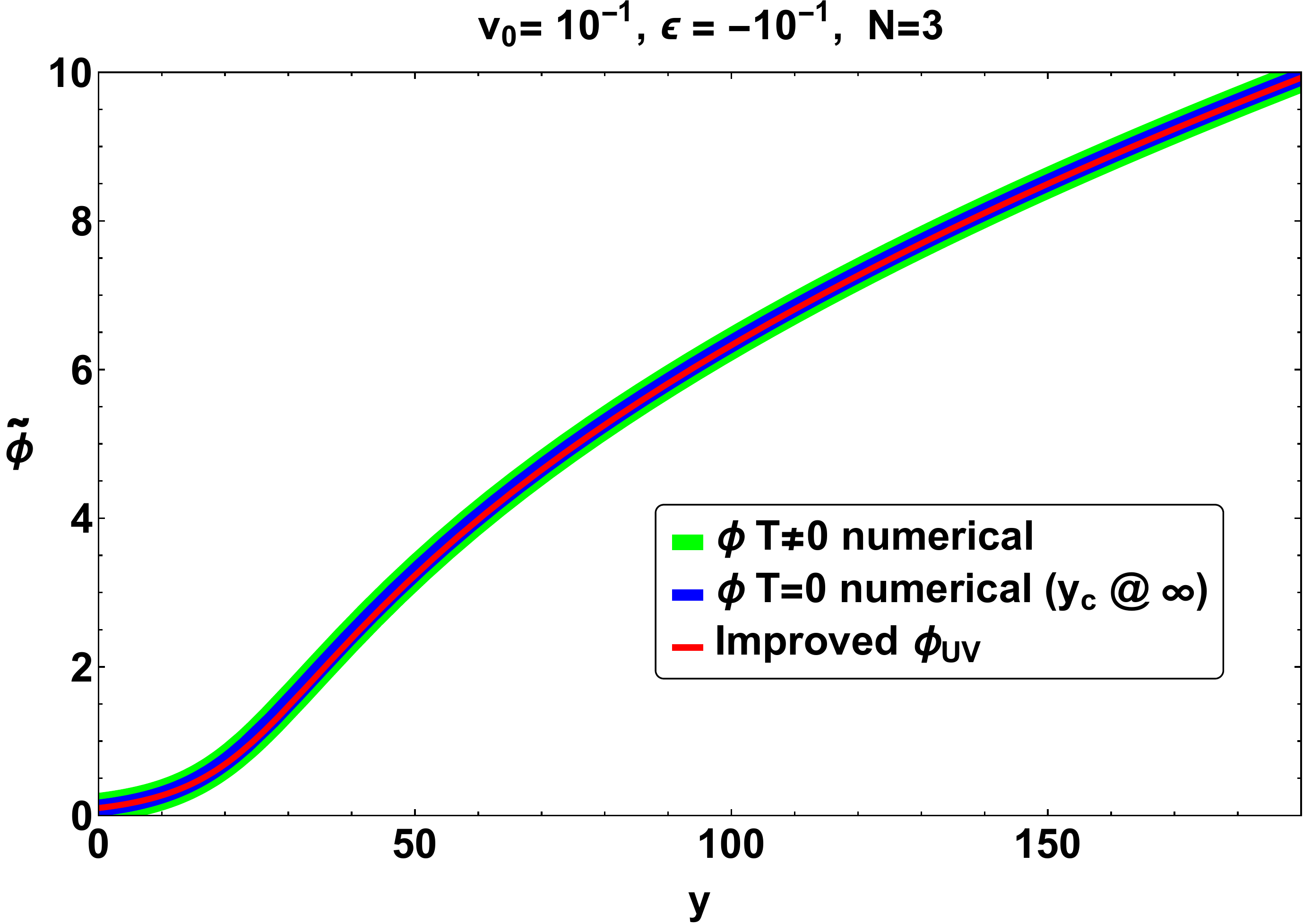}
	\caption{A comparison between the $\phi$ solution in the presence of a horizon and then with no horizon (numerical and analytical).	Since the boundary condition for $\phi$ at the horizon is the equation of motion itself to leading order, a horizon can be placed at any point along this trajectory which nominally effects $\phi$ numerically, and only at next-to-leading-order analytically.}	\label{fig:phi_hor_plot}
\end{figure}

We further note that the solution for the scalar field in the finite temperature scenario where there is a black hole horizon also allows an approximate analytic solution.  The near horizon limit of the $\tilde{\phi}$ equation of motion produces the boundary condition
\beq
\dot{\tilde{\phi}} |_{y_h}= \frac{3}{2 } \frac{\partial \log V (\tilde{\phi})}{\partial \tilde{\phi}} |_{y_h}
\eeq
but this is the same relation employed as an approximation in the UV region of the zero temperature solution.  This means that in order to accommodate a horizon, $\tilde{\phi}$ never moves far from the approximate analytic solution studied above.  A comparison between $\tilde{\phi}$ in the presence of a horizon function and with no horizon function are plotted in figure \ref{fig:phi_hor_plot}. 

\end{document}